\documentclass[aps,showkeys,superscriptaddress,groupedaddress,nofootinbib]{revtex4-2}  
\usepackage{cancel}
\usepackage{graphicx} 
\usepackage[T1]{fontenc}
\usepackage{lmodern} 
\usepackage{dcolumn}   
\usepackage{bm}        
\usepackage{amssymb}   
\usepackage{amsmath}
\usepackage{bbold}
\usepackage{siunitx}
\usepackage{array}
\usepackage{float}
\usepackage{makecell}
\usepackage{braket}
\usepackage{longtable}
\usepackage{supertabular,booktabs}
\usepackage{soul}
\usepackage{dutchcal}
\usepackage{bm}  

\def \bea {\begin{eqnarray}}
\def \eea {\end{eqnarray}}

\usepackage{mathrsfs}  

\usepackage{titlesec} 
\usepackage[colorlinks,citecolor=blue,urlcolor=blue,hypertexnames=true]{hyperref}
\usepackage{subfigure}

\usepackage[usenames,dvipsnames,svgnames,table]{xcolor} 

\newcommand{\bml}{\begin{subequations}}
\newcommand{\eml}{\end{subequations}}
\newcommand{\bfig}{\begin{figure}}
\newcommand{\efig}{\end{figure}}

\newcommand{\bmat}{\begin{pmatrix}}
\newcommand{\emat}{\end{pmatrix}}
\usepackage{slashed,booktabs, multirow, float,
amsfonts, bbold, mathtools, sidecap, tikz, bm,enumitem}
\usepackage{multirow}
\usepackage{bbding}
\usepackage{titlesec}
\usepackage{hyperref}
\usepackage{wasysym}
\usepackage{amssymb}
\usepackage{pifont}

\renewcommand{\leq}{\leqslant}

\usepackage[dvipsnames, usenames]{xcolor}

\definecolor{linkcolor}{rgb}{0.55, 0.13, .32}

\definecolor{oucrimsonred}{rgb}{0.6, 0.0, 0.0}
\definecolor{persianblue}{rgb}{0.11, 0.22, 0.73}
\definecolor{forestgreen}{rgb}{0.13,0.35,0.13}
\definecolor{lightgray}{rgb}{0.83, 0.83, 0.83}
 \hypersetup{colorlinks, citecolor=oucrimsonred, linkcolor=persianblue, urlcolor=oucrimsonred}
\definecolor{cornellred}{rgb}{0.7, 0.11, 0.11}
\definecolor{navyblue}{rgb}{0.0, 0.0, 0.5}
\definecolor{amethyst}{rgb}{0.6, 0.4, 0.8}
\definecolor{yellow}{rgb}{1.0, 1.0, 0.0}
\definecolor{firebrick}{rgb}{0.7, 0.13, 0.13}
\definecolor{tangerineyellow}{rgb}{1.0, 0.8, 0.0}
\definecolor{deepfuchsia}{rgb}{0.76, 0.33, 0.76}
\definecolor{amber}{rgb}{1.0, 0.75, 0.0}
\definecolor{VioletRed4}{rgb}{0.55, 0.13, .32}
\definecolor{indiagreen}{rgb}{0.07, 0.53, 0.03}
\definecolor{VioletRed4}{rgb}{0.55, 0.13, .32}

\usepackage{tcolorbox}

\definecolor{oucrimsonred}{rgb}{0.6, 0.0, 0.0}
\definecolor{persianblue}{rgb}{0.11, 0.22, 0.73}
\definecolor{forestgreen}{rgb}{0.13,0.35,0.13}
\definecolor{lightgray}{rgb}{0.83, 0.83, 0.83}
 \hypersetup{colorlinks, citecolor=oucrimsonred, linkcolor=persianblue, urlcolor=oucrimsonred}
\definecolor{cornellred}{rgb}{0.7, 0.11, 0.11}
\definecolor{navyblue}{rgb}{0.0, 0.0, 0.5}
\definecolor{amethyst}{rgb}{0.6, 0.4, 0.8}
\definecolor{yellow}{rgb}{1.0, 1.0, 0.0}
\definecolor{firebrick}{rgb}{0.7, 0.13, 0.13}
\definecolor{tangerineyellow}{rgb}{1.0, 0.8, 0.0}
\definecolor{deepfuchsia}{rgb}{0.76, 0.33, 0.76}
\definecolor{amber}{rgb}{1.0, 0.75, 0.0}
\definecolor{VioletRed4}{rgb}{0.55, 0.13, .32}
\definecolor{indiagreen}{rgb}{0.07, 0.53, 0.03}
\definecolor{VioletRed4}{rgb}{0.55, 0.13, .32}
\newcommand{\nn}{\nonumber}

\definecolor{oucrimsonred}{rgb}{0.6, 0.0, 0.0}
\newcommand\vertarrowbox[3][6ex]{%
  \begin{array}[t]{@{}c@{}} #2 \\
  \left\uparrow\vcenter{\hrule height #1}\right.\kern-\nulldelimiterspace\\
  \makebox[0pt]{\scriptsize#3}
  \end{array}%
}

\definecolor{mtcolor}{rgb}{.8,.3,.1}

\definecolor{violachiaro}{rgb}{1,0.6,1}

\definecolor{gbcolor}{rgb}{.43,.22,.12}
 
\definecolor{gbcolor2}{rgb}{.9,.2,.6}
\definecolor{gbcolor3}{rgb}{.3,.2,.6}

\definecolor{verdechiaro}{rgb}{0.6,1,0.6}
\definecolor{giallochiaro}{rgb}{1,1,0.6}
\definecolor{bluscuro}{rgb}{0.15, 0.2, 0.9}
\definecolor{verdes}{rgb}{0.1, 0.5, 0.1}%
\definecolor{tangerineyellow}{rgb}{1.0, 0.8, 0.0}
\definecolor{smokyblack}{rgb}{0.06, 0.05, 0.03}

\definecolor{americanrose}{rgb}{1.0, 0.01, 0.24}
\definecolor{cobalt}{rgb}{0.0, 0.28, 0.67}
\definecolor{brandeisblue}{rgb}{0.0, 0.44, 1.0}
\definecolor{mycolor}{rgb}{0.0, 0.0, 0.5}
\definecolor{oxfordblue}{rgb}{0.0, 0.13, 0.28}
\definecolor{azure}{rgb}{0.0, 0.5, 1.0}
\definecolor{turquoiseblue}{rgb}{0.0, 1.0, 0.94}
\newtcolorbox{mynewbox}[1]{colback=white!5!white,colframe=azure!75!black,fonttitle=\bfseries,title=#1}
\newtcolorbox{mybox}{colback=mycolor!5!white,colframe=azure!75!black}
\newtcolorbox{mynamedbox}[1]{colback=mycolor!5!white,colframe=azure!75!black,title=#1}
\definecolor{venetianred}{rgb}{0.78, 0.03, 0.08}
\newtcolorbox{mynamedbox1}[1]{colback=venetianred!5!white,colframe=venetianred!80!black,title=#1}
\newtcolorbox{mynamedbox2}[1]{colback=azure!5!white,colframe=azure!80!black,title=#1}

\definecolor{rossocorsa}{rgb}{0.83, 0.0, 0.0}

\tikzset{->-/.style={decoration={
  markings,
  mark=at position #1 with {\arrow{>}}},postaction={decorate}}}
\tikzset{-<-/.style={decoration={
  markings,
  mark=at position #1 with {\arrow{<}}},postaction={decorate}}} 

\def\L*{{\cal L}_*}
\def\L{\mathcal{L}}

\def\nn{\nonumber}

\def\<{\langle}
\def\>{\rangle}

\def\neq{\not\equiv}

\def\cs2{c_{s}^{2}}

\newcounter{subsubsubsection}[subsubsection]
\renewcommand\thesubsubsubsection{\thesubsubsection.\arabic{subsubsubsection}}

\titleformat{\subsubsubsection}
  {\normalfont\normalsize\bfseries}{\thesubsubsubsection}{1em}{}
\titlespacing*{\subsubsubsection}
{0pt}{3.25ex plus 1ex minus .2ex}{1.5ex plus .2ex}

\makeatletter
\renewcommand\paragraph{\@startsection{paragraph}{5}{\z@}%
  {3.25ex \@plus1ex \@minus.2ex}%
  {-1em}%
  {\normalfont\normalsize\bfseries}}
\renewcommand\subparagraph{\@startsection{subparagraph}{6}{\parindent}%
  {3.25ex \@plus1ex \@minus .2ex}%
  {-1em}%
  {\normalfont\normalsize\bfseries}}
\def\toclevel@subsubsubsection{4}
\def\toclevel@paragraph{5}
\def\toclevel@paragraph{6}
\def\l@subsubsubsection{\@dottedtocline{4}{7em}{4em}}
\def\l@paragraph{\@dottedtocline{5}{10em}{5em}}
\def\l@subparagraph{\@dottedtocline{6}{14em}{6em}}
\makeatother

\usepackage{titlesec} 
\usepackage[colorlinks,citecolor=blue,urlcolor=blue,hypertexnames=true]{hyperref}
\usepackage{subfigure}

\usepackage[usenames,dvipsnames,svgnames,table]{xcolor}

\usepackage{tikzsymbols}
\usepackage{natbib}
\usepackage{float}
\usepackage{tikz,xcolor,hyperref}
\usepackage{slashed}
\definecolor{lime}{HTML}{A6CE39}
\DeclareRobustCommand{\orcidicon}{
	\begin{tikzpicture}
	\draw[lime, fill=lime] (0,0) 
	circle [radius=0.2] 
	node[white] {{\fontfamily{qag}\selectfont \tiny ID}};
	\draw[white, fill=white] (-0.0625,0.095) 
	circle [radius=0.007];
	\end{tikzpicture}
	\hspace{-2mm}
}

\usepackage[usenames,dvipsnames,svgnames,table]{xcolor}  
\usepackage{tikzsymbols}
\usepackage{natbib}
\usepackage{float}
\usepackage{slashed}
\usepackage{bbold}
\usepackage{tikz}
\usepackage{adjustbox}
\usepackage{tcolorbox}
\usepackage{enumitem}
\usepackage{amsfonts}

\setlist[itemize,1]{label=$\times$}
\setlist[itemize,2]{label=$\checkmark$}
\setlist[itemize,3]{label=$\diamond$}
\setlist[itemize,4]{label=$\bullet$}

\definecolor{lime}{HTML}{A6CE39}
\DeclareRobustCommand{\orcidicon}{
	\begin{tikzpicture}
	\draw[lime, fill=lime] (0,0) 
	circle [radius=0.2] 
	node[white] {{\fontfamily{qag}\selectfont \tiny ID}};
	\draw[white, fill=white] (-0.0625,0.095) 
	circle [radius=0.007];
	\end{tikzpicture}
	\hspace{-2mm}
}

\foreach \x in {A, ..., Z}{\expandafter\xdef\csname orcid\x\endcsname{\noexpand\href{https://orcid.org/\csname orcidauthor\x\endcsname}
			{\noexpand\orcidicon}}
}

\preprint{ECTP-2026-10}
\preprint{WLCAPP-2026-10}

\begin{document}

\title{Quantum-Deformed Phase-Space Geometry and Emergent Inflation in Effective Four-Dimensional Spacetime}

\author{Swapnil~K.~Singh\orcidA$^{1}$, Saleh~O.~Allehabi\orcidB$^{2}$, Azzah~A.~Alshehri\orcidC$^{3}$, Mahmoud~Nasar\orcidD$^{4}$,
Ashraf~F.~El-Sherif\orcidE$^{5}$, Abdel~Nasser~Tawfik\orcidF$^{2}$}
\email{Corresponding author: atawfik@acu.edu.eg; 400778@iu.edu.sa; atawfik@bnl.gov}

\affiliation{$^1$BMS College of Engineering, Bangalore, Karnataka, 560019, India} 
\affiliation{$^2$Department of Physics, Faculty of Science, Islamic University of Madinah, Madinah 42351, Saudi Arabia} 
\affiliation{$^3$Department of Science and Technology, University College at Nairiyah, University of Hafr Al Batin, Nairiyah 31981, Saudi Arabia} 
\affiliation{$^4$Department of Physics, Faculty of Science, Benha University, 13518 Benha, Egypt} 
\affiliation{$^5$Basic Science Department, Faculty of Engineering, Ahram Canadian University, 12556 Giza, Egypt}  

\begin{abstract}
A phase-space approach to quantum-deformed gravity has been developed. After its reduction to an effective four-dimensional spacetime structure, we employ it to reanalyze the dynamics of cosmic inflation and quantum gravity. The construction commences on a cotangent bundle, where the gravitational Hamiltonian is deformed by a zero-homogeneous scalar determined by projective momentum directions and quantum phase-space properties. This leads to an anisotropic Hamilton geometry on a non-null conic domain, from which an effective spacetime metric is obtained through a section-pullback procedure. In the homogeneous and isotropic sector, the pullback consistently reduces to a conformally deformed FLRW geometry governed by a scalar deformation field. We derive the corresponding modified Einstein, Klein--Gordon, geodesic-deviation, and Raychaudhuri equations. This allows for the construction of the inflationary background dynamics, slow-roll regime, and number of e-folds, as well as scalar and tensor perturbations. The resulting framework indicates that leading inflationary corrections arise from the phase-space deformation and its time dependence, while the canonical quantization of cosmological perturbations remains standard after suitable background redefinitions. In this way, the model provides a covariant and controlled link between quantum-deformed phase-space geometry and effective four-dimensional inflationary dynamics. We conclude that the present construction demonstrates that the effects of quantum gravity can be consistently encoded as a deformation of projective phase-space geometry, from which an effective spacetime metric emerges only after a section-dependent reduction. The resulting theory presumably modifies gravitational dynamics through a constrained, derivative scalar response while preserving the classical limit and standard perturbative structure, thereby providing a geometrically coherent and phenomenologically viable realization of emergent spacetime.
\end{abstract}

\keywords{Cosmic inflation; Quantum deformed gravity; Phase-space (Hamilton) geometry; Effective spacetime; Cosmological perturbations}

\maketitle
\tableofcontents

\section{Introduction}
\label{sec:introduction}

The inflationary paradigm has become a central component of modern cosmology because it provides a coherent explanation for several otherwise unexplained features of the observed Universe \cite{Ryden_2016,lemoine2007inflationary,Barrow1981}. A phase of accelerated expansion in the very early Universe accounts for the large-scale homogeneity, isotropy, and near-flatness of spacetime, while among others resolving the horizon and monopole problems \cite{1980ApJ...241L..59K,LINDE1983177,PhysRevLett.48.1220,PhysRevD.23.347,LINDE1982389}. At the same time, inflation furnishes a mechanism for generating primordial perturbations: quantum fluctuations of the relevant fields are stretched from microscopic to astrophysical scales and subsequently seed the anisotropies of the cosmic microwave background and the formation of large-scale structure \cite{PhysRevB.21.702,1093mnras195,Liddle:1994yx,Chibisov:1982nx,Lukash:1980iv}. In this sense, inflation is not only a dynamical solution to several classical cosmological problems, but also the standard bridge between quantum fluctuations in the early Universe and present-day observational cosmology. Despite these successes, the physical origin of inflation remains unsettled. In the standard framework one postulates a scalar inflaton field with a suitable potential and then studies the resulting background and perturbative dynamics \cite{guth1998inflationary,STAROBINSKY198099,TawfikDahab2017}. While this approach is phenomenologically successful, it does not by itself explain why such a field should exist, why its potential should have the required structure, or how inflation should be embedded into a theory that consistently combines gravity with quantum physics. 

The inflationary epoch lies precisely in the regime where both quantum effects and strong spacetime curvature are relevant, so any fully satisfactory account of inflation should ultimately be compatible with a framework for quantum gravity \cite{Padmanabhan:1984vv,Tsamis:1996qq,Ellis:2002we}. This motivates a re-examination of the geometric assumptions underlying gravity. In general relativity, in its semiclassical approximation, the gravitational interaction is encoded in the pseudo-Riemannian geometry of a four-dimensional spacetime manifold \cite{EinsteinGR}. That framework is extraordinarily successful in the classical regime, but it is not obvious that it must remain fundamental when quantum effects become important \cite{Boulware:1974sr}. More general geometric structures have long been considered as possible extensions of the Riemannian and Lorentzian framework \cite{Gibbons:2007iu}. Among these, Finsler geometry is particularly significant because it allows the fundamental geometric properties to depend not only on position but also on direction in the tangent bundle \cite{Riemann1,Riemann2,Finsler,Bao,Bucataru}. This leads naturally to anisotropic geometries in which the metric structure depends on the direction of propagation and not merely on the spacetime point. Such direction-dependent geometries are not mathematically exotic constructions without physical relevance. They appear naturally in effective descriptions of wave propagation in media, modified dispersion relations, generalized Lorentz symmetry, kinetic theory, and several extensions of gravitation \cite{Asanov,Pfeifer:2019wus,Cerveny,Klimes,MARKVORSEN2016208,Yajima2009,Perlick,Rubilar:2007qm,TAVAKOL198523,Tavakol1986,Schreck:2015seb,Kostelecky:2010hs,Kostelecky:2003fs,Kostelecky:2011qz,Bogoslovsky1994,Bogoslovsky,Raetzel:2010je,Amelino-Camelia:2014rga,Lobo:2020qoa,Gibbons:2007iu,Lammerzahl:2018lhw,Rutz,Pfeifer-Wohlfarthgravity,kinetic-gas,Minguzzi:2014fxa}. In the spacetime context, substantial progress has been made in formulating Finslerian generalizations with well-defined causal cones, observer structure, variational principles, and gravitational dynamics \cite{Beem,Pfeifer:2011tk,Minguzzi:2014aua,Lammerzahl:2018lhw,Javaloyes:2018lex,Hohmann:2018rpp,Bernal2020}. These developments strongly suggest that anisotropic geometry is a serious candidate for describing effective gravitational phenomena beyond the strictly classical regime.

A particularly important lesson from Finsler (distance and direction) and Hamilton (distance and momentum) geometry is that once the geometry depends on directions, the natural arena is no longer the base manifold alone. The relevant geometric objects are defined on the tangent (Finsler) or cotangent (Hamilton) bundle and satisfy definite homogeneity properties with respect to the fiber coordinates \cite{Bao,Bucataru,Hohmann:2018rpp}. This immediately points toward projective bundle constructions, in which one works with rays in tangent or cotangent space rather than with vectors or covectors of arbitrary scale. The positively projectivized tangent or cotangent bundle is then the natural setting for homogeneous geometric fields and for variational principles that are insensitive to redundant positive rescalings. From this perspective, one is led to the idea that the correct semiclassical arena for quantum gravity effects may be phase space rather than spacetime alone.

This viewpoint is reinforced by a broad class of approaches in which quantum-gravity effects manifest themselves first as deformations of kinematics, dispersion relations, or uncertainty structures. Generalized uncertainty principles, modified phase-space commutators, deformed relativistic symmetries, Born-type reciprocity ideas, and several related constructions all indicate that the interplay between position and momentum may be more fundamental than a direct quantization of the spacetime metric itself \cite{Benczik:2002tt,Todorinov:2020jtq,phdthesisXun,Tawfik:2023onh,Tawfik:2023orl}. In such scenarios, one expects that quantum-gravity corrections should appear first at the level of phase-space geometry and only afterwards induce an effective spacetime description including one in four-dimensional Riemannian geometry.

Motivated by these considerations, a sequence of earlier works proposed a geometric quantization program in which the metric sector of general relativity is extended to phase space and quantum effects are encoded through deformations of the underlying relativistic phase-space structure \cite{Tawfik:2025icy,Tawfik:2025rel,Tawfik:2025kae,Tawfik:2024itt,NasserTawfik:2024afw,Tawfik:2023orl,Tawfik:2023ugm,Tawfik:2023hdi}. The essential idea is that quantum corrections should not be introduced by promoting the spacetime metric directly to an operator, but rather by deforming the phase-space kinematics and then constructing the associated effective geometry. This naturally leads to Hamiltonian and Hamilton--Finsler type structures on the cotangent bundle, with the resulting spacetime geometry emerging only after an appropriate reduction.

The present paper develops this idea in a more systematic and cosmologically relevant setting. We formulate a quantum-deformed Hamilton geometry on the cotangent bundle $T^{\ast}M$, with the deformation encoded in a projectively well-defined scalar depending on momentum directions rather than momentum magnitudes. The geometry is therefore defined on a non-null conic domain of $T^{\ast}M$ and, more fundamentally, on the positively projectivized cotangent bundle. The effective spacetime metric is obtained only after choosing a section of this projective cotangent bundle and pulling back the anisotropic Hamiltonian metric to the base manifold. This pullback step is essential: it is the first stage at which one recovers an ordinary pseudo-Riemannian metric on spacetime. Only after this step one can consistently impose additional symmetry reductions, such as the homogeneous and isotropic conformal truncation relevant for Friedman-Lemaitre-Robertson-Walker (FLRW) cosmology, i.e., evolution of Universe, for instance.

A central feature of the construction is that the reduced spacetime dynamics is controlled by a scalar deformation function inherited from the underlying phase-space geometry. This scalar is not introduced as an independent matter degree of freedom. Rather, it arises from the pullback of the anisotropic Hamiltonian structure and from the corresponding reduction to the cosmological sector. In this way the theory modifies the Einstein equations, the inflaton dynamics, and the perturbation equations without postulating new dynamical fields unrelated to the underlying geometry. In this regard, it should be emphasized that the modifications are geometric in origin and remain continuously connected to general relativity in the undeformed limit.

The inflationary epoch provides a natural arena in which this framework can be tested. Inflation is sensitive both to gravitational dynamics and to quantum fluctuations, so it is an ideal setting in which a phase-space based quantum-gravity deformation may leave observable traces. In the formalism developed here, the pullback of the quantum-deformed Hamiltonian geometry induces corrections to the Friedmann equations, to the effective inflaton equation of motion, and to the scalar and tensor perturbation sectors. At the same time, the canonical structure of the perturbation variables remains intact after the reduction, so the deformation acts through the background geometry rather than through an {\it ad hoc} modification of the perturbative commutation relations. This makes the framework both geometrically controlled and phenomenologically accessible.

The conceptual message of the paper is therefore broader than a particular inflationary model. The results support a view in which spacetime geometry is not fundamental, but emerges as a reduced description of a more primitive phase-space geometry. In the present setting, the cotangent-bundle Hamiltonian structure is primary, the section-induced pseudo-Riemannian metric is secondary, and the conformal FLRW geometry is a further symmetry reduction of that secondary structure. Inflation then becomes a concrete physical regime in which the consequences of this hierarchy can be studied explicitly.

The paper is organized as follows. Section~\ref{sec:formalism} develops the quantum-deformed Hamiltonian formalism on the cotangent bundle, introduces the projective structure, constructs the pullback geometry, and derives the corresponding effective spacetime tensors and actions. Three distinct geometric levels and effective spacetime metric are derived in Section \ref{sec:EffctST}.  The Levi--Civita connections along with various curvature tensors are presented in Section \ref{sec:CivitaCrvt}. This is succeeded by the effective Einstein--Hilbert action in Section \ref{sec:actions}. The derivation of the Hamilton--Finsler nonlinear and metric-compatible $d$-connection is detailed in Section \ref{sec:HF_connection}. The geodesic deviation within Hamilton--Finsler geometry is summarized in Section \ref{sec:geodesic_deviation}. The formalism concludes with the Raychaudhuri equation in quantum-deformed Hamilton--Finsler geometry, as outlined in Section \ref{sec:Raychaudhuri}. The analytical results are detailed in Section \ref{sec:anlRsl}. This begins with the dynamics of the inflationary background in Section \ref{sec:inflation}. The examination of the inflaton sector concerning pullback and conformal geometries is presented in Section \ref{sec:inflaton_sector}. The derivation of the background Einstein and Klein--Gordon equations is found in Section \ref{sec:allIA_final}, which is subsequently followed by the derivation of the slow-roll regime and the number of e-Folds for power-law potential in Section \ref{sec:QCeFoldsPowerLaw_final}. The analysis of scalar and tensor perturbations on the effective FLRW background is addressed in Section \ref{sec:results_final}. Section \ref{sec:powerlaw_final} focuses on power-law inflation and the time dependence of deformation. The canonical quantization of cosmological perturbations is articulated in Section \ref{sec:canonical_quantization_final}. The second broader implication is devoted to quantum gravity, as discussed in Section \ref{sec:QGImplications}. The final conclusions and prospective outlook are thoroughly described in Section \ref{sec:Conclusion}.

\section{Formalism}
\label{sec:formalism}

As introduced, the construction is formulated on phase space and only afterwards reduced to an effective pseudo-Riemannian description on spacetime. As shall be illustrated, this order is dictated by the geometry itself. Once the gravitational response is allowed to depend on momentum directions, the fundamental object is no longer a tensor field on the base manifold $M$, but an anisotropic tensor field defined on a conic subbundle of the cotangent bundle $T^{\ast}M$. The appropriate mathematical framework is therefore Hamilton geometry on $T^{\ast}M$, together with the positively projectivized cotangent bundle, which removes the redundant positive rescaling in the fibers and supplies the correct homogeneous arena for a well-defined variational theory \cite{Bao,Bucataru,cosmoFinsler,Hohmann:2018rpp}. The quantum input is introduced at the level of phase-space kinematics rather than through a direct operator quantization of the spacetime metric. This represents a significant expansion of the geometric quantization method that was initially proposed by one of the authors (AT) \cite{Tawfik:2025icy,Tawfik:2025rel,Tawfik:2025kae,Tawfik:2024itt,NasserTawfik:2024afw,Tawfik:2023orl,Tawfik:2023ugm,Tawfik:2023hdi}. In this sense, let us emphasize that the theory is a semiclassical gravitational theory induced by a deformed quantum phase space rather than a direct quantization of Lorentzian geometry \cite{Mackey1949DukeJ,Mackey1949PNAS,Mackey1952,Mackey1958,aharonov2008quantum,GelfandNaimark,Freidel:2013zga,Freidel:2014qna,Freidel:2015pka,Freidel:2015uug,Polchinski:1998rq}. The central physical idea is that the modular quantum structure modifies the kinematics on phase space first, and only then induces an effective spacetime geometry after a controlled reduction \cite{Bertozzini2010}.\footnote{This order of construction is essential. If one starts directly from a spacetime metric ansatz, then the homogeneity properties that characterize Hamilton geometry on $T^{\ast}M$ are obscured, and the modular origin of the deformation is lost.}

Let $(M,g)$ be a connected four-dimensional Lorentzian manifold with signature $\mathrm{sign}(g_{\mu\nu})=(-,+,+,+)$. The proper phase space is the cotangent bundle $\mathcal{P}=T^{\ast}M$ 
equipped with canonical local coordinates $(x^{\mu},p_{\mu})$ and the canonical symplectic form reads
\begin{equation}
\omega=dp_{\mu}\wedge dx^{\mu}.
\label{eq:formalism_symplectic_form}
\end{equation}
A generic tangent vector to phase space is to expressed as
\begin{equation}
\mathbb{K}=(k^{\mu},\tilde{k}_{\mu})\in T_{(x,p)}\mathcal{P},
\label{eq:formalism_K_def}
\end{equation}
where $k^{\mu}$ is tangent to the base directions and $\tilde{k}_{\mu}$ is tangent to the cotangent-fiber directions \cite{Hamilton1982,RELANCIO2025105626,PhysRevD.92.084053,Albuquerque:2023icp}. Let us  denote the corresponding pair of quantum phase-space operators, distance and momentum, by
\begin{equation}
\hat{\mathbb{X}}=(\hat{x}^{\mu},\hat{p}_{\mu}).
\label{eq:formalism_Xhat_def}
\end{equation}
In this regard, let us recall Weyl operators
\begin{equation}
W_{\mathbb{K}} = \exp\!\left(2\pi i\,\omega(\mathbb{K},\hat{\mathbb{X}})\right) = \exp\!\left(2\pi i\bigl(\tilde{k}_{\mu}\hat{x}^{\mu}-k^{\mu}\hat{p}_{\mu}\bigr)\right),
\label{eq:formalism_weyl_operator_final}
\end{equation}
which satisfy Weyl multiplication law
\begin{equation}
W_{\mathbb{K}}W_{\mathbb{K}'} = e^{\pi i\,\omega(\mathbb{K},\mathbb{K}')}\,W_{\mathbb{K}+\mathbb{K}'}.
\label{eq:formalism_weyl_product_final}
\end{equation}
The choice of quantum configuration space is the choice of maximal commutative $^{\ast}$-subalgebra of the Heisenberg algebra or equivalently the polarization \cite{Mackey1949DukeJ,Mackey1949PNAS,Mackey1952,Mackey1958,Freidel:2013zga,Freidel:2014qna,Freidel:2015pka,Freidel:2015uug}. In the modular setting such a polarization can be encoded by a self-dual lattice $\Lambda$ and by neutral bilinear form $\eta$ of split signature on phase space \cite{aharonov2008quantum,GelfandNaimark,Freidel:2013zga,Freidel:2014qna,Freidel:2015pka,Freidel:2015uug}. Concretely, the lattice sector is then taken to be a rank-$2d$ integral symplectic lattice $\Lambda\subset \mathcal{P}_{x}\simeq T^{\ast}_{x}M\oplus T_{x}M$, where  $d=4$, satisfying the following integrality condition:
\begin{equation}
\omega(\lambda,\lambda')\in \mathbb{Z}, \qquad \lambda,\lambda'\in\Lambda,
\label{eq:formalism_lattice_integrality}
\end{equation}
and the self-duality with respect to the symplectic pairing as well. The neutral bilinear form $\eta$ is taken to be compatible with the polarization splitting, so that it has signature $(4,4)$ on the eight-dimensional phase space.\footnote{At the level of the present semiclassical construction, $\Lambda$ and $\eta$ are treated as polarization data specifying the modular sector. They are not varied dynamically in the action. A more complete theory may promote them to background fields or collective variables, but that extension is not required for the present derivation.}

To make the quantum sector concrete, let us fix a modular polarization $(\Lambda,\eta)$ and let $T_{\Lambda}:=\mathcal{P}_{x}/\Lambda$ be the corresponding modular torus \cite{PhysRevD.101.106021}. The associated Hilbert space is taken to be the space of square-integrable sections of the prequantum line bundle over $T_{\Lambda}$,
\begin{equation}
\mathcal{H}_{(\Lambda,\eta)} = L^{2}(T_{\Lambda},L_{\Lambda}),
\label{eq:formalism_modular_hilbert}
\end{equation}
Here $L_{\Lambda}\to T_{\Lambda}$ is the prequantum line bundle whose connection is chosen so that its holonomies reproduce the lattice-twisted Weyl relations determined by $(\omega,\eta)$ \cite{Holonomy1955}. In particular, the quasi-periodicity of sections is fixed by the compatibility between the symplectic form and the polarization bilinear structure with quasi-periodicity determined by the pair $(\omega,\eta)$ \cite{Yoshida2024}. The Weyl operators act on $\mathcal{H}_{(\Lambda,\eta)}$ through the standard Heisenberg representation compatible with the lattice identifications. The semiclassical states introduced below are taken to be wave-packet sections in $\mathcal{H}_{(\Lambda,\eta)}$, sharply localized in the chosen modular cell and centered on the effective spacetime quantities, distances and momenta, $(x,[p])$. This is sufficient for the semiclassical symbol calculus used in the sequel.

The undeformed Hamiltonian is introduced at this point as
\begin{equation}
H_{0}(x,p)=\frac12\,Q(x,p),
\label{eq:formalism_H0_final}
\end{equation}
where $Q(x,p):=g^{\mu\nu}(x)p_{\mu}p_{\nu}$ so that the homogeneity properties lead to 
\begin{equation}
H_{0}(x,\lambda p)=\lambda^{2}H_{0}(x,p), \qquad \lambda > 0.
\label{eq:formalism_H0_hom_final}
\end{equation}
i.e., $H_{0}$ is positively two-homogeneous in the fiber variables. Since the deformation is required to preserve the Hamiltonian homogeneity class, it cannot depend on the absolute scale of the momentum but only on the corresponding momentum ray. This immediately implies that the deformation factor must be zero-homogeneous in $p$.

The appropriate non-null conic domain is defined as
\begin{equation}
\mathcal{A}^{\ast}_{0} = \left\{(x,p)\in T^{\ast}M\setminus\{0\}
\;\middle|\; Q(x,p)\neq 0\right\},
\label{eq:formalism_A0star_final}
\end{equation}
which is invariant under positive rescaling $p \mapsto \lambda p$ with $\lambda>0$. The exclusion of the null sector is required because the normalized projective momentum introduced below is singular at $Q=0$. In this regard, the null sector is physically important, especially for massless propagation, but it requires a separate limiting treatment because the normalization $u_{\mu}=p_{\mu}/\sqrt{|Q|}$ is undefined on $Q=0$. In the present paper we restrict the construction to $\mathcal{A}^{\ast}_{0}$ and interpret the null regime as a boundary sector to be analyzed separately. The positively projectivized cotangent bundle is then given as
\begin{equation}
PT^{\ast} M_{+} = \mathcal{A}^{\ast}_{0}/\!\sim, \qquad (x,p)\sim(x,\lambda p), \qquad \lambda > 0.
\label{eq:formalism_PTstar_final}
\end{equation}
Every scalar field $\Phi$ that genuinely descends to $PT^{\ast}M_{+}$ therefore satisfies following Euler's zero-homogeneity condition:
\begin{equation}
p_{\mu}\frac{\partial \Phi}{\partial p_{\mu}}=0.
\label{eq:formalism_Euler_zero_hom_final}
\end{equation}

Let us now introduce the normalized representative of the projective momentum ray,
\begin{equation}
u_{\mu}(x,[p]) = \frac{p_{\mu}}{\sqrt{|Q(x,p)|}}, \qquad u^{\mu}= g^{\mu\nu} u_{\nu}, \qquad \varepsilon := \mathrm{sign}(Q) \in \{+1,-1\},
\label{eq:formalism_u_def_final}
\end{equation}
where $\varepsilon$ records the causal character of the non-null momentum with respect to the background metric. One then finds that
\begin{equation}
g^{\mu\nu} u_{\mu} u_{\nu}=\varepsilon, \qquad u_{\mu}(x,[\lambda p]) = u_{\mu}(x,[p]),
\label{eq:formalism_u_projective_final}
\end{equation}
so that $u_{\mu}$ is a well-defined covector field on $PT^{\ast}M_{+}$. We now define the projector orthogonal to the projective ray as
\begin{equation}
\Pi^{\mu}_{\nu} = \delta^{\mu}_{\nu} - \varepsilon\, u^{\mu}u_{\nu},
\label{eq:formalism_projector_final}
\end{equation}
which obviously satisfies
\begin{equation}
\Pi^{\mu}_{\nu}u^{\nu}=0, \qquad u_{\mu}\Pi^{\mu}_{\nu}=0,
\qquad \Pi^{\mu}_{\rho}\Pi^{\rho}_{\nu}=\Pi^{\mu}_{\nu}.
\label{eq:formalism_projector_props_final}
\end{equation}
The decomposition encoded by $\Pi^{\mu}_{\nu}$ well control the differential structure of every zero-homogeneous scalar built from the momentum ray.

As emphasized, the quantum correction is not introduced phenomenologically. It is extracted from a semiclassical family of normalized states
\begin{equation}
\left\{\ket{\Psi_{x,[p]}}\right\}_{(x,[p])\in M\times PT^{\ast}M_{+}}
\label{eq:formalism_state_family}
\end{equation}
which are parametrized by spacetime position and projective momentum direction. More concretely, we take $\ket{\Psi_{x,[p]}}$ to be modular semiclassical wave-packet states constructed from the chosen polarization sector, centered at $x$ and peaked on the projective momentum ray $[p]$. Therefore, we find that
\begin{equation}
\bra{\Psi_{x,[p]}}\hat{x}^{\mu}\ket{\Psi_{x,[p]}}=x^{\mu},
\qquad
\bra{\Psi_{x,[p]}}\hat{p}_{\mu}\ket{\Psi_{x,[p]}}=\varpi(x,[p])\,u_{\mu}(x,[p]),
\label{eq:formalism_state_peaking}
\end{equation}
for some positive scale function $\varpi(x,[p])$, while the covariance of the state is adapted to the chosen modular cell.\footnote{The precise microscopic realization of these states is not required for the semiclassical derivation. They are modular analogues of coherent or wave-packet states in the Weyl representation, with the crucial additional property that their dependence on momentum is only through the projective class $[p]$.} These states define the following symbol map:
\begin{equation}
\mathfrak{s}_{x,[p]}(\hat{\mathcal{O}}) := \bra{\Psi_{x,[p]}}\hat{\mathcal{O}}\ket{\Psi_{x,[p]}},
\label{eq:formalism_symbol_map_final}
\end{equation}
on the operator algebra. The essential covariance requirement is that the state family depends on the momentum variable only through the following projective class:
\begin{equation}
\ket{\Psi_{x,[\lambda p]}}=\ket{\Psi_{x,[p]}}, \qquad \lambda > 0.
\label{eq:formalism_projective_state_covariance_final}
\end{equation}
Obviously, this is the quantum counterpart of the classical homogeneity requirement and implies that every semiclassical correction extracted from these states is automatically zero-homogeneous in $p$.

The effective Hamiltonian is then defined as the expectation value
\begin{equation}
H_{\mathrm{eff}}(x,p) := \mathfrak{s}_{x,[p]}(\hat{H}_{\beta}),
\label{eq:formalism_Heff_final}
\end{equation}
where $\hat{H}_{\beta}$ represents the modularly deformed quantum Hamiltonian. At this point we impose the structural assumption that the deformation preserves the same quadratic Hamiltonian class as the undeformed theory, so that no additional independent two-homogeneous scalar survives in the effective sector beyond the rescaling of $Q(x,p)$. Because the undeformed part is homogeneous of degree two and because the state family depends only on projective momentum directions, the ratio $H_{\mathrm{eff}}/H_{0}$ is zero-homogeneous and thus descends to $PT^{\ast}M_{+}$. Hence there exists a scalar field $\Omega_{\beta}(x,[p])$ such that
\begin{equation}
H_{\mathrm{eff}}(x,p)=\Omega_{\beta}(x,[p])\,H_{0}(x,p).
\label{eq:formalism_Heff_factorization_final}
\end{equation}
This multiplicative form is taken as the defining lowest-order effective truncation of the modular backreaction. Higher-order two-homogeneous invariants are not forbidden in principle, but are neglected consistently at the present order in the semiclassical expansion.

Now we identify $H_{\mathrm{eff}}$ with the deformed classical Hamiltonian and find that
\begin{equation}
H_{\beta}(x,p) = \Omega_{\beta}(x,[p])\, H_{0}(x,p) = \frac12\, \Omega_{\beta}(x,[p])\, Q(x,p).
\label{eq:formalism_Hbeta_final}
\end{equation}
The undeformed limit is imposed by
\begin{equation}
\Omega_{\beta}(x,[p]) = 1 + \beta\,\Xi(x,[p];\eta,\Lambda) + \mathcal{O}(\beta^{2}),
\label{eq:formalism_Omega_expand_final}
\end{equation}
where $\beta$ is a dimensionless deformation parameter and $\Xi$ is a dimensionless zero-homogeneous scalar,
\begin{equation}
p_{\mu}\frac{\partial \Xi}{\partial p_{\mu}}=0.
\label{eq:formalism_Xi_zero_hom_final}
\end{equation}
The physical meaning of $\Omega_{\beta}$ is the measure of the leading semiclassical backreaction of the modular quantum sector on the classical Hamiltonian. Accordingly, the function $\Xi$ is not a matter source, but a phase-space response scalar induced by the polarization.

In order for $H_{\beta}$ to define a regular Hamilton space, we assume the following regularity properties throughout the non-null conic domain $\mathcal{A}^{\ast}_{0}$:
\begin{equation}
H_{\beta} \in C^{\infty}(\mathcal{A}^{\ast}_{0}),
\qquad \det\!\left(\frac{\partial^{2}H_{\beta}}{\partial p_{\mu}\partial p_{\nu}}\right)\neq 0, \quad \text{on } \mathcal{A}^{\ast}_{0},
\label{eq:formalism_regularity_assumption}
\end{equation}
and the signature of the inverse Hamilton metric is assumed to remain Lorentzian for sufficiently small $|\beta|$. These assumptions guarantee that the perturbed Hamiltonian geometry is regular and that the fiber Hessian defines a non-degenerate phase-space metric.\footnote{This is the cotangent-bundle analogue of the regularity condition in Finsler and Hamilton geometry. Without non-degeneracy of the fiber Hessian the nonlinear connection, $d$-connection, and curvature tensors below would not be well defined.}

Our task is therefore to derive the most general admissible leading-order scalar $\Xi$ from the modular data $(\omega,\eta,\Lambda)$. For this purpose let us define the characteristic function of the semiclassical state family by
\begin{equation}
\Phi_{x,[p]}(\mathbb{K}) := \bra{\Psi_{x,[p]}}W_{\mathbb{K}}\ket{\Psi_{x,[p]}}.
\label{eq:formalism_characteristic_function_final}
\end{equation}
In the modular setting the self-dual lattice identifies compact phase directions, and therefore any smooth scalar response induced by the modular sector admits a Fourier expansion in the corresponding modular angles. Restricting to the spacetime sector selected by the polarization, let $\lambda_{(a)\mu}(x)$, where $a=1,\dots,N$, be local spacetime representatives of the projected lattice generators in that sector. They are not independent new fields, but the spacetime covector images of selected generators of $\Lambda$ after restriction to the chosen polarization sector. Here $N$ denotes the number of retained generating lattice directions in the effective spacetime sector; in a minimal truncation one may take $N\leq 4$, while in a more refined treatment one may include additional harmonics and redundant generators subject to the lattice relations. We now define the modular phases
\begin{equation}
\theta_{a}(x,[p]) := \lambda_{(a)}^{\mu}(x)u_{\mu}, \qquad \lambda_{(a)}^{\mu}:=g^{\mu\nu}\lambda_{(a)\nu}.
\label{eq:formalism_theta_final}
\end{equation}
Since $u_{\mu}$ depends only on $[p]$, each $\theta_{a}$ is a well-defined scalar on $PT^{\ast}M_{+}$. The most general real smooth modular response therefore has the local Fourier representation,
\begin{equation}
\Xi_{\mathrm{mod}}(x,[p]) = \sum_{n\in\mathbb{Z}^{N}}
\mathcal{C}_{n}(x)\, e^{2\pi i\,n^{a}\theta_{a}(x,[p])},
\qquad \mathcal{C}_{-n}(x)=\overline{\mathcal{C}_{n}(x)}.
\label{eq:formalism_Xi_mod_fourier_final}
\end{equation}
The validity of this expansion requires only that the response function be smooth on the compact modular torus associated with the chosen lattice cell. In particular, the Fourier coefficients $\mathcal{C}_{n}(x)$ are assumed to decay sufficiently fast so that the series converges absolutely together with its first two derivatives on the sector of interest.\footnote{For a $C^{\infty}$ modular response on a compact torus, the Fourier coefficients decay faster than any power, which is sufficient for the formal manipulations below. The minimal truncation used later simply retains the leading harmonics.} If one imposes invariance under reversal of the modular cell orientation, i.e., 
\begin{equation}
u_{\mu}\mapsto -u_{\mu},
\label{eq:formalism_u_reversal_final}
\end{equation}
then only the even Fourier sector survives and the expansion, Eq. \eqref{eq:formalism_Xi_mod_fourier_final}, reduces to
\begin{equation}
\Xi_{\mathrm{mod}}(x,[p]) = \mathcal{C}_{0}(x) + \sum_{n\neq 0}
\widetilde{\mathcal{C}}_{n}(x)\, \cos\!\bigl(2\pi n^{a}\theta_{a}(x,[p])\bigr).
\label{eq:formalism_Xi_mod_cosine_final}
\end{equation}
Retaining the leading harmonics associated with the generating lattice directions yields the minimal modular contribution
\begin{equation}
\Xi_{\mathrm{mod}}(x,[p]) = \sum_{a=1}^{N} c_{a}(x) \left[
1-\cos\!\left(2\pi \lambda_{(a)}^{\mu}(x)u_{\mu}\right)\right].
\label{eq:formalism_Xi_mod_minimal_final}
\end{equation}
The subtraction of the constant term isolates the genuinely periodic part of the response and makes the undeformed limit explicit.

The neutral polarization bilinear form $\eta$ produces a second, non-periodic but anisotropic, zero-homogeneous contribution. Let $e^{A}_{\mu}(x)$, with internal indices $A,B=0,1,2,3$, be a local coframe adapted to the chosen modular polarization sector, and let $\eta^{(\mathrm{pol})}_{AB}$ denote the pullback of the polarization bilinear form to this four-dimensional sector. Now, we introduce the definition
\begin{equation}
h_{\mu\nu}(x) = e^{A}_{\mu}(x)e^{B}_{\nu}(x)\eta^{(\mathrm{pol})}_{AB}(x), \qquad h^{\mu\nu} := g^{\mu\alpha}g^{\nu\beta} h_{\alpha\beta}.
\label{eq:formalism_h_tensor_final}
\end{equation}
By construction $h_{\mu\nu}$ is not an independent spacetime metric but the spacetime imprint of the chosen polarization sector. Since $u_{\mu}$ is the only available covector on projective cotangent space, the lowest-order anisotropic scalar compatible with covariance and projective homogeneity reads
\begin{equation}
\Xi_{\mathrm{aniso}}(x,[p])=\alpha\,h^{\mu\nu}(x)u_{\mu}u_{\nu},
\label{eq:formalism_Xi_aniso_final}
\end{equation}
with $\alpha$ being dimensionless. The minimal leading-order scalar compatible with covariance on $M$, projective homogeneity on $PT^{\ast}M_{+}$, modular periodicity, locality in $x$, and a smooth undeformed limit is therefore
\begin{equation}
\Xi(x,[p];\eta,\Lambda) = \alpha\,h^{\mu\nu}(x)u_{\mu}u_{\nu} + \sum_{a=1}^{N} c_{a}(x) \left[1 - \cos\!\left(2\pi\,\lambda_{(a)}^{\mu}(x)u_{\mu}\right) \right].
\label{eq:formalism_Xi_final_final}
\end{equation}
This expression is not an arbitrary ansatz. It is the first nontrivial truncation of the most general modular Fourier response consistent with the structural constraints of covariance, homogeneity, smoothness, and the existence of a classical undeformed limit. Its physical interpretation is transparent: the first term measures a direction-dependent anisotropic distortion induced by the polarization metric, whereas the second term quantifies the periodic response of the modular lattice cell.

The derivative structure of $u_{\mu}$ is required in order to compute the fiber Hessian of the deformed Hamiltonian. Since
\begin{equation}
u_{\rho}=p_{\rho}|Q|^{-1/2}, \qquad \frac{\partial Q}{\partial p_{\mu}}=2p^{\mu}, \qquad \frac{\partial |Q|}{\partial p_{\mu}} = 2 \varepsilon p^{\mu},
\label{eq:formalism_Q_derivatives_final}
\end{equation}
one finds that
\bea
\frac{\partial u_{\rho}}{\partial p_{\mu}} &=& \frac{\delta_{\rho}^{\mu}}{\sqrt{|Q|}} - \frac{p_{\rho}}{2|Q|^{3/2}} \frac{\partial |Q|}{\partial p_{\mu}} 
%
%
%
= \frac{1}{\sqrt{|Q|}}\Pi_{\rho}{}^{\mu}.
\label{eq:formalism_du_dp_final}
\eea
Obviously raising the first index yields that
\begin{equation}
\frac{\partial u^{\rho}}{\partial p_{\mu}} = \frac{1}{\sqrt{|Q|}}\Pi^{\rho\mu}.
\label{eq:formalism_duup_dp_final}
\end{equation}
Differentiating once more leads to
\bea
\frac{\partial^{2}u^{\rho}}{\partial p_{\mu}\partial p_{\nu}}
&=& -\frac{\varepsilon p^{\nu}}{|Q|^{3/2}}\Pi^{\rho\mu} + \frac{1}{\sqrt{|Q|}}
\frac{\partial \Pi^{\rho\mu}}{\partial p_{\nu}}.
\label{eq:formalism_second_u_intermediate_final}
\eea
Hence, we get
\bea
\frac{\partial \Pi^{\rho\mu}}{\partial p_{\nu}}
&=& - \frac{\varepsilon}{\sqrt{|Q|}} \left(\Pi^{\rho\nu}u^{\mu}+ u^{\rho}\Pi^{\mu\nu}
\right).
\label{eq:formalism_projector_derivative_final}
\eea
By substituting Eq. \eqref{eq:formalism_projector_derivative_final} into Eq.  \eqref{eq:formalism_second_u_intermediate_final}, the closed identity reads
\begin{equation}
\frac{\partial^{2}u^{\rho}}{\partial p_{\mu}\partial p_{\nu}}
= -\frac{\varepsilon}{|Q|} \left(u^{\nu}\Pi^{\rho\mu} + u^{\mu}\Pi^{\rho\nu} + u^{\rho}\Pi^{\mu\nu} \right).
\label{eq:formalism_d2u_dp2_final}
\end{equation}

Let us now define
\begin{equation}
\Xi_{|\rho}:=\frac{\partial \Xi}{\partial u^{\rho}},
\qquad \Xi_{|\rho\sigma}:=\frac{\partial^{2}\Xi}{\partial u^{\rho}\partial u^{\sigma}}.
\label{eq:formalism_Xi_u_defs_final}
\end{equation}
Then Eq. \eqref{eq:formalism_Xi_final_final} implies that
\bea
\Xi_{|\rho} &=& 2 \alpha\, h_{\rho\nu}u^{\nu} + 2 \pi \sum_{a=1}^{N} c_{a}(x) \sin(2\pi\theta_{a})\, \lambda_{(a)\rho},
\label{eq:formalism_Xi_u_1_final}
\\
\Xi_{|\rho\sigma} &=& 2 \alpha\, h_{\rho\sigma} + (2 \pi)^{2}
\sum_{a=1}^{N} c_{a}(x) \cos(2 \pi \theta_{a})\, \lambda_{(a)\rho}\lambda_{(a)\sigma}.
\label{eq:formalism_Xi_u_2_final}
\eea
Using the chain rule together with Eq.  \eqref{eq:formalism_duup_dp_final} and Eq.  \eqref{eq:formalism_d2u_dp2_final}, one finds that
\bea
\frac{\partial \Xi}{\partial p_{\mu}}
&=& \frac{1}{\sqrt{|Q|}}\Pi^{\rho\mu}\Xi_{|\rho},
\label{eq:formalism_Xi_p_1_final}\\
\frac{\partial^{2}\Xi}{\partial p_{\mu}\partial p_{\nu}}
&=& \frac{1}{|Q|}\Pi^{\rho\mu}\Pi^{\sigma\nu}\Xi_{|\rho\sigma} - \frac{\varepsilon}{|Q|} \left(u^{\nu}\Pi^{\rho\mu} + u^{\mu}\Pi^{\rho\nu} + u^{\rho}\Pi^{\mu\nu} \right)\Xi_{|\rho}.
\label{eq:formalism_Xi_p_2_final}
\eea
Because $\Omega_{\beta}$ is zero-homogeneous and $H_{0}$ is two-homogeneous, the deformed Hamiltonian preserves the homogeneity class,
\begin{equation}
H_{\beta}(x,\lambda p)=\lambda^{2}H_{\beta}(x,p),
\qquad \lambda > 0.
\label{eq:formalism_Hbeta_hom_final}
\end{equation}
As discussed, the Hamilton metric can be derived from the fiber Hessian as follows:
\begin{equation}
\widetilde{g}^{\mu\nu}(x,p) = \frac{\partial^{2} H_{\beta}(x,p)}{\partial p_{\mu}\partial p_{\nu}}.
\label{eq:formalism_gtilde_def_final}
\end{equation}

\subsection{Distinct Geometric Levels and Effective Spacetime Metric}
\label{sec:EffctST}

Differentiating Eq. \eqref{eq:formalism_Hbeta_final} results in
\bea
\frac{\partial H_{\beta}}{\partial p_{\mu}} &=& \frac12 \frac{\partial\Omega_{\beta}}{\partial p_{\mu}}\, Q
+ \frac12 \Omega_{\beta} \frac{\partial Q}{\partial p_{\mu}}
 = \frac12 \frac{\partial\Omega_{\beta}}{\partial p_{\mu}}\, Q + \Omega_{\beta}p^{\mu},
\label{eq:formalism_dHdp_final}
\eea
where $\partial Q/\partial p_{\mu}=2p^{\mu}$. Another differentiation yields the exact formula
\begin{equation}
\widetilde{g}^{\mu\nu}(x,p) = \Omega_{\beta} g^{\mu\nu} + p^{\mu} \frac{\partial\Omega_{\beta}}{\partial p_{\nu}} + p^{\nu} \frac{\partial\Omega_{\beta}}{\partial p_{\mu}} + \frac12 Q
\frac{\partial^{2} \Omega_{\beta}}{\partial p_{\mu} \partial p_{\nu}}.
\label{eq:formalism_gtilde_full_final}
\end{equation}
It is obvious that this expression exhibits the essential structural point that the modular correction is not generically conformal on phase space. While the first term is conformal to the undeformed metric, the last three terms are non-conformal derivative contributions arising from the projective momentum dependence of $\Omega_{\beta}$. Therefore conformality cannot hold at the fundamental Hamiltonian level except under additional restrictions; it can emerge only after further symmetry reduction.

Let us now introduce perturbative expansion of $\widetilde{g}^{\mu\nu}$ to the first order in $\beta$, 
\bea
\begin{array}{ccl}
\widetilde{g}^{\mu\nu}(x,p) &=& g^{\mu\nu}(x) + \beta\, \Delta^{\mu\nu}(x,p) + \mathcal{O}(\beta^{2}), \\
\Delta^{\mu\nu}(x,p) &=& \Xi\, g^{\mu\nu} + p^{\mu} \frac{\partial \Xi}{\partial p_{\nu}} +  p^{\nu}\frac{\partial \Xi}{\partial p_{\mu}} + \frac12  Q \frac{\partial^{2}\Xi}{\partial p_{\mu}\partial p_{\nu}}.
\end{array}
\label{eq:formalism_gtilde_expand_final}   
\eea
This perturbative expansion is justified by the smooth undeformed limit $\Omega_{\beta} \to 1$ as $\beta \to 0$, together with the assumption that the deformation remains small on the physical sector under consideration. Substituting Eq.  \eqref{eq:formalism_Xi_p_1_final} and Eq.  \eqref{eq:formalism_Xi_p_2_final} into Eq. \eqref{eq:formalism_gtilde_expand_final} and using $p^{\mu}=\sqrt{|Q|}\,u^{\mu}$ with some trivial algebra, one obtains 
\begin{equation}
\Delta^{\mu\nu} = \Xi\,g^{\mu\nu} + \frac12 \left(u^{\mu}\Pi^{\rho\nu} + u^{\nu} \Pi^{\rho\mu} - u^{\rho} \Pi^{\mu\nu}\right)\Xi_{|\rho} + \frac{\varepsilon}{2} \Pi^{\rho\mu} \Pi^{\sigma\nu}\Xi_{|\rho\sigma}.
\label{eq:formalism_Delta_explicit_final}
\end{equation}
Equation \eqref{eq:formalism_Delta_explicit_final} is the explicit leading anisotropic correction to the Hamilton metric generated by modular quantum kinematics. For sufficiently small $|\beta|$, standard perturbation theory implies that the inverse exists and is given by
\begin{equation}
\widetilde{g}_{\mu\nu}(x,p) = g_{\mu\nu}(x) - \beta\, \Delta_{\mu\nu}(x,p) + \mathcal{O}(\beta^{2}), \;\;\; \mathtt{where}\;\; \Delta_{\mu\nu} := g_{\mu\alpha} g_{\nu\beta} \Delta^{\alpha\beta}.
\label{eq:formalism_gtilde_inverse_final}
\end{equation}

At this stage three distinct geometric levels must be carefully separated. The first level is the full anisotropic Hamilton metric $\widetilde{g}^{\mu\nu}(x,p)$ on the conic domain $\mathcal{A}^{\ast}_{0}\subset T^{\ast}M$. The second level arises only after choosing a section
\begin{equation}
\sigma:M\rightarrow PT^{\ast}M_{+}, \qquad x \mapsto [p(x)],
\label{eq:formalism_sigma_final}
\end{equation}
where $p(x)$ is a representative covector field of the projective class $[p(x)]$. In the present theory $\sigma$ has a clear physical meaning: it selects, at each spacetime point, the projective momentum ray that defines the effective observer sector used to pull back the anisotropic geometry to spacetime.\footnote{Thus $\sigma$ is not a gauge artifact. It is an auxiliary but physically meaningful choice of effective congruence. In the cosmological applications developed below, the section $\sigma$ is fixed to the comoving FLRW congruence, so that $p(x)$ is aligned with the cosmological rest-frame momentum direction. No independent variation of $\sigma$ is considered in that sector. In the present paper it is treated as an externally specified effective congruence. As emphasized, in cosmological applications it is specialized to the comoving FLRW rest-frame momentum ray. In more general situations it may be chosen by symmetry, by phenomenological input, or by an additional dynamical principle imposed on the congruence.} 

The section-induced pseudo-Riemannian metric is then given as
\begin{equation}
\widetilde{g}^{\sigma}_{\mu\nu}(x) = \widetilde{g}_{\mu\nu}(x,p(x)).
\label{eq:formalism_pullback_metric_final}
\end{equation}
In general this pullback is not conformal. By when writing
\begin{equation}
\widetilde{g}^{\sigma}_{\mu\nu}(x) = g_{\mu\nu}(x) - \beta\,\Delta^{\sigma}_{\mu\nu}(x) + \mathcal{O}(\beta^{2}), \qquad
\Delta^{\sigma}_{\mu\nu}(x) := \Delta_{\mu\nu}(x,p(x)),
\label{eq:formalism_pullback_expand_final}
\end{equation}
the most general FLRW-compatible homogeneous and isotropic decomposition with respect to the cosmological four-velocity $U_{\mu}$ is then given as
\begin{equation}
\Delta^{\sigma}_{\mu\nu} = A(x)\, g_{\mu\nu} + B(x)\, U_{\mu} U_{\nu}.
\label{eq:formalism_AB_decomposition_final}
\end{equation}
A consistent isotropic conformal reduction requires that
\begin{equation}
B(x) = 0, \qquad \Delta^{\sigma}_{\mu\nu} = A(x)\, g_{\mu\nu}.
\label{eq:formalism_isotropic_condition_final}
\end{equation}
Only under this additional symmetry restriction does one reach the third level, namely the conformal sector,
\begin{equation}
\widetilde{g}^{\mathrm{conf}}_{\mu\nu}(x) = C(x)\, g_{\mu\nu}(x).
\label{eq:formalism_conformal_sector_final}
\end{equation}
The conformal sector is therefore not fundamental. It is the final isotropic truncation of the section-induced geometry and must not be confused with the underlying Hamiltonian geometry on phase space. The physical interpretation is that the anisotropic phase-space response is first compressed into a section-dependent spacetime geometry and only then further reduced, by symmetry, to a cosmological scalar deformation.

The scalar conformal factor can be written as
\begin{equation}
C(x) = 1 + \epsilon(x),
\label{eq:formalism_C_def_final}
\end{equation}
where $\epsilon(x)$ must be kept distinct from $\varepsilon = \mathrm{sign}(Q)$. To first nontrivial order one sets
\begin{equation}
\epsilon(x) = \beta\, \Xi_{\sigma}(x) + \kappa\, \mathfrak{A}_{\sigma}(x) + \mathcal{O}(\beta^{2}, \kappa^{2}, \beta \kappa),
\label{eq:formalism_epsilon_def_final}
\end{equation}
where $\Xi_{\sigma}(x) := \Xi(x,[p(x)];\eta,\Lambda)$. The first term in Eq. \eqref{eq:formalism_epsilon_def_final} is the derived pullback of the modular Hamiltonian deformation. The second term is included as an additional first-order effective scalar invariant at the same perturbative order, rather than as a fully microscopic consequence of the modular sector. To avoid confusion with the conic domain $\mathcal{A}_{0}^{\ast}$, the section-dependent acceleration invariant is denoted by $\mathfrak{A}_{\sigma}(x)$ and defined as follows:
\begin{equation}
\mathfrak{A}_{\sigma}(x) := \frac{g^{\mu\nu}\, \nabla_{\tau}\, u_{\mu}\, \nabla_{\tau}\, u_{\nu}}{\mathscr{F}^{2}}, \;\;\; \mathtt{where} \;\; \mathscr{F} = \left(\frac{c^{7}}{\hbar G}\right)^{1/2}.
\label{eq:formalism_Asigma_final}
\end{equation}
Unlike $\Xi_{\sigma}$, which is derived from the modular Hamiltonian deformation, the invariant $\mathfrak{A}_{\sigma}$ should be regarded as an additional effective contribution at the same perturbative order in the reduced spacetime theory. It is therefore not tied uniquely to the modular sector.

The scale $\mathscr{F}$ is motivated by maximal-force and maximal-acceleration considerations \cite{Tawfik:2023rrm,Tawfik:2023kxq,Caianiello:1981jq,caianiello1984maximal,Brandt:1988sh}. Since $u_{\mu}$ is originally defined on projective cotangent space, its spacetime covariant derivative along the chosen section is computed by the chain rule,
\begin{equation}
\nabla_{\alpha} u_{\mu}\bigl(x,[p(x)]\bigr) = \frac{\partial u_{\mu}}{\partial x^{\alpha}}\Big|_{[p]} + \frac{\partial u_{\mu}}{\partial p_{\rho}}\Big|_{[p]}\nabla_{\alpha}p_{\rho}(x).
\label{eq:formalism_chainrule_u_final}
\end{equation}
With $\nabla_{\tau} := U^{\alpha} \nabla_{\alpha}$, one finds that
\begin{equation}
\nabla_{\tau} u_{\mu} = U^{\alpha} \nabla_{\alpha} u_{\mu},
\label{eq:formalism_nabla_tau_u_final}
\end{equation}
and therefore $\mathfrak{A}_{\sigma}(x)$ is well defined on the section-induced geometry and dimensionless by construction.

The matter sector enters only after the reduction to effective spacetime. Accordingly, in the field equations below the quantity $\langle T_{\mu\nu}\rangle$ is interpreted as the effective semiclassical spacetime energy-stress tensor of matter in the same reduced sector, not as a direct phase-space tensor on $T^{\ast}M$. In particular, for the inflationary analysis this will be specialized to the effective matter source appropriate to the chosen FLRW congruence.

Notice that the undeformed limit of the theory appears to be immediate. For instance, if $\beta \to 0,\; \kappa \to 0,\;$ and  $\Omega_{\beta} \to 1$, one finds that
\begin{equation}
\widetilde{g}^{\mu\nu}(x,p) \to g^{\mu\nu}(x), \qquad \widetilde{g}^{\sigma}_{\mu\nu}(x) \to g_{\mu\nu}(x), \qquad \widetilde{g}^{\mathrm{conf}}_{\mu\nu}(x) \to g_{\mu\nu}(x).
\label{eq:formalism_classical_limit_final}
\end{equation}
Thus general relativity is straightforwardly recovered as the strict undeformed limit. Furthermore, the logical chain of the construction leads to 
\begin{equation}
(\omega,\eta,\Lambda) \Longrightarrow \{\ket{\Psi_{x,[p]}}\} \Longrightarrow \Omega_{\beta}(x,[p]) \Longrightarrow
H_{\beta}(x,p) \Longrightarrow \widetilde{g}^{\mu\nu}(x,p) \Longrightarrow \widetilde{g}^{\sigma}_{\mu\nu}(x) \Longrightarrow \widetilde{g}^{\mathrm{conf}}_{\mu\nu}(x).
\label{eq:formalism_logic_chain_final}
\end{equation}
In particular, we find that the conformal sector is not the starting point of the theory but the final isotropic truncation of a fully anisotropic Hamiltonian geometry induced by modular quantum phase-space kinematics. 

The intended implication of this new approach is broader than a cosmological conformal toy model: the formalism provides a controlled route from a modular quantum phase-space structure to direction-dependent effective gravity, with possible applications to cosmology, tidal observables, focusing and defocusing effects, and modified gravitational propagation. 

Having determined the metric structures at the anisotropic, pulled-back, and conformally reduced levels, the corresponding Levi--Civita connections and curvature tensors is constructed Section \ref{sec:CivitaCrvt}.

\subsection{Levi--Civita Connections and Curvature Tensors}
\label{sec:CivitaCrvt}

After the pullback and the isotropic conformal truncation, one obtains an ordinary Lorentzian metric on $M$ of the form given in Eq. \eqref{eq:formalism_conformal_sector_final}, where $C(x)$ is expressed in Eq. \eqref{eq:formalism_C_def_final} with  $\epsilon(x)$ is the scalar conformal deformation, Section~\ref{sec:formalism}. In this regard, let us emphasize that Eq. \eqref{eq:formalism_conformal_sector_final} is exact. The perturbative expansion is introduced only afterwards, when one wishes to isolate the leading semiclassical correction around the undeformed background metric $g_{\mu\nu}$. The inverse conformal metric is therefore exactly
\begin{equation}
\widetilde{g}_{\mathrm{conf}}^{\mu\nu}(x) = C^{-1}(x)\, g^{\mu\nu}(x).
\label{eq:inverse_conformal_exact_final}
\end{equation}
For $|\epsilon| \ll 1$, the expansion to the first order leads to
\begin{equation}
\widetilde{g}_{\mathrm{conf}}^{\mu\nu}(x) = (1 - \epsilon\, g^{\mu\nu}(x) + \mathcal{O}(\epsilon^{2}).
\label{eq:inverse_conformal_final}
\end{equation}
Likewise, the metric itself may be expanded as
\begin{equation}
\widetilde{g}^{\mathrm{conf}}_{\mu\nu}(x) = g_{\mu\nu}(x) + \epsilon(x)\,g_{\mu\nu}(x) + \mathcal{O}(\epsilon^{2}),
\label{eq:metric_conformal_expand_final}
\end{equation}
but it is advantageous to keep Eq. \eqref{eq:metric_conformal_expand_final} is exact form until the curvature tensors are computed.

Let us now derive the Levi--Civita connection of the background metric $g_{\mu\nu}$,
\begin{equation}
\Gamma^{\lambda}_{\mu\nu} = \frac12 g^{\lambda\sigma}
\left(\partial_{\mu} g_{\nu\sigma} + \partial_{\nu} g_{\mu\sigma} - \partial_{\sigma} g_{\mu\nu}\right),
\label{eq:Gamma_classical_final}
\end{equation}
while the Levi--Civita connection of the conformally related metric $\widetilde{g}^{\mathrm{conf}}_{\mu\nu}$ is given as 
\begin{equation}
\widetilde{\Gamma}^{\lambda}_{\mu\nu} = \frac12 \widetilde{g}_{\mathrm{conf}}^{\lambda\sigma} \left(\partial_{\mu} \widetilde{g}^{\mathrm{conf}}_{\nu\sigma} + \partial_{\nu} \widetilde{g}^{\mathrm{conf}}_{\mu\sigma} - \partial_{\sigma} \widetilde{g}^{\mathrm{conf}}_{\mu\nu}\right).
\label{eq:Gamma_tilde_final}
\end{equation}
By using $\partial_{\mu} \widetilde{g}^{\mathrm{conf}}_{\nu\sigma} = (\partial_{\mu}C)\, g_{\nu\sigma} + C\, \partial_{\mu} g_{\nu\sigma}$ and Eq.  \eqref{eq:inverse_conformal_exact_final}, one finds that the exact expression of Eq. \eqref{eq:Gamma_tilde_final} reads
\begin{equation}
\widetilde{\Gamma}^{\lambda}_{\mu\nu} = \Gamma^{\lambda}_{\mu\nu} + \frac12\, C^{-1} \left(\delta^{\lambda}_{\nu} \partial_{\mu} C + \delta^{\lambda}_{\mu} \partial_{\nu} C - g_{\mu\nu} g^{\lambda\sigma} \partial_{\sigma} C\right).
\label{eq:Gamma_tilde_exact_final}
\end{equation}
As intended this formula is exact for any smooth positive conformal factor $C$. To extract the leading semiclassical correction one expands around the undeformed sector $C=1$, namely around the regime in which $|\epsilon| \ll 1$. Since $C^{-1} = 1 - \epsilon + \mathcal{O}(\epsilon^{2})$ and $\partial_{\mu} C = \partial_{\mu}\epsilon$, the Levi--Civita connection becomes
\begin{equation}
\widetilde{\Gamma}^{\lambda}_{\mu\nu} = \Gamma^{\lambda}_{\mu\nu} + \delta\Gamma^{\lambda}_{\mu\nu} + \mathcal{O}(\epsilon^{2}),
\label{eq:Gamma_tilde_expand_final}
\end{equation}
where the first-order correction is given as
\begin{equation}
\delta\Gamma^{\lambda}_{\mu\nu} = \frac12 \left(\delta^{\lambda}_{\nu} \nabla_{\mu} \epsilon + \delta^{\lambda}_{\mu} \nabla_{\nu}\epsilon - g_{\mu\nu} \nabla^{\lambda} \epsilon\right).
\label{eq:deltaGamma_final}
\end{equation}
Here $\nabla_{\mu} \epsilon = \partial_{\mu} \epsilon$ because $\epsilon$ is a scalar and $\nabla^{\lambda} \epsilon = g^{\lambda\sigma} \nabla_{\sigma} \epsilon$ is the corresponding contravariant derivative obtained by index raising with the background metric.

The sign convention for curvature is fixed by
\begin{equation}
\left[\widetilde{\nabla}_{\mu}, \widetilde{\nabla}_{\nu}\right] V^{\rho} = \widetilde{R}^{\rho}_{\sigma\mu\nu} V^{\sigma},
\label{eq:Riemann_sign_convention_final}
\end{equation}
so that the modified Riemann curvature tensor reads
\begin{equation}
\widetilde{R}^{\rho}_{\sigma\mu\nu} = \partial_{\mu}\widetilde{\Gamma}^{\rho}_{\nu\sigma} - \partial_{\nu}\widetilde{\Gamma}^{\rho}_{\mu\sigma} + \widetilde{\Gamma}^{\rho}_{\mu\lambda} \widetilde{\Gamma}^{\lambda}_{\nu\sigma} - \widetilde{\Gamma}^{\rho}_{\nu\lambda} \widetilde{\Gamma}^{\lambda}_{\mu\sigma}.
\label{eq:Riemann_tilde_final}
\end{equation}
Substituting Eq. \eqref{eq:Gamma_tilde_expand_final} into Eq. \eqref{eq:Riemann_tilde_final} and discarding quadratic terms in $\delta\Gamma$, one obtains that
\bea
\begin{array}{ccl}
\widetilde{R}^{\rho}_{\sigma\mu\nu} &=& R^{\rho}_{\sigma\mu\nu} + \delta R^{\rho}_{\sigma\mu\nu} + \mathcal{O}(\epsilon^{2}), \\
\delta R^{\rho}_{\sigma\mu\nu} &=& \nabla_{\mu} \delta\Gamma^{\rho}_{\nu\sigma} - \nabla_{\nu}\delta\Gamma^{\rho}_{\mu\sigma}.
\end{array}
\label{eq:RmnTns_final}   
\eea
Using Eq. \eqref{eq:deltaGamma_final} and the fact that the second derivatives commute on scalars, Eq. \eqref{eq:RmnTns_final} can be rewritten explicitly as
\begin{equation}
\delta R^{\rho}_{\sigma\mu\nu} = \frac12 \left(\delta^{\rho}_{\nu} \nabla_{\mu} \nabla_{\sigma} \epsilon - \delta^{\rho}_{\mu}\nabla_{\nu} \nabla_{\sigma} \epsilon - g_{\nu\sigma} \nabla_{\mu} \nabla^{\rho} \epsilon + g_{\mu\sigma} \nabla_{\nu} \nabla^{\rho} \epsilon\right).
\label{eq:deltaRiemann_explicit_final}
\end{equation}
This explicit formula is the first-order conformal correction to the Riemann curvature tensor in mixed-index form and is consistent with the convention given in Eq. \eqref{eq:Riemann_sign_convention_final}.

Contracting the Riemann curvature tensor, $\widetilde{g}^{\delta\sigma} (\widetilde{g}_{\delta\rho} \widetilde{R}^{\rho}_{\sigma\mu\nu})$, results in the Ricci curvature tensor,
\bea
\begin{array}{ccl}
\widetilde{R}_{\mu\nu} 
&=& R_{\mu\nu} + \delta R_{\mu\nu} + \mathcal{O}(\epsilon^{2}), \\
\delta R_{\mu\nu} &=& \nabla_{\rho} \delta\Gamma^{\rho}_{\mu\nu} - \nabla_{\nu}\delta\Gamma^{\rho}_{\rho\mu}.
\end{array}
\label{eq:deltaRicci_final0}
\eea
Substituting Eq. \eqref{eq:deltaGamma_final}  into Eq. \eqref{eq:deltaRicci_final0} yields
\bea
\delta R_{\mu\nu} = - \nabla_{\mu} \nabla_{\nu} \epsilon - \frac12 g_{\mu\nu} \Box \epsilon,\;\;\; \mathtt{where}\;\;\Box \epsilon := g^{\rho\sigma} \nabla_{\rho} \nabla_{\sigma} \epsilon.
\label{eq:deltaRicci_final}
\eea
Equations \eqref{eq:RmnTns_final} and \eqref{eq:deltaRicci_final0} are thus written in fully parallel form: both are expressed as covariant derivatives of $\delta\Gamma$, and the explicit result for the Ricci tensor follows after contraction.

The Ricci scalar can be derived as follows:
\begin{equation}
\widetilde{R} = \widetilde{g}_{\mathrm{conf}}^{\mu\nu} \widetilde{R}_{\mu\nu}.
\label{eq:RicciScalar_def_final}
\end{equation}
By using Eq. \eqref{eq:inverse_conformal_final} and Eq. \eqref{eq:deltaRicci_final0}, one finds that
\bea
\widetilde{R} &=&  (1-\epsilon) g^{\mu\nu} \left(R_{\mu\nu} + \delta R_{\mu\nu}\right) + \mathcal{O}(\epsilon^{2}) \nn \\
&=& R - 3 \Box \epsilon - \epsilon R + \mathcal{O}(\epsilon^{2}).
\label{eq:RicciScalar_final}
\eea
In this derivation, we use $g^{\mu\nu} \delta R_{\mu\nu} = - \Box \epsilon - \frac12 g^{\mu\nu} g_{\mu\nu} \Box \epsilon = - 3 \Box \epsilon$ in four spacetime dimensions.

Now, the Einstein tensor of the conformally reduced metric can be constructed as follows:
\begin{equation}
\widetilde{G}_{\mu\nu} = \widetilde{R}_{\mu\nu} -  \frac12 \widetilde{g}^{\mathrm{conf}}_{\mu\nu} \widetilde{R}.
\label{eq:Einstein_tilde_def_final}
\end{equation}
Substituting Eqs. \eqref{eq:metric_conformal_expand_final}, \eqref{eq:deltaRicci_final0}, and \eqref{eq:RicciScalar_final} into Eq. \eqref{eq:Einstein_tilde_def_final} and keeping only first-order terms in $\epsilon$, one obtains
\bea
\widetilde{G}_{\mu\nu} &=& \left(R_{\mu\nu} + \delta R_{\mu\nu}\right) - \frac12 (1 + \epsilon) g_{\mu\nu} \left(R - \epsilon  R - 3 \Box \epsilon \right) + \mathcal{O}(\epsilon^{2}) \nn \\
&=& G_{\mu\nu} - \nabla_{\mu} \nabla_{\nu} \epsilon + g_{\mu\nu} \Box \epsilon + \mathcal{O}(\epsilon^{2}),
\label{eq:Einstein_final}
\eea
where Eq. \eqref{eq:deltaRicci_final} has been used in the last step. Therefore $\delta G_{\mu\nu}$ can be determined from the subtraction of $G_{\mu\nu}$ from $\widetilde{G}_{\mu\nu}$, 
\begin{equation}
\delta G_{\mu\nu}  = - \nabla_{\mu} \nabla_{\nu} \epsilon + g_{\mu\nu} \Box \epsilon + \mathcal{O}(\epsilon^{2}).
\label{eq:deltaG_final}
\end{equation}

Using the conformal scalar introduced in Section~\ref{sec:formalism}, namely, 
\begin{equation}
\epsilon(x) = \beta\, \Xi_{\sigma}(x) + \kappa\, \mathfrak{A}_{\sigma}(x) + \mathcal{O}(\beta^{2},\kappa^{2},\beta\kappa).
\label{eq:epsilon_def_final_reused}
\end{equation}
Accordingly, one finds that 
\bea
\nabla_{\mu} \epsilon &=& \beta\, \nabla_{\mu} \Xi_{\sigma} + \kappa\, \nabla_{\mu} \mathfrak{A}_{\sigma} + \mathcal{O}(\beta^{2},\kappa^{2},\beta\kappa),
\label{eq:grad_epsilon_final} \\
\nabla_{\mu} \nabla_{\nu} \epsilon &=& \beta\, \nabla_{\mu} \nabla_{\nu} \Xi_{\sigma} + \kappa\, \nabla_{\mu} \nabla_{\nu} \mathfrak{A}_{\sigma} + \mathcal{O}(\beta^{2},\kappa^{2},\beta\kappa).
\label{eq:hessian_epsilon_final}
\eea
Thus, the effective semiclassical Einstein field equations can be constructed as
\begin{equation}
\widetilde{G}_{\mu\nu} = 8 \pi G\, \langle T_{\mu\nu}\rangle,
\label{eq:EFE_tilde_final}
\end{equation}
where $\widetilde{G}_{\mu\nu}$ is determined by the pullback of the projective Hamilton deformation followed by the isotropic conformal truncation. To first order in the deformation parameters, Eq.~\eqref{eq:EFE_tilde_final} becomes
\bea
G_{\mu\nu} - \beta\, \nabla_{\mu} \nabla_{\nu} \Xi_{\sigma} - \kappa\, \nabla_{\mu} \nabla_{\nu} \mathfrak{A}_{\sigma} + g_{\mu\nu} \left(\beta\, \Box \Xi_{\sigma} + \kappa\, \Box \mathfrak{A}_{\sigma}\right)
&=&  8 \pi G\, \langle T_{\mu\nu}\rangle  
\mathcal{O}(\beta^{2}, \kappa^{2}, \beta \kappa).
\label{eq:EFE_first_order_final}
\eea
Therefore, after the phase-space deformation has been pulled back to spacetime and then consistently reduced to the conformal FLRW-compatible sector, the semiclassical corrections enter the Einstein field equations through second derivatives of the section-dependent modular scalar $\Xi_{\sigma}$ and the acceleration scalar $\mathfrak{A}_{\sigma}$. 

Section~\ref{sec:actions} discusses the derivation of the corresponding quantum-corrected Einstein--Hilbert action.

\subsection{Effective Einstein--Hilbert Action}
\label{sec:actions}

Before imposing the conformal truncation, the natural gravitational action associated with the section-induced effective spacetime metric is the Einstein--Hilbert functional built from the pullback metric $\widetilde{g}^{\sigma}_{\mu\nu}(x)$. Referring to the section $\sigma:M \to PT^{\ast}M_{+}$, one has
\begin{equation}
\widetilde{S}_{\mathrm{EH}}[\sigma] = \frac{1}{16 \pi G} \int_{M}  d^{4}x\, \sqrt{-\widetilde{g}^{\sigma}}\, \widetilde{R}[\sigma].
\label{eq:SEH_sigma_final}
\end{equation}
This action is geometrically primary in the spacetime sector because it is the direct pullback of the anisotropic Hamiltonian geometry to an ordinary Lorentzian metric on the base manifold. In particular, $\widetilde{S}_{\mathrm{EH}}[\sigma]$ is the action that follows directly from the section-induced metric $\widetilde{g}^{\sigma}_{\mu\nu}$, before any isotropic truncation is imposed.

The conformal action used in the homogeneous and isotropic sector is obtained only after the further reduction
\begin{equation}
\widetilde{g}^{\sigma}_{\mu\nu}(x) \longrightarrow \widetilde{g}^{\mathrm{conf}}_{\mu\nu}(x) = C(x)\, g_{\mu\nu}(x),
\label{eq:SEH_sigma_to_conf_final}
\end{equation}
with $C(x)=1+\epsilon(x)$, as introduced in Eq.~\eqref{eq:formalism_C_def_final}. Then, the corresponding Einstein--Hilbert action reads
\begin{equation}
\widetilde{S}_{\mathrm{EH}}^{\mathrm{conf}} = \frac{1}{16 \pi G}
\int_{M} d^{4}x\, \sqrt{-\widetilde{g}^{\mathrm{conf}}}\, \widetilde{R}^{\mathrm{conf}}.
\label{eq:SEH_conf_final}
\end{equation}
Thus $\widetilde{S}_{\mathrm{EH}}[\sigma]$ and $\widetilde{S}_{\mathrm{EH}}^{\mathrm{conf}}$ should not be viewed as competing definitions, but as the primary and symmetry-reduced actions associated with two different geometric levels of the same construction.

In four-dimensional  spacetime the determinant of a conformally related metric scales as
\begin{equation}
\sqrt{-\widetilde{g}^{\mathrm{conf}}} = C^{2} \sqrt{-g}.
\label{eq:sqrtgtilde_exact_final}
\end{equation}
Using $C=1+\epsilon$, one obtains the first-order expansion
\begin{equation}
\sqrt{-\widetilde{g}^{\mathrm{conf}}} = (1 + 2 \epsilon) \sqrt{-g} + \mathcal{O}(\epsilon^{2}).
\label{eq:sqrtgtilde_expand_final}
\end{equation}
From Eq.~\eqref{eq:RicciScalar_final} one also finds that
\begin{equation}
\widetilde{R}^{\mathrm{conf}} = R - \epsilon R - 3 \Box \epsilon + \mathcal{O}(\epsilon^{2}),
\label{eq:Rconf_expand_final}
\end{equation}
where $\Box\epsilon = g^{\mu\nu} \nabla_{\mu} \nabla_{\nu} \epsilon$. Multiplying Eq. \eqref{eq:sqrtgtilde_expand_final} and Eq. \eqref{eq:Rconf_expand_final} leads to
\bea
\sqrt{-\widetilde{g}^{\mathrm{conf}}}\, \widetilde{R}^{\mathrm{conf}} 
&=& \sqrt{-g} \left(R + \epsilon R - 3 \Box \epsilon\right) + \mathcal{O}(\epsilon^{2}).
\label{eq:sqrtgR_final}
\eea
The term $\sqrt{-g}\, \Box\epsilon$ is a total divergence. Therefore, 
\begin{equation}
\sqrt{-g}\, \Box \epsilon = \partial_{\mu}\! \left(\sqrt{-g}\, \nabla^{\mu} \epsilon\right),
\label{eq:total_divergence_boxepsilon_final}
\end{equation}
and under the standard assumption of compactly supported variations or equivalently after supplementing the action with the appropriate boundary term, it does not contribute to the bulk field equations. Hence the conformally reduced Einstein--Hilbert action becomes
\begin{equation}
\widetilde{S}_{\mathrm{EH}}^{\mathrm{conf}} = \frac{1}{16 \pi G} 
\int_{M} d^{4}x\, \sqrt{-g} \left[R + \epsilon R\right] + \mathcal{O}(\epsilon^{2}).
\label{eq:SEH_firstorder_final}
\end{equation}

At this point it is important to distinguish the origin of the two contributions entering $\epsilon$. The term $\Xi_{\sigma}$ is the section pullback of the modular Hamiltonian deformation derived in Section~\ref{sec:formalism}, whereas $\mathfrak{A}_{\sigma}$ is an additional first-order effective invariant retained at the same perturbative order in the spacetime reduction. Thus
\bea
\begin{array}{ccl}
\epsilon(x) &=&  \beta\, \Xi_{\sigma}(x) + \kappa\, \mathfrak{A}_{\sigma}(x) + \mathcal{O}(\beta^{2},\kappa^{2},\beta\kappa), \\
\Xi_{\sigma}(x) &=&\Xi(x,[p(x)]; \eta,\Lambda).
\end{array}
\label{eq:epsilon_recall_actions_final} 
\eea
Substituting Eq.~\eqref{eq:epsilon_recall_actions_final} into Eq.~\eqref{eq:SEH_firstorder_final} leads to
\bea
\widetilde{S}_{\mathrm{EH}}^{\mathrm{conf}} &=& \frac{1}{16 \pi G} \int_{M} d^{4}x\, \sqrt{-g} \left[R + \beta\, \Xi_{\sigma} R +  \kappa\, \mathfrak{A}_{\sigma} R\right] + \mathcal{O}(\beta^{2},\kappa^{2},\beta\kappa).
\label{eq:SEH_final}
\eea
The first correction correction term, $\beta\, \Xi_{\sigma}R$, is the derived modular contribution obtained from the phase-space Hamiltonian deformation after pullback and isotropic reduction. The second correction term, $\kappa\, \mathfrak{A}_{\sigma}R$, should instead be interpreted as an additional effective contribution retained at the same order in the reduced spacetime theory. It is therefore useful to regard Eq.~\eqref{eq:SEH_final} as an effective first-order action consisting of a derived modular sector and an extended phenomenological sector.

At first order there is no kinetic term quadratic in $\nabla \epsilon$. Such terms arise only at second order in the conformal expansion. It is also essential to distinguish the actions Eq.  \eqref{eq:SEH_sigma_final} and Eq. \eqref{eq:SEH_final}. Equation \eqref{eq:SEH_sigma_final} is the geometrically primary section-dependent action obtained directly from the pullback of the anisotropic Hamiltonian geometry, whereas Eq.~\eqref{eq:SEH_final} is the cosmological conformal truncation appropriate to the homogeneous and isotropic sector.

The conformal truncation is sufficient for the reduced spacetime description, but it suppresses the momentum-direction dependence that is intrinsic to the underlying phase-space geometry. For this reason one must now return to the full anisotropic Hamiltonian framework in order to identify the nonlinear connection, the metric-compatible distinguished connection, and the corresponding phase-space curvature tensors. These objects encode the genuinely Hamilton--Finsler part of the geometry and are required if one wishes to understand parallel transport, curvature, geodesic deviation, and anisotropic tidal response before the symmetry reduction to spacetime.

The Hamilton--Finsler nonlinear and metric-compatible $d$-connection are discussed in Section \ref{sec:HF_connection}.

\subsection{Hamilton--Finsler Nonlinear and Metric-Compatible $d$-Connection}
\label{sec:HF_connection}

Let us now return to the full anisotropic phase-space geometry before pullback. The deformed Hamiltonian, Eq.~\eqref{eq:formalism_Hbeta_final}, is given as
\begin{equation}
H_{\beta}(x,p) = \frac12\, \Omega_{\beta}(x,[p])\, Q(x,p),
\label{eq:Hbeta_HF_final}
\end{equation}
and defines a regular Hamilton space on $\mathcal{A}^{\ast}_{0}\subset T^{\ast}M$, with regularity understood in the sense of Eq.~\eqref{eq:formalism_regularity_assumption}. Its fundamental tensor is the fiber Hessian
\begin{equation}
\widetilde{g}^{\mu\nu}(x,p) = \frac{\partial^{2} H_{\beta}(x,p)}{\partial p_{\mu} \partial p_{\nu}},
\label{eq:Hamilton_metric_recall_final}
\end{equation}
together with the inverse tensor $\widetilde{g}_{\mu\nu}(x,p)$. The purpose of the present section is to construct the intrinsic transport and curvature data of this Hamilton space before any spacetime pullback is imposed.

The Hamiltonian vector field $X_{H}$ is defined intrinsically by
\begin{equation}
\iota_{X_{H}} \omega = dH_{\beta}.
\label{eq:Hamilton_vector_field_def_final}
\end{equation}
Using the canonical symplectic form $\omega = dp_{\mu} \wedge dx^{\mu}$, one obtains
\begin{equation}
X_{H} = \frac{\partial H_{\beta}}{\partial p_{\mu}} \frac{\partial}{\partial x^{\mu}} - \frac{\partial H_{\beta}}{\partial x^{\mu}}\frac{\partial}{\partial p_{\mu}}.
\label{eq:Hamilton_vector_field_final}
\end{equation}
The integral curves of $X_H$ satisfy the Hamilton equations
\begin{equation}
\dot{x}^{\mu} = \frac{\partial H_{\beta}}{\partial p_{\mu}},
\qquad
\dot{p}_{\mu} = - \frac{\partial H_{\beta}}{\partial x^{\mu}},
\label{eq:ham_eqs_final}
\end{equation}
which define the canonical Hamiltonian flow on $T^{\ast}M$.

To define the nonlinear connection on the cotangent bundle, one introduces the semispray coefficients associated with the regular Hamiltonian $H_{\beta}$,
\begin{equation}
\mathcal{G}^{\mu}(x,p) := \frac12\, \widetilde{g}^{\mu\nu} \left[\frac{\partial H_{\beta}}{\partial x^{\nu}} - \frac{\partial^{2}H_{\beta}}{\partial x^{\rho} \partial p_{\nu}} \frac{\partial H_{\beta}}{\partial p_{\rho}}\right].
\label{eq:Gmu_final}
\end{equation}
This expression corresponds to the canonical semispray associated with a regular two-homogeneous Hamiltonian on the cotangent bundle \cite{BUCATARU2007269}, ensuring compatibility with the Hamiltonian flow and the induced nonlinear connection structure. Since $H_{\beta}$ is two-homogeneous in momentum $p$, the quantity $\partial H_{\beta}/\partial p_{\rho}$ is one-homogeneous, $\partial^{2} H_{\beta}/(\partial x^{\rho}\partial p_{\nu})$ is also one-homogeneous, while $\widetilde{g}^{\mu\nu}$ is zero-homogeneous. Consequently $\mathcal{G}^{\mu}$ is two-homogeneous in the fiber variables, as required for a canonical semispray coefficient on the cotangent bundle.

The canonical nonlinear connection associated with $H_{\beta}$ is defined by
\begin{equation}
\widetilde{N}_{\mu\nu}(x,p) := \frac{\partial \mathcal{G}_{\mu}}{\partial p_{\nu}}, \qquad \mathcal{G}_{\mu} := \widetilde{g}_{\mu\rho} \mathcal{G}^{\rho}.
\label{eq:Nlower_final}
\end{equation}
Equivalently, after raising one index, one obtains
\begin{equation}
\widetilde{N}^{\mu}_{\nu} = \widetilde{g}^{\mu\rho} \widetilde{N}_{\rho\nu}.
\label{eq:Nupper_final}
\end{equation}
This nonlinear connection defines the horizontal-vertical splitting,
\begin{equation}
TT^{\ast}M = \mathcal{H} \oplus \mathcal{V},
\label{eq:HVsplitting_final}
\end{equation}
with the adapted basis
\begin{equation}
\widetilde{\delta}_{\mu} = \frac{\partial}{\partial x^{\mu}} - \widetilde{N}^{\nu}_{\mu}\, \bar{\partial}_{\nu},
\qquad
\bar{\partial}_{\nu} := \frac{\partial}{\partial p_{\nu}},
\label{eq:adapted_basis_final}
\end{equation}
and the dual cobasis
\begin{equation}
dx^{\mu}, \qquad \delta p_{\mu} = dp_{\mu} + \widetilde{N}^{\nu}_{\mu} dx^{\nu}.
\label{eq:adapted_cobasis_final}
\end{equation}
The notation $\bar{\partial}_{\nu}$ fixes unambiguously that differentiation is taken with respect to the cotangent fiber coordinates $p_{\nu}$. Geometrically, the nonlinear connection specifies how the phase-space tangent bundle is split into horizontal directions interpreted as generalized spacetime transport and vertical directions interpreted as transport in momentum space.

On this Hamilton space one introduces a metric-compatible distinguished connection, or $d$-connection, adapted to the splitting Eq. \eqref{eq:HVsplitting_final}. Its horizontal and vertical coefficients respectively are taken in the canonical Hamilton form as follows:
\bea
\begin{array}{ccl}
\widetilde{H}^{\mu}_{\nu\lambda} &=& \frac12\, \widetilde{g}^{\mu\sigma} \left(\widetilde{\delta}_{\nu} \widetilde{g}_{\sigma\lambda} + \widetilde{\delta}_{\lambda} \widetilde{g}_{\sigma\nu} - \widetilde{\delta}_{\sigma} \widetilde{g}_{\nu\lambda}\right), \\
\widetilde{C}_{\rho}^{\mu\nu} &=& \frac12\, \widetilde{g}_{\rho\sigma} \left(\bar{\partial}^{\mu} \widetilde{g}^{\sigma\nu}
+ \bar{\partial}^{\nu} \widetilde{g}^{\sigma\mu} - \bar{\partial}^{\sigma} \widetilde{g}^{\mu\nu}\right),
\end{array}
\label{eq:Hcoeff_final} 
\eea
where $\bar{\partial}^{\mu} := g^{\mu\nu} \bar{\partial}_{\nu}$. The coefficients $\widetilde{H}^{\mu}_{\nu\lambda}$ govern horizontal transport, while $\widetilde{C}_{\rho}^{\mu\nu}$ encode the vertical momentum-space response. They are chosen so that horizontal and vertical metric compatibility hold the following conditions:
\bea
\begin{array}{ccl}
\widetilde{\delta}_{\lambda} \widetilde{g}_{\mu\nu} - \widetilde{H}^{\rho}_{\mu\lambda} \widetilde{g}_{\rho\nu} - \widetilde{H}^{\rho}_{\nu\lambda} \widetilde{g}_{\mu\rho} &=& 0, \\
\bar{\partial}_{\lambda} \widetilde{g}_{\mu\nu} - \widetilde{C}_{\mu \lambda}^{\rho} \widetilde{g}_{\rho\nu} - \widetilde{C}_{\nu \lambda}^{\rho} \widetilde{g}_{\mu\rho} &=& 0.
\end{array}
\label{eq:hmetriccompatibility_final}  
\eea
These are the cotangent-bundle analogues of the usual Levi--Civita metric-compatibility conditions, split according to the nonlinear connection. In particular, the $d$-connection preserves the Hamilton metric separately under horizontal and vertical differentiation, which is the natural analogue of metric compatibility in anisotropic phase-space geometry.

The curvature of the nonlinear connection is defined as
\begin{equation}
\widetilde{\mathcal{R}}^{\mu}_{\nu\lambda} = \widetilde{\delta}_{\lambda}\widetilde{N}^{\mu}_{\nu} - \widetilde{\delta}_{\nu}\widetilde{N}^{\mu}_{\lambda}.
\label{eq:N_curvature_final}
\end{equation}
Correspondingly, the horizontal, vertical, and mixed curvature tensors of the $d$-connection are respectively given as follows:
\bea
\begin{array}{ccl}
\widetilde{R}^{\mu}_{\nu\lambda\rho} &=& \widetilde{\delta}_{\lambda} \widetilde{H}^{\mu}_{\nu\rho} - \widetilde{\delta}_{\rho}\widetilde{H}^{\mu}_{\nu\lambda} + \widetilde{H}^{\mu}_{\sigma\lambda} \widetilde{H}^{\sigma}_{\nu\rho} - \widetilde{H}^{\mu}_{\sigma\rho} \widetilde{H}^{\sigma}_{\nu\lambda}, \\
\widetilde{S}_{\rho}^{\mu\nu\lambda} &=& \bar{\partial}^{\lambda}\widetilde{C}_{\rho}^{\mu\nu} - \bar{\partial}^{\nu}\widetilde{C}_{\rho}^{\mu\lambda} + \widetilde{C}_{\sigma}^{\mu\lambda}\widetilde{C}_{\rho}^{\sigma\nu} - \widetilde{C}_{\sigma}^{\mu\nu}\widetilde{C}_{\rho}^{\sigma\lambda},\\
\widetilde{P}^{\mu\rho}_{\nu\lambda} &=& \bar{\partial}^{\rho} \widetilde{H}^{\mu}_{\nu\lambda} - \widetilde{\delta}_{\lambda} \widetilde{C}_{\nu}^{\mu\rho} + \widetilde{C}_{\sigma}^{\mu\rho} \widetilde{H}^{\sigma}_{\nu\lambda} - \widetilde{H}^{\mu}_{\sigma\lambda} \widetilde{C}_{\nu}^{\sigma\rho}.
\end{array}
\label{eq:Rhorizontal_final}
\eea
These curvature tensors capture the full anisotropic phase-space geometry prior to pullback and conformal reduction. In particular, $\widetilde{R}^{\mu}_{\nu\lambda\rho}$ controls the horizontal spacetime-like curvature, $\widetilde{S}_{\rho}^{\mu\nu\lambda}$ measures purely vertical fiber curvature, and $\widetilde{P}^{\mu\rho}_{\nu\lambda}$ describes mixed horizontal-vertical coupling. The mixed tensor is especially important because it encodes the direct coupling between spacetime transport and momentum-space anisotropy, and therefore measures the extent to which the geometry departs from an ordinary cotangent lift of pseudo-Riemannian spacetime.

To connect this structure with the perturbative Hamiltonian geometry of Section~\ref{sec:formalism}, we use the following expansions:
\bea
\widetilde{g}^{\mu\nu}(x,p) &=& g^{\mu\nu}(x) + \beta\,\Delta^{\mu\nu}(x,p) + \mathcal{O}(\beta^{2}),
\label{eq:gtilde_beta_expand_final_reused}\\
\widetilde{g}_{\mu\nu}(x,p) &=& g_{\mu\nu}(x) - \beta\,\Delta_{\mu\nu}(x,p) + \mathcal{O}(\beta^{2}).
\label{eq:gtilde_inverse_final_reused}
\eea
Since the undeformed metric $g_{\mu\nu}(x)$ is independent of the momentum variables, the background geometry on $T^{\ast}M$ is simply the cotangent lift of the spacetime Levi--Civita geometry. Consequently the horizontal coefficients expand as the follows:
\begin{equation}
\widetilde{H}^{\mu}_{\nu\lambda} = \Gamma^{\mu}_{\nu\lambda}
 + \beta\,\mathcal{H}^{\mu}_{\nu\lambda}(x,p) + \mathcal{O}(\beta^{2}),
\label{eq:H_expand_final}
\end{equation}
where $\mathcal{H}^{\mu}_{\nu\lambda} = \frac12\, g^{\mu\sigma} \left(\delta_{\nu} \Delta_{\sigma\lambda} + \delta_{\lambda} \Delta_{\sigma\nu} - \delta_{\sigma} \Delta_{\nu\lambda}\right)$, and the undeformed horizontal derivative is given as
\begin{equation}
\delta_{\mu} = \frac{\partial}{\partial x^{\mu}} - \Gamma^{\rho}_{\mu\nu}p_{\rho}\bar{\partial}^{\nu}.
\label{eq:delta_background_final}
\end{equation}
Similarly, the vertical coefficients start only at first order in $\beta$,
\begin{equation}
\widetilde{C}_{\rho}^{\mu\nu} = \beta\, \mathcal{C}_{\rho}^{\mu\nu}(x,p) + \mathcal{O}(\beta^{2}),
\label{eq:C_expand_final}
\end{equation}
with $\mathcal{C}_{\rho}^{\mu\nu} = \frac12\, g_{\rho\sigma}
\left(\bar{\partial}^{\mu} \Delta^{\sigma\nu} + \bar{\partial}^{\nu}\Delta^{\sigma\mu} - \bar{\partial}^{\sigma} \Delta^{\mu\nu}
\right)$. The nonlinear connection itself has the expansion
\begin{equation}
\widetilde{N}^{\mu}_{\nu} = \Gamma^{\mu}_{\nu\rho}p^{\rho} + \beta\,\mathcal{N}^{\mu}_{\nu}(x,p) + \mathcal{O}(\beta^{2}),
\label{eq:N_expand_final}
\end{equation}
where $\mathcal{N}^{\mu}_{\nu}$ is obtained from Eqs.~\eqref{eq:Gmu_final} - \eqref{eq:Nupper_final} by retaining the terms linear in $\Delta^{\mu\nu}$ and its derivatives. Thus the undeformed geometry is recovered as the cotangent lift of the background Levi--Civita geometry, while the first-order terms $\mathcal{N}^{\mu}_{\nu}$, $\mathcal{H}^{\mu}_{\nu\lambda}$, and $\mathcal{C}_{\rho}^{\mu\nu}$ quantify the genuinely Hamiltonian deformation induced by the modular sector.

In the undeformed limit one has
\begin{equation}
\widetilde{N}^{\mu}_{\nu} \to \Gamma^{\mu}_{\nu\rho} p^{\rho},
\qquad \widetilde{H}^{\mu}_{\nu\lambda} \to \Gamma^{\mu}_{\nu\lambda}, \qquad \widetilde{C}_{\rho}^{\mu\nu} \to 0,
\label{eq:undeformed_HF_limit_final}
\end{equation}
so the anisotropic cotangent-bundle geometry reduces to the ordinary cotangent lift of the background pseudo-Riemannian geometry. 

The derivation of the geodesic deviation structure related to Hamilton--Finsler geometry, as presented in this manuscript, is elaborated upon in Section~\ref{sec:geodesic_deviation}.

\subsection{Geodesic Deviation in Hamilton--Finsler Geometry}
\label{sec:geodesic_deviation}

The Hamilton equations, Eq.~\eqref{eq:ham_eqs_final}, define the Hamiltonian geodesic flow on $T^{\ast}M$. In order to study the relative behavior of neighboring phase-space trajectories, one introduces a deviation vector field along a reference congruence in the tangent bundle $TT^{\ast}M$. With respect to the adapted horizontal-vertical basis introduced in Eq.~\eqref{eq:adapted_basis_final}, the deviation vector is decomposed as follows:
\begin{equation}
\mathbb{X}(\tau,s) = \xi^{\mu}(\tau,s)\, \widetilde{\delta}_{\mu} + \zeta_{\mu}(\tau,s)\, \bar{\partial}^{\mu},
\label{eq:deviation_split_final}
\end{equation}
where $\xi^{\mu}$ is the horizontal deviation and $\zeta_{\mu}$ is the vertical deviation, while $\bar{\partial}^{\mu}:=g^{\mu\nu}\bar{\partial}_{\nu}$. The tangent to the reference congruence is expressed as
\begin{equation}
\mathbb{U} = U^{\mu}\widetilde{\delta}_{\mu} + V_{\mu}\bar{\partial}^{\mu},
\label{eq:reference_congruence_final}
\end{equation}
The horizontal component is expressed as
\begin{equation}
U^{\mu}=\frac{dx^{\mu}}{d\tau}.
\label{eq:horizontal_velocity_final}
\end{equation}
In the cosmological reduction only the horizontal transport survives, but the full phase-space formula must first be established before any projection is imposed. 

The horizontal covariant derivative along the congruence is defined by
\begin{equation}
\nabla_{\tau} = U^{\mu} \widetilde{\delta}_{\mu}.
\label{eq:nabla_tau_final}
\end{equation}
Its action on the horizontal deviation is
\begin{equation}
\nabla_{\tau} \xi^{\mu} = U^{\nu} \widetilde{\delta}_{\nu} \xi^{\mu} + \widetilde{H}^{\mu}_{\nu\lambda} U^{\nu} \xi^{\lambda}.
\label{eq:nabla_tau_xi_final}
\end{equation}
Differentiating once more and using the commutator relations of the adapted basis together with the curvature tensors introduced in Eqs.~\eqref{eq:Rhorizontal_final}, one obtains the Jacobi equation on phase space,
\begin{equation}
\nabla_{\tau}^{2} \xi^{\mu} = - \widetilde{R}^{\mu}_{\nu\lambda\rho}(x,p)\, U^{\nu} U^{\rho} \xi^{\lambda} - \widetilde{P}^{\mu\rho}_{\nu\lambda}(x,p)\, U^{\nu} U^{\lambda} \zeta_{\rho}.
\label{eq:HF_deviation_final}
\end{equation}
The sign in front of the curvature term is just fixed by the Riemann convention adopted earlier. With the standard Jacobi identity for a geodesic congruence, one finds that
\begin{equation}
[\widetilde{\nabla}_{\mu}, \widetilde{\nabla}_{\nu}] V^{\rho} = \widetilde{R}^{\rho}_{\sigma\mu\nu} V^{\sigma}.
\end{equation}
The second term in Eq.~\eqref{eq:HF_deviation_final} depends on the vertical deviation $\zeta_{\rho}$ and is the genuine mixed Hamilton--Finsler correction. It disappears only after the purely horizontal reduction.

Using the perturbative expansion of the horizontal curvature tensor, we obtain
\bea
\begin{array}{ccl}
\widetilde{R}^{\mu}_{\nu\lambda\rho}(x,p) &=& R^{\mu}_{\nu\lambda\rho}(g) + \beta\, \delta R^{\mu}_{\nu\lambda\rho}(x,p) + \mathcal{O}(\beta^{2}),\\
\delta R^{\mu}_{\nu\lambda\rho} &=&  \nabla_{\lambda}\mathcal{H}^{\mu}_{\nu\rho} - \nabla_{\rho} \mathcal{H}^{\mu}_{\nu\lambda}.
\end{array}
\label{eq:R_expand_dev_final} 
\eea
Hence Eq.~\eqref{eq:HF_deviation_final} can be reexpressed as
\begin{equation}
\nabla_{\tau}^{2} \xi^{\mu} = - R^{\mu}_{\nu\lambda\rho}(g)\, U^{\nu} U^{\rho} \xi^{\lambda} - \beta\, \delta R^{\mu}_{\nu\lambda\rho}\, U^{\nu} U^{\rho} \xi^{\lambda} - \widetilde{P}^{\mu\rho}_{\nu\lambda} U^{\nu} U^{\lambda} \zeta_{\rho} + \mathcal{O}(\beta^{2}).
\label{eq:HF_deviation_pert_final}
\end{equation}

After pullback along a section and in the reduction where vertical variations are neglected, one recovers the ordinary spacetime Jacobi equation
\begin{equation}
\frac{D^{2}\xi^{\mu}}{D\tau^{2}} = - \widetilde{R}^{\mu}_{\nu\lambda\rho}(x)\, U^{\nu} U^{\rho} \xi^{\lambda},
\label{eq:jacobi_spacetime_final}
\end{equation}
where $\widetilde{R}^{\mu}_{\nu\lambda\rho}(x)$ is the Riemann tensor of the conformally reduced metric $\widetilde{g}^{\mathrm{conf}}_{\mu\nu}=C\,g_{\mu\nu}$.
With the assumption that $C=e^{2\phi}$ and $\phi=\frac12 \ln C$, the exact conformal transformation of the Riemann tensor in mixed-index form reads
\bea
\widetilde{R}^{\mu}_{\nu\lambda\rho} &=&  R^{\mu}_{\nu\lambda\rho} + \delta^{\mu}_{\lambda} \nabla_{\nu}  \nabla_{\rho} \phi - \delta^{\mu}_{\rho} \nabla_{\nu}  \nabla_{\lambda} \phi + g_{\nu\rho} \nabla^{\mu} \nabla_{\lambda} \phi - g_{\nu\lambda} \nabla^{\mu} \nabla_{\rho} \phi \nn \\
&+& \delta^{\mu}_{\rho} \nabla_{\nu} \phi\, \nabla_{\lambda} \phi -
 \delta^{\mu}_{\lambda} \nabla_{\nu} \phi\, \nabla_{\rho} \phi - g_{\nu\rho} \nabla^{\mu} \phi\, \nabla_{\lambda} \phi + g_{\nu\lambda} \nabla^{\mu} \phi\, \nabla_{\rho} \phi \nn \\
&+& \left(\delta^{\mu}_{\lambda} g_{\nu\rho} - \delta^{\mu}_{\rho}g_{\nu\lambda}\right) (\nabla\phi)^{2}.
\label{eq:Riemann_conformal_exact_final}
\eea
To first order in the conformal deformation, one has
\begin{equation}
\phi = \frac12 \epsilon + \mathcal{O}(\epsilon^{2}),
\label{eq:phi_linear_final}
\end{equation}
and all terms quadratic in $\nabla\phi$ can be neglected. Therefore, the Riemann curvature tensor reads
\begin{equation}
\widetilde{R}^{\mu}_{\nu\lambda\rho} = R^{\mu}_{\nu\lambda\rho} + \frac12 \left[\delta^{\mu}_{\lambda} \nabla_{\nu} \nabla_{\rho} \epsilon - \delta^{\mu}_{\rho} \nabla_{\nu} \nabla_{\lambda}  \epsilon + g_{\nu\rho} \nabla^{\mu} \nabla_{\lambda} \epsilon - g_{\nu\lambda} \nabla^{\mu} \nabla_{\rho} \epsilon\right] + \mathcal{O}(\epsilon^{2}).
\label{eq:Riemann_conformal_linear_final}
\end{equation}
Substituting Eq. \eqref{eq:Riemann_conformal_linear_final} into Eq. \eqref{eq:jacobi_spacetime_final} the linearized spacetime Jacobi equation can be reexpressed as 
\bea
\frac{D^{2} \xi^{\mu}}{D\tau^{2}} &=& - R^{\mu}_{\nu\lambda\rho}\, U^{\nu} U^{\rho} \xi^{\lambda} \nn \\
& & - \frac12 \left[\delta^{\mu}_{\lambda} \nabla_{\nu} \nabla_{\rho} \epsilon - \delta^{\mu}_{\rho} \nabla_{\nu} \nabla_{\lambda} \epsilon + g_{\nu\rho} \nabla^{\mu} \nabla_{\lambda} \epsilon - g_{\nu\lambda} \nabla^{\mu} \nabla_{\rho} \epsilon\right] U^{\nu} U^{\rho} \xi^{\lambda} + \mathcal{O}(\epsilon^{2}).
\label{eq:jacobi_spacetime_linear_final}
\eea
This is the observable tidal-deformation equation in the effective spacetime geometry. It shows that the modular/Hamilton--Finsler correction enters through second derivatives of the scalar deformation $\epsilon$, and therefore modifies geodesic focusing, tidal stretching, and the stability of neighboring worldlines.

The next natural step is the Raychaudhuri equation, Section \ref{sec:Raychaudhuri}. This is required because the Jacobi equation governs the relative acceleration of neighboring curves, whereas the Raychaudhuri equation governs the evolution of the expansion, shear, and vorticity of an entire congruence. In other words, the Jacobi equation controls tidal deformation locally, while the Raychaudhuri equation controls focusing and defocusing globally. This makes it the appropriate tool for studying singularity formation, caustics, accelerated cosmological expansion, the modification of gravitational collapse, and the possibility of defocusing mechanisms induced by the phase-space deformation.

\subsection{Raychaudhuri Equation in Quantum-Deformed Hamilton--Finsler Geometry}
\label{sec:Raychaudhuri}

Throughout this section, $U^{\mu}$ denotes a timelike congruence normalized with respect to the conformally reduced metric $\widetilde{g}^{\mathrm{conf}}_{\mu\nu}$. When needed, background-normalized quantities with respect to $g_{\mu\nu}$ are related by the conformal rescaling $\widetilde{U}^{\mu}=C^{-1/2} U^{\mu}$. The Raychaudhuri equation follows from the Jacobi equation after projecting to the purely horizontal sector. Let the distortion tensor of the congruence be defined by
\begin{equation}
\nabla_{\tau}\xi^{\mu} = \mathcal{B}^{\mu}_{\nu} \xi^{\nu}.
\label{eq:Btensor_final}
\end{equation}
Differentiating once more, $\nabla_{\tau}\left(\mathcal{B}^{\mu}_{\nu} \xi^{\nu}\right)$, leads to
\bea
\nabla_{\tau}^{2} \xi^{\mu} 
&=& \left(\nabla_{\tau} \mathcal{B}^{\mu}_{\nu} + \mathcal{B}^{\mu}_{\rho} \mathcal{B}^{\rho}_{\nu}\right) \xi^{\nu}.
\label{eq:Briccati_derivation_final}
\eea
Comparison with the purely horizontal part of Eq.~\eqref{eq:HF_deviation_final} yields the Riccati equation \cite{bittanti1991riccati}
\bea
\begin{array}{ccl}
\nabla_{\tau} \mathcal{B}^{\mu}_{\nu} + \mathcal{B}^{\mu}_{\rho}\mathcal{B}^{\rho}_{\nu} &=& - \mathcal{R}^{\mu}_{\nu}, \\ 
\mathcal{R}^{\mu}_{\nu} &=& \widetilde{R}^{\mu}_{\alpha\nu\beta} U^{\alpha} U^{\beta}.
\end{array}
\label{eq:Briccati_main_final}
\eea
The sign of the curvature term is fixed by the Jacobi equation and is therefore fully consistent with the Riemann convention adopted in Section~\ref{sec:CivitaCrvt}. In the full phase-space theory an additional mixed-curvature source proportional to $\widetilde{P}^{\mu\rho}_{\nu\lambda} U^{\nu} U^{\lambda} \zeta_{\rho}$ may be retained, but it disappears in the purely horizontal reduction relevant for the cosmological pullback.

For a timelike congruence in the effective spacetime, define the spatial projector by
\bea
\begin{array}{ccl}
h_{\mu\nu} &=& \widetilde{g}^{\mathrm{conf}}_{\mu\nu} + U_{\mu} U_{\nu}, \\ 
U_{\mu} &=& \widetilde{g}^{\mathrm{conf}}_{\mu\nu}U^{\nu}, \\
\widetilde{g}^{\mathrm{conf}}_{\mu\nu} U^{\mu} U^{\nu} &=& - 1,
\end{array}
\label{eq:h_projector_final}
\eea
which satisfies
\begin{equation}
h_{\mu\nu} U^{\nu}=0, \qquad h_{\mu}^{\rho} h_{\rho\nu} = h_{\mu\nu}, \qquad h^{\mu}_{\mu} = 3.
\label{eq:h_projector_props_final}
\end{equation}
The distortion tensor is decomposed as follows:
\begin{equation}
\mathcal{B}_{\mu\nu} = \frac13 \Theta\, h_{\mu\nu} + \Sigma_{\mu\nu} + \Omega_{\mu\nu},
\label{eq:B_decomp_final}
\end{equation}
where $\Theta$ is the expansion scalar, $\Sigma_{\mu\nu}$ is the shear tensor, and $\Omega_{\mu\nu}$ is the vorticity tensor. Explicitly,
\bea
\Theta &=& \mathcal{B}^{\mu}_{\mu},
\label{eq:Theta_final} \\
\Sigma_{\mu\nu} = \mathcal{B}_{(\mu\nu)} - \frac13 \Theta\, h_{\mu\nu}, \qquad \Sigma_{\mu\nu} &=& \Sigma_{\nu\mu},
 \Sigma^{\mu}_{\mu} = 0, \qquad \Sigma_{\mu\nu} U^{\nu} = 0,
\label{eq:Sigma_final} \\
\Omega_{\mu\nu} = \mathcal{B}_{[\mu\nu]} \qquad
\Omega_{\mu\nu} &=& - \Omega_{\nu\mu}, \qquad \Omega_{\mu\nu} U^{\nu} = 0.
\label{eq:Omega_final}
\eea
Thus $\Theta$ measures the local rate of volume expansion or contraction of the congruence, $\Sigma_{\mu\nu}$ assesses its distortion at fixed volume, while $\Omega_{\mu\nu}$ evaluates the local rotation.
These tensors satisfy the standard contraction identity
\begin{equation}
\mathcal{B}_{\mu\nu} \mathcal{B}^{\nu\mu} = \frac13 \Theta^{2}
+ \Sigma_{\mu\nu} \Sigma^{\mu\nu} - \Omega_{\mu\nu} \Omega^{\mu\nu}.
\label{eq:Bsquare_final}
\end{equation}
Taking the trace of Eq.~\eqref{eq:Briccati_main_final} gives the exact Raychaudhuri equation for a timelike geodesic congruence,
\begin{equation}
\nabla_{\tau} \Theta = - \frac13 \Theta^{2} - \Sigma_{\mu\nu} \Sigma^{\mu\nu} + \Omega_{\mu\nu} \Omega^{\mu\nu} - \widetilde{R}_{\mu\nu} U^{\mu} U^{\nu}.
\label{eq:Raychaudhuri_exact_final}
\end{equation}
For non-geodesic congruences the acceleration term must be added,
\begin{equation}
\nabla_{\tau} \Theta = - \frac13 \Theta^{2} - \Sigma_{\mu\nu} \Sigma^{\mu\nu} + \Omega_{\mu\nu} \Omega^{\mu\nu} - \widetilde{R}_{\mu\nu} U^{\mu} U^{\nu} + \widetilde{\nabla}_{\mu} a^{\mu},
\label{eq:Raychaudhuri_nongeodesic_final}
\end{equation}
where $a^{\mu}=U^{\nu} \widetilde{\nabla}_{\nu} U^{\mu}$.

Now in order to express the curvature term through the conformal deformation, let us introduce
\begin{equation}
\widetilde{g}^{\mathrm{conf}}_{\mu\nu} = C\, g_{\mu\nu}, \qquad C = e^{2\phi}, \qquad \phi = \frac12\ln C.
\label{eq:phi_conformal_final}
\end{equation}
The exact conformal Ricci tensor in four dimensions is then expressed as
\begin{equation}
\widetilde{R}_{\mu\nu} = R_{\mu\nu} - 2 \nabla_{\mu} \nabla_{\nu}\phi - g_{\mu\nu} \Box \phi + 2 \nabla_{\mu} \phi\, \nabla_{\nu} \phi - 2 g_{\mu\nu}(\nabla\phi)^{2}.
\label{eq:Ricci_conformal_exact_final}
\end{equation}
If $\widetilde{U}^{\mu}=e^{-\phi} U^{\mu}$ is normalized with respect to $\widetilde{g}^{\mathrm{conf}}_{\mu\nu}$ while $U^{\mu}$ is normalized with respect to $g_{\mu\nu}$, then we find that
\begin{equation}
\widetilde{R}_{\mu\nu} \widetilde{U}^{\mu} \widetilde{U}^{\nu} = C^{-1} \left[R_{\mu\nu} U^{\mu} U^{\nu} - 2 \ddot{\phi} + \Box \phi
+ 2 \dot{\phi}^{2} + 2 (\nabla\phi)^{2}\right],
\label{eq:Ricci_UU_exact_final}
\end{equation}
where
\begin{equation}
\dot{\phi} := U^{\mu} \nabla_{\mu} \phi, \qquad \ddot{\phi} := U^{\mu} U^{\nu} \nabla_{\mu} \nabla_{\nu} \phi.
\label{eq:phi_dot_defs_final}
\end{equation}
Substitution into Eq.~\eqref{eq:Raychaudhuri_exact_final} yields
\bea
\widetilde{\nabla}_{\tau} \widetilde{\Theta} &=& - \frac13 \widetilde{\Theta}^{2} - \widetilde{\Sigma}_{\mu\nu} \widetilde{\Sigma}^{\mu\nu} + \widetilde{\Omega}_{\mu\nu} \widetilde{\Omega}^{\mu\nu} 
 - C^{-1} \left[R_{\mu\nu} U^{\mu} U^{\nu} - 2 \ddot{\phi}
+ \Box \phi + 2 \dot{\phi}^{2} + 2 (\nabla\phi)^{2} \right].
\label{eq:Raychaudhuri_conformal_exact_final}
\eea

To obtain the perturbative regime, we introduce
\begin{equation}
C=1+\epsilon, \qquad \phi = \frac12 \ln(1 + \epsilon) = \frac12 \epsilon + \mathcal{O}(\epsilon^{2}), \qquad C^{-1} = 1 - \epsilon + \mathcal{O}(\epsilon^{2}).
\label{eq:epsilon_phi_final}
\end{equation}
Neglecting terms quadratic in derivatives of $\phi$, Eq.~\eqref{eq:Ricci_UU_exact_final} can be reexpressed as
\begin{equation}
\widetilde{R}_{\mu\nu} \widetilde{U}^{\mu} \widetilde{U}^{\nu} = 
R_{\mu\nu} U^{\mu} U^{\nu} - \ddot{\epsilon} + \frac12 \Box \epsilon - \epsilon\, R_{\mu\nu} U^{\mu} U^{\nu} + \mathcal{O}(\epsilon^{2}).
\label{eq:RicciUU_linear_final}
\end{equation}
Accordingly, the Raychaudhuri equation in the perturbative regime reads
\bea
\widetilde{\nabla}_{\tau} \widetilde{\Theta} &=&  - \frac13 \widetilde{\Theta}^{2} - \widetilde{\Sigma}_{\mu\nu} \widetilde{\Sigma}^{\mu\nu} + \widetilde{\Omega}_{\mu\nu} \widetilde{\Omega}^{\mu\nu} - R_{\mu\nu} U^{\mu} U^{\nu} + \ddot{\epsilon} - \frac12 \Box \epsilon + \epsilon\, R_{\mu\nu} U^{\mu} U^{\nu} + \mathcal{O}(\epsilon^{2}).
\label{eq:Raychaudhuri_linear_final}
\eea
In this derivation, we used $\epsilon = \beta\, \Xi_{\sigma} + \kappa\, \mathfrak{A}_{\sigma} + \mathcal{O}(\beta^{2},\kappa^{2},\beta\kappa)$. Furthermore, all first-order corrections are controlled by derivatives of the pullback modular scalar $\Xi_{\sigma}$ and of the normalized-momentum acceleration invariant $\mathfrak{A}_{\sigma}$.

A particularly important case is that of hypersurface-orthogonal congruences, for which $\widetilde{\Omega}_{\mu\nu}=0$. In an exactly homogeneous and isotropic FLRW sector $\widetilde{\Sigma}_{\mu\nu}=0$ is also obtained. Then Eq.~\eqref{eq:Raychaudhuri_linear_final} reduces to
\begin{equation}
\widetilde{\nabla}_{\tau} \widetilde{\Theta} + \frac13 \widetilde{\Theta}^{2} = - R_{\mu\nu} U^{\mu} U^{\nu} + \ddot{\epsilon} - \frac12 \Box \epsilon + \epsilon\, R_{\mu\nu} U^{\mu} U^{\nu} + \mathcal{O}(\epsilon^{2}).
\label{eq:Raychaudhuri_FLRW_final}
\end{equation}
For a comoving FLRW congruence one has $\widetilde{\Theta} = 3 \widetilde{H}$, so the modified Hubble evolution equation is given as
\begin{equation}
3 \dot{\widetilde{H}} + 3 \widetilde{H}^{2} = - R_{\mu\nu} U^{\mu} U^{\nu} + \ddot{\epsilon} - \frac12 \Box \epsilon + \epsilon\, R_{\mu\nu} U^{\mu} U^{\nu} + \mathcal{O}(\epsilon^{2}).
\label{eq:Raychaudhuri_Hubble_final}
\end{equation}

In the undeformed limit, i.e., $\beta \to 0,\;\; \kappa \to 0,\;\;
\epsilon \to 0$, one recovers
\begin{equation}
\widetilde{g}^{\mathrm{conf}}_{\mu\nu} \to g_{\mu\nu}, \qquad \widetilde{\Theta} \to \Theta, \qquad \widetilde{\Sigma}_{\mu\nu} \to \Sigma_{\mu\nu}, \qquad \widetilde{\Omega}_{\mu\nu} \to \Omega_{\mu\nu}.
\label{eq:undeformed_kinematics_final}
\end{equation}
Hence the standard general-relativistic Raychaudhuri equation is obtained as
\begin{equation}
\nabla_{\tau} \Theta = -\frac13 \Theta^{2} - \Sigma_{\mu\nu} \Sigma^{\mu\nu} + \Omega_{\mu\nu} \Omega^{\mu\nu} - R_{\mu\nu} U^{\mu} U^{\nu}.
\label{eq:Raychaudhuri_GR_final}
\end{equation}

The geometric content of the construction can now be summarized as follows. The Hamilton--Finsler deformation first modifies phase-space geodesic deviation through the horizontal curvature correction and the mixed curvature source. After pullback and conformal reduction, it modifies the effective spacetime Jacobi equation through second derivatives of the scalar deformation $\epsilon$. Finally, through the Raychaudhuri equation, it modifies the focusing properties of timelike congruences. The intended implications are therefore broader than cosmic inflation alone. They include modified focusing and defocusing conditions, possible changes in singularity formation, corrections to gravitational collapse, changes in tidal observables, and the possibility of an effective geometric contribution to accelerated cosmological expansion. In other words, cosmic inflation is one possible application, but not the only one. More generally, the formalism provides a phase-space origin for corrections to the classical focusing equation and thereby to the causal and dynamical structure of effective spacetime.

For the cosmological analysis that follows, we restrict the discussion to the homogeneous and isotropic FLRW sector and fix the section $\sigma$ to the comoving congruence. The null sector of phase space does not enter at the level of the background dynamics considered here. The effective gravitational dynamics are therefore governed by the conformally reduced metric with scalar deformation
\begin{equation}
\epsilon(x) = \beta\, \Xi_{\sigma}(x) + \kappa\, \mathfrak{A}_{\sigma}(x),
\end{equation}
which will be used to derive the modified Friedmann and inflationary equations.

In the inflationary analysis below, $\langle T_{\mu\nu}\rangle$ will be specialized to the effective energy-stress tensor of a homogeneous scalar-field sector in the comoving FLRW frame, so that all dynamical modifications arise from the geometric deformation encoded in $\epsilon(x)$.

\section{Analytical Results}
\label{sec:anlRsl}

\subsection{Inflationary Background Dynamics}
\label{sec:inflation}

In this section the inflationary sector of the reduced effective spacetime theory induced by the modular Hamiltonian deformation is derived explicitly. The aim is to determine how the scalar spacetime deformation \eqref{eq:formalism_epsilon_def_final} modifies the inflaton action, the Klein--Gordon equation, and the cosmological background Einstein equations after pullback and isotropic reduction. The significance of this analysis is twofold. First, it provides a concrete cosmological realization of the phase-space quantization framework developed in the previous sections. Second, it shows that the phase-space modular structure produces inflationary corrections through the induced spacetime geometry itself, without requiring any direct coupling of the inflaton to phase-space fiber variables. The inflationary sector therefore furnishes a physically testable realization of the proposed quantization scheme.

Let us first start with the section $\sigma$ which is fixed to the comoving FLRW congruence, as prescribed at the end of Section~\ref{sec:formalism}. Accordingly, the inflationary background is studied entirely in the homogeneous and isotropic timelike sector, and the null sector of phase space does not enter at the level of the background dynamics considered here. Moreover, $\langle T_{\mu\nu}\rangle$ is specialized to the effective energy-stress tensor of a homogeneous scalar field in the comoving FLRW frame.

In Section \ref{sec:inflaton_sector}, the inflaton sector on pullback and conformal geometries are detailed.

\subsubsection{Inflaton Sector on Pullback and Conformal Geometries}
\label{sec:inflaton_sector}

The inflationary matter sector must be formulated in a way consistent with the three geometric layers introduced in Section~\ref{sec:formalism}: \begin{enumerate} 
\item the full anisotropic Hamilton metric $\widetilde{g}^{\mu\nu}(x,p)$ on the conic phase-space domain $\mathcal{A}_{0}^{\ast} \subset T^{\ast}M$, 
\item the section-pullback Lorentzian metric $\widetilde{g}^{\sigma}_{\mu\nu}(x) := \widetilde{g}_{\mu\nu}(x,p(x))$, and 
\item the homogeneous and isotropic conformal truncation $
\widetilde{g}^{\mathrm{conf}}_{\mu\nu}(x) = C(x)\, g_{\mu\nu}(x)$.
\end{enumerate}
The cosmological inflation model is therefore not defined directly on the anisotropic tensor $\widetilde{g}^{\mu\nu}(x,p)$. Rather, it is defined first on the section-pullback metric $\widetilde{g}^{\sigma}_{\mu\nu}(x)$, and only afterwards reduced to the FLRW-compatible conformal metric $\widetilde{g}^{\mathrm{conf}}_{\mu\nu}(x)$. This order is forced by the geometry: the pullback metric is the first ordinary pseudo-Riemannian object on spacetime, while the conformal ansatz becomes meaningful only after the isotropy condition has been imposed.

Let $\varphi:M\to\mathbb{R}$ be a real scalar inflaton field. On an ordinary Lorentzian spacetime $(M,g)$, the minimally coupled scalar action is
\begin{equation}
S_{\varphi}[g,\varphi] = - \int d^{4}x\, \sqrt{-g} \left[\frac12\, g^{\mu\nu} \nabla_{\mu} \varphi\, \nabla_{\nu} \varphi + V(\varphi)\right],
\label{eq:classical_inflaton_action_pub}
\end{equation}
where $V(\varphi)$ is a smooth scalar potential. Now, the variation with respect to $\varphi$ yields
\begin{equation}
\nabla_{\mu} \nabla^{\mu} \varphi - V'(\varphi) = 0.
\label{eq:KG_classical_pub}
\end{equation}

In the present framework, minimal coupling is preserved in the reduced spacetime sense: the inflaton couples only to the effective spacetime metric, while the quantum-gravity corrections enter through the metric deformation induced by the Hamiltonian phase-space construction. In particular, there is no direct coupling of the inflaton to cotangent-fiber coordinates in the reduced first-order spacetime theory. At the section-pullback level, the geometrically primary inflaton action is therefore expressed as
\begin{equation}
\widetilde{S}_{\varphi}[\sigma;\varphi] = - \int d^{4}x\, \sqrt{-\widetilde{g}^{\sigma}} \left[\frac12\, \widetilde{g}_{\sigma}^{\mu\nu}\, \nabla^{(\sigma)}_{\mu}\varphi\, \nabla^{(\sigma)}_{\nu}\varphi + V(\varphi)\right],
\label{eq:inflaton_action_pullback_pub}
\end{equation}
where $\widetilde{g}_{\sigma}^{\mu\nu}:=\bigl(\widetilde{g}^{\sigma}\bigr)^{-1\,\mu\nu}$ and $\nabla^{(\sigma)}$ is the Levi--Civita connection of $\widetilde{g}^{\sigma}_{\mu\nu}$. Equation~\eqref{eq:inflaton_action_pullback_pub} is the primary matter action obtained from the pullback of the anisotropic Hamiltonian geometry. The conformal cosmological sector is not introduced independently; it arises from Eq.~\eqref{eq:inflaton_action_pullback_pub} only after the isotropy condition required by FLRW symmetry has been imposed.

Recall that the pullback metric has the following perturbative form:
\bea
\begin{array}{ccl}
\widetilde{g}^{\sigma}_{\mu\nu} &=& g_{\mu\nu} - \beta\,  \Delta^{\sigma}_{\mu\nu} + \mathcal{O}(\beta^{2}), \\
\Delta^{\sigma}_{\mu\nu}(x) &:=& \Delta_{\mu\nu}(x,p(x)).
\end{array}
\label{eq:pullback_metric_pert_pub}
\eea
Compatibility with FLRW symmetry requires the following decomposition:
\begin{equation}
\Delta^{\sigma}_{\mu\nu} = A(x)\, g_{\mu\nu} + B(x)\, U_{\mu} U_{\nu}.
\label{eq:isotropy_decomposition_pub}
\end{equation}
The homogeneous and isotropic sector is obtained by imposing $B(x)=0$, so that
\begin{equation}
\Delta^{\sigma}_{\mu\nu} = A(x)\, g_{\mu\nu}.
\label{eq:pullback_isotropic_pub}
\end{equation}
Only under this condition may the pullback metric be reduced consistently to the conformal form
\begin{equation}
\widetilde{g}^{\mathrm{conf}}_{\mu\nu} = C(x)\, g_{\mu\nu}, \qquad C(x) = 1 + \epsilon(x), \qquad |\epsilon| \ll 1.
\label{eq:conformal_metric_inflation_pub}
\end{equation}

The scalar deformation field $\epsilon(x)$ is inherited from the pullback of the modular projective scalar together with the section-dependent acceleration invariant,
\bea
\begin{array}{ccl}
\epsilon(x) &=& \beta\, \Xi_{\sigma}(x) + \kappa\, \mathfrak{A}_{\sigma}(x) + \mathcal{O}(\beta^{2},\kappa^{2},\beta\kappa), \\
\Xi_{\sigma}(x) &=& \Xi\!\bigl(x,[p(x)];\eta,\Lambda\bigr), \\
\mathfrak{A}_{\sigma}(x) &=& \frac{g^{\mu\nu} \nabla_{\tau} u_{\mu} \nabla_{\tau} u_{\nu}}{\mathscr{F}^{2}}.
\end{array}
\label{eq:epsilon_inflation_pub}
\eea
As emphasized previously $\mathscr{F}=(c^{7}/\hbar G)^{1/2}$. Whereas $\Xi_{\sigma}$ is the derived modular contribution, $\mathfrak{A}_{\sigma}$ is an additional effective invariant retained at the same perturbative order in the reduced theory.

Because $u_{\mu}$ is fundamentally defined on the projectivized cotangent bundle, the derivative in Eq.~\eqref{eq:formalism_chainrule_u_final} must be understood through the section-dependent chain rule, Eq. \eqref{eq:formalism_nabla_tau_u_final}.
%
Thus $\mathfrak{A}_{\sigma}$ is an ordinary spacetime scalar induced by the chosen section and the pulled-back projective Hamiltonian geometry. In the reduced first-order spacetime theory it is treated as prescribed geometric background data rather than as an independently varied dynamical field.

With the isotropic conformal reduction, the inflaton action becomes
\begin{equation}
\widetilde{S}^{\mathrm{conf}}_{\varphi} = - \int d^{4}x\, \sqrt{-\widetilde{g}^{\mathrm{conf}}} \left[\frac12\, \widetilde{g}^{\mathrm{conf}\,\mu\nu}\, \nabla_{\mu} \varphi\, \nabla_{\nu} \varphi + V(\varphi)
\right].
\label{eq:inflaton_action_conformal_pub}
\end{equation}
After the conformal reduction, all covariant derivatives are taken with respect to the Levi--Civita connection of the background metric $g_{\mu\nu}$, and the explicit $\sigma$-superscript on $\nabla$ is dropped.

To first order in $\epsilon$, we have the following expansions:
\bea
\begin{array}{ccl}
\widetilde{g}^{\mathrm{conf}}_{\mu\nu} &=& (1 + \epsilon) g_{\mu\nu}, \\ 
\widetilde{g}_{\mathrm{conf}}^{\mu\nu} &=& (1 - \epsilon) g^{\mu\nu} + \mathcal{O}(\epsilon^{2}), \\ 
\sqrt{-\widetilde{g}^{\mathrm{conf}}} &=& (1 + 2 \epsilon) \sqrt{-g} + \mathcal{O}(\epsilon^{2}).
\end{array}
\label{eq:conf_expansions_inflation_pub}
\eea
Substituting Eq.~\eqref{eq:conf_expansions_inflation_pub} into Eq.~\eqref{eq:inflaton_action_conformal_pub} leads to
\bea
\widetilde{S}^{\mathrm{conf}}_{\varphi} 
&=& - \int d^{4}x\, \sqrt{-g} \left[\frac12\, g^{\mu\nu} \nabla_{\mu} \varphi \nabla_{\nu} \varphi + V(\varphi) + \frac12\, \epsilon\, g^{\mu\nu} \nabla_{\mu} \varphi \nabla_{\nu} \varphi + 2 \epsilon\, V(\varphi)\right] + \mathcal{O}(\epsilon^{2}).
\label{eq:inflaton_action_expanded_conformal_pub}
\eea
The conformal truncation therefore produces a universal first-order rescaling of both the kinetic and the potential sectors.

It is convenient to define
\bea
\begin{array}{ccl}
\mathcal{K}(x) &=& 1 + \epsilon(x), \\
\mathcal{V}(x,\varphi) &=& \left(1 + 2 \epsilon(x)\right) V(\varphi),
\end{array}
\label{eq:K_and_Veff_pub}
\eea
so that the reduced inflaton action takes the following compact form:
\begin{equation}
\widetilde{S}^{\mathrm{conf}}_{\varphi} = - \int d^{4}x\, \sqrt{-g}
\left[\frac12\, \mathcal{K}(x)\, g^{\mu\nu} \nabla_{\mu} \varphi\nabla_{\nu} \varphi + \mathcal{V}(x,\varphi)\right] + \mathcal{O}(\epsilon^{2}).
\label{eq:inflaton_action_KV_form_pub}
\end{equation}
All spacetime dependence in $\mathcal{K}$ and $\mathcal{V}$ is inherited from the pulled-back geometric scalars $\Xi_{\sigma}$ and $\mathfrak{A}_{\sigma}$. There is no direct inflaton coupling to fiber momenta in the reduced spacetime action.

At the level of the reduced first-order effective theory, $\epsilon(x)$ is treated as an externally prescribed background scalar inherited from the pullback construction. Its dependence on the section, the lattice data, and the normalized projective momentum belongs to the geometric input of the reduction and is not varied independently inside the reduced spacetime action. Thus the reduced variational prescription is the following: the action $\widetilde{S}^{\mathrm{conf}}_{\varphi}$ is varied with respect to the inflaton field $\varphi$ and, when needed, with respect to the background metric $g_{\mu\nu}$, while $\epsilon(x)$, $\Xi_{\sigma}(x)$, and $\mathfrak{A}_{\sigma}(x)$ are held fixed as externally specified geometric data at first order.

The reduced conformal action Eq.  \eqref{eq:inflaton_action_KV_form_pub} is the appropriate starting point for deriving the background Einstein field equations, the effective energy-stress tensor, the modified Klein--Gordon equation, and finally the FLRW inflationary dynamics, Section \ref{sec:allIA_final}.

\subsubsection{Background Einstein and Klein--Gordon equations}
\label{sec:allIA_final}

The total first-order effective action in the conformal cosmological sector is given as follows:
\begin{equation}
\widetilde{S}^{\mathrm{conf}} = \widetilde{S}_{\mathrm{EH}}^{\mathrm{conf}} + \widetilde{S}^{\mathrm{conf}}_{\varphi},
\label{eq:total_action_conf_pub}
\end{equation}
where the reduced Einstein--Hilbert sector is derived from
\begin{equation}
\widetilde{S}_{\mathrm{EH}}^{\mathrm{conf}} = \frac{1}{16 \pi G}  \int d^{4}x\, \sqrt{-g}\, (1 + \epsilon)\, R + \mathcal{O}(\epsilon^{2}).
\label{eq:EH_conf_for_inflation_pub}
\end{equation}
Because $\epsilon$ is treated as fixed background data in the reduced variational problem, variation of Eq.~\eqref{eq:EH_conf_for_inflation_pub} with respect to $g^{\mu\nu}$ gives
\begin{equation}
(1 + \epsilon) G_{\mu\nu} - \nabla_{\mu} \nabla_{\nu} \epsilon + g_{\mu\nu} \Box \epsilon = 8 \pi G\, T_{\mu\nu}^{\mathrm{eff}} + \mathcal{O}(\epsilon^{2}),
\label{eq:EinsteinEqExact1order}
\end{equation}
where $\Box\epsilon:=g^{\mu\nu}\nabla_{\mu}\nabla_{\nu}\epsilon$ and the effective matter tensor is defined as 
\begin{equation}
T_{\mu\nu}^{\mathrm{eff}}:=- \frac{2}{\sqrt{-g}} \frac{\delta \widetilde{S}^{\mathrm{conf}}_{\varphi}}{\delta g^{\mu\nu}}
 \Bigg|_{\epsilon\ \mathrm{fixed}}.
\label{eq:Teff_definition_pub}
\end{equation}
Equation~\eqref{eq:EinsteinEqExact1order} is the variationally consistent first-order gravitational field equation of the reduced theory. The algebraic factor $(1 + \epsilon) G_{\mu\nu}$ must be retained explicitly at the same perturbative order as the derivative corrections.

Because $\epsilon$ is external, $T_{\mu\nu}^{\mathrm{eff}}$ is not the stress tensor of an autonomous matter sector. Instead, the reduced Bianchi identity implies a modified conservation law. Taking the divergence of Eq.~\eqref{eq:EinsteinEqExact1order}, using $\nabla^{\mu} G_{\mu\nu}=0$ and the scalar identity
\begin{equation}
\nabla^{\mu} \nabla_{\mu} \nabla_{\nu} \epsilon - \nabla_{\nu} \Box \epsilon = R_{\nu}{}^{\rho} \nabla_{\rho} \epsilon,
\label{eq:scalar_commutator_identity_inflation}
\end{equation}
one finds that
\begin{equation}
8 \pi G\, \nabla^{\mu} T_{\mu\nu}^{\mathrm{eff}} = G_{\mu\nu} \nabla^{\mu} \epsilon + \mathcal{O}(\epsilon^{2}).
\label{eq:modified_conservation_law_pub}
\end{equation}
Thus the matter sector is conserved only in the modified sense appropriate to the fixed-background deformation, exactly as expected in the reduced conformal truncation.

We next compute $T_{\mu\nu}^{\mathrm{eff}}$ from Eq.~\eqref{eq:inflaton_action_KV_form_pub}. Starting from
\begin{equation}
\widetilde{S}^{\mathrm{conf}}_{\varphi} = - \int d^{4}x\, \sqrt{-g} 
\left[\frac12 (1 + \epsilon)\, g^{\alpha\beta} \nabla_{\alpha} \varphi \nabla_{\beta} \varphi + (1 + 2 \epsilon) V(\varphi)\right] + \mathcal{O}(\epsilon^{2}),
\label{eq:matter_action_for_metric_variation_pub}
\end{equation}
its variation at fixed $\epsilon$ reads
\bea
\delta \widetilde{S}^{\mathrm{conf}}_{\varphi} &=& - \int d^{4}x\, 
\delta(\sqrt{-g}) \left[\frac12 (1 + \epsilon)\, g^{\alpha\beta} \nabla_{\alpha} \varphi \nabla_{\beta} \varphi + (1 + 2 \epsilon) V(\varphi)\right] \nn \\
& & - \int d^{4}x\, \sqrt{-g}\, \frac12 (1 + \epsilon)\, \delta  g^{\alpha\beta} \nabla_{\alpha} \varphi \nabla_{\beta} \varphi.
\label{eq:variation_matter_action_inflation}
\eea
Using $\delta\sqrt{-g}=-\frac12\sqrt{-g}\,g_{\mu\nu}\delta g^{\mu\nu}$, one obtains that
\bea
\delta \widetilde{S}^{\mathrm{conf}}_{\varphi} &=&  \frac12  \int d^{4}x\, \sqrt{-g}\, \left\{g_{\mu\nu}\left[\frac12 (1 + \epsilon)\, g^{\alpha\beta} \nabla_{\alpha} \varphi \nabla_{\beta} \varphi + (1 + 2 \epsilon) V(\varphi)\right] - (1 + \epsilon) \nabla_{\mu} \varphi \nabla_{\nu} \varphi \right\} \delta g^{\mu\nu}.
\label{eq:variation_matter_action_inflation_final}
\eea
Hence, the energy-stress tensor reads
\begin{equation}
T_{\mu\nu}^{\mathrm{eff}} = (1 + \epsilon)\, \nabla_{\mu} \varphi\nabla_{\nu} \varphi - g_{\mu\nu} \left[\frac12 (1 + \epsilon)\, g^{\alpha\beta} \nabla_{\alpha} \varphi \nabla_{\beta} \varphi + (1 + 2 \epsilon) V(\varphi)\right] + \mathcal{O}(\epsilon^{2}).
\label{eq:Tmunu_inflation_pub}
\end{equation}

The inflaton equation follows from variation of Eq.~\eqref{eq:inflaton_action_KV_form_pub} with respect to $\varphi$ at fixed $\epsilon$, 
\begin{equation}
\nabla_{\mu} \left(\mathcal{K}\, \nabla^{\mu} \varphi\right) - \frac{\partial \mathcal{V}}{\partial\varphi} = 0.
\label{eq:KG_preexpanded_inflation_pub}
\end{equation}
Substituting Eq.~\eqref{eq:K_and_Veff_pub} inn the preceding expression gives
\begin{equation}
\nabla_{\mu} \left[(1 + \epsilon) \nabla^{\mu} \varphi\right] - (1 + 2 \epsilon) V'(\varphi) = \mathcal{O}(\epsilon^{2}).
\label{eq:KG_intermediate_precise_pub}
\end{equation}
Expanding the derivative leads to
\begin{equation}
(1 + \epsilon) \Box \varphi + (\nabla_{\mu} \epsilon) \nabla^{\mu} \varphi - (1 + 2 \epsilon) V'(\varphi) = \mathcal{O}(\epsilon^{2}).
\label{eq:KG_before_division_precise_pub}
\end{equation}
Dividing by $(1+\epsilon)$ and expanding consistently to first order yields
\begin{equation}
\Box \varphi + (\nabla_{\mu} \epsilon) \nabla^{\mu} \varphi - (1 + \epsilon) V'(\varphi) = \mathcal{O}(\epsilon^{2}).
\label{eq:KG_expanded_final_pub}
\end{equation}
This is the correct first-order Klein--Gordon equation in the conformal truncation. The effective potential force is multiplied by $1+\epsilon$ not by $1+2\epsilon$ after the consistent first-order reduction.

We now specialize to a spatially flat FLRW background for the metric $g_{\mu\nu}$,
\begin{equation}
ds^{2} = - dt^{2} + a^{2}(t)\, \delta_{ij}\, dx^{i}dx^{j},
\label{eq:FLRW_metric_background_pub}
\end{equation}
where the Hubble constant is defined as $H:=\dot{a}/a$. Although the physical conformal metric is $\widetilde{g}^{\mathrm{conf}}_{\mu\nu} = (1 + \epsilon) g_{\mu\nu}$, the first-order reduced dynamics is most transparently expressed in terms of the background FLRW metric $g_{\mu\nu}$ together with the scalar deformation $\epsilon(t)$. For a homogeneous inflaton and homogeneous deformation, $\varphi=\varphi(t)$ and $\epsilon=\epsilon(t)$. Therefrom one finds that
\begin{equation}
\Box \varphi = - \ddot{\varphi} - 3 H \dot{\varphi}, \qquad \nabla_{\mu} \epsilon\, \nabla^{\mu} \varphi = - \dot{\epsilon}\, \dot{\varphi}.
\label{eq:boxphi_and_cross_inflation}
\end{equation}
Substituting Eq.~\eqref{eq:boxphi_and_cross_inflation} into Eq.~\eqref{eq:KG_expanded_final_pub} yields the homogeneous inflaton equation
\begin{equation}
\ddot{\varphi} + (3 H + \dot{\epsilon}) \dot{\varphi} + (1 + \epsilon) V'(\varphi) = \mathcal{O}(\epsilon^{2}).
\label{eq:KG_FLRW_final_pub}
\end{equation}
The phase-space deformation therefore modifies both the friction term and the effective potential force acting on the inflaton.

We next derive the background Einstein equations. For the spatially flat FLRW metric one expresses the temporal and spacial components as
\begin{equation}
G_{00} = 3 H^{2}, \qquad G_{ij} = - (2 \dot{H} + 3 H^{2}) g_{ij}.
\label{eq:FLRW_Einstein_tensor_components}
\end{equation}
Since $\epsilon = \epsilon(t)$, one also finds that $\nabla_{0} \nabla_{0} \epsilon = \ddot{\epsilon}$ and $\Box \epsilon = - \ddot{\epsilon} - 3 H \dot{\epsilon}$, therefore
\begin{equation}
- \nabla_{0} \nabla_{0} \epsilon + g_{00} \Box \epsilon = 3 H \dot{\epsilon}.
\label{eq:time_component_epsilon_combination}
\end{equation}
Accordingly, the $00$-component of Eq.~\eqref{eq:EinsteinEqExact1order} becomes
\begin{equation}
3 (1 + \epsilon) H^{2} + 3 H \dot{\epsilon} = 8 \pi G\, T_{00}^{\mathrm{eff}} + \mathcal{O}(\epsilon^{2}).
\label{eq:G00_intermediate_pub}
\end{equation}
For the spatial sector, the homogeneity implies that
\bea
\begin{array}{ccl}
\nabla_{i} \nabla_{j} \epsilon &=& - H g_{ij} \dot{\epsilon},\\
-\nabla_{i} \nabla_{j} \epsilon + g_{ij} \Box \epsilon &=& - (\ddot{\epsilon} + 2 H \dot{\epsilon})\, g_{ij}.
\end{array}
\label{eq:spatial_hessian_epsilon}
\eea
Hence the $ij$-components of Eq.~\eqref{eq:EinsteinEqExact1order} are given as 
\begin{equation}
- (1 + \epsilon)(2 \dot{H} + 3 H^{2}) g_{ij} - (\ddot{\epsilon} + 2 H \dot{\epsilon}) g_{ij} = 8 \pi G\, T_{ij}^{\mathrm{eff}} + \mathcal{O}(\epsilon^{2}).
\label{eq:Gij_intermediate_pub}
\end{equation}
We now compute the effective matter components. Since $g^{\alpha\beta} \nabla_{\alpha} \varphi \nabla_{\beta} \varphi
 = - \dot{\varphi}^{2}$, the temporal and spacial components of Eq.~\eqref{eq:Tmunu_inflation_pub} respectively read
\bea
T_{00}^{\mathrm{eff}} &=& \frac12 (1 + \epsilon) \dot{\varphi}^{2} 
+ (1 + 2 \epsilon) V(\varphi) + \mathcal{O}(\epsilon^{2}),
\label{eq:T00_final_pub} \\
T_{ij}^{\mathrm{eff}} &=& g_{ij} \left[\frac12 (1 + \epsilon \dot{\varphi}^{2} - (1 + 2 \epsilon) V(\varphi)\right] + \mathcal{O}(\epsilon^{2}).
\label{eq:Tij_final_pub}
\eea
It is therefore natural to respectively define the effective energy density and pressure by
\bea
\begin{array}{ccl}
\rho_{\mathrm{eff}} &:=& \frac12 (1 + \epsilon) \dot{\varphi}^{2} 
+ (1 + 2 \epsilon) V(\varphi), \\
p_{\mathrm{eff}} &:=& \frac12 (1 + \epsilon) \dot{\varphi}^{2} - (1 + 2 \epsilon) V(\varphi),
\end{array}
\label{eq:rho_p_eff_inflation}
\eea
so that $T_{00}^{\mathrm{eff}}=\rho_{\mathrm{eff}}$ and 
$T_{ij}^{\mathrm{eff}}=p_{\mathrm{eff}}\,g_{ij}$.

The modified Friedmann equations take the form
\bea
3 (1 + \epsilon) H^{2} + 3 H \dot{\epsilon} &=& 8 \pi G\, \rho_{\mathrm{eff}} + \mathcal{O}(\epsilon^{2}),
\label{eq:Friedmann_00_final_pub} \\
(1 + \epsilon)(2 \dot{H} + 3 H^{2}) + \ddot{\epsilon} + 2 H \dot{\epsilon} &=& - 8 \pi G\, p_{\mathrm{eff}} + \mathcal{O}(\epsilon^{2}).
\label{eq:Friedmann_ii_final_pub}
\eea
Substituting Eq.~\eqref{eq:rho_p_eff_inflation} into Eqs.~\eqref{eq:Friedmann_00_final_pub} and \eqref{eq:Friedmann_ii_final_pub} gives explicitly
\bea
3 (1 + \epsilon) H^{2} + 3 H \dot{\epsilon} &=&  8 \pi G \left[\frac12 (1 + \epsilon) \dot{\varphi}^{2} + (1 + 2 \epsilon) V(\varphi)\right] + \mathcal{O}(\epsilon^{2}),
\label{eq:Friedmann_00_explicit_final_pub} \\
(1 + \epsilon)(2 \dot{H} + 3 H^{2}) + \ddot{\epsilon} + 2 H \dot{\epsilon} &=&  - 8 \pi G \left[\frac12 (1 + \epsilon) \dot{\varphi}^{2} - (1 + 2 \epsilon) V(\varphi)\right] + \mathcal{O}(\epsilon^{2}).
\label{eq:Friedmann_ii_explicit_final_pub}
\eea

A particularly useful combination can be obtained by subtracting Eq.~\eqref{eq:Friedmann_ii_final_pub} from Eq.~\eqref{eq:Friedmann_00_final_pub}. Then, using $\rho_{\mathrm{eff}} + p_{\mathrm{eff}} = (1 + \epsilon) \dot{\varphi}^{2}$, one obtains the modified Raychaudhuri-type equation
\begin{equation}
- 2 (1 + \epsilon) \dot{H} + H \dot{\epsilon} - \ddot{\epsilon} = 8 \pi G\, (1 + \epsilon) \dot{\varphi}^{2} + \mathcal{O}(\epsilon^{2}).
\label{eq:modified_Hdot_equation_inflation}
\end{equation}
Equivalently, the cosmic acceleration equation may be written as
\begin{equation}
\frac{\ddot{a}}{a} = - \frac{4 \pi G}{3 (1 + \epsilon)} \bigl(\rho_{\mathrm{eff}} + 3 p_{\mathrm{eff}}\bigr) - \frac{\ddot{\epsilon} + H \dot{\epsilon}}{2 (1 + \epsilon)} + \mathcal{O}(\epsilon^{2}).
\label{eq:modified_acceleration_equation_inflation}
\end{equation}
Equation~\eqref{eq:modified_acceleration_equation_inflation} seems to make the inflationary content of the deformation transparent: accelerated expansion is controlled not only by the effective equation of state of the inflaton, but also by the derivative terms in $\epsilon$, which act as a genuinely geometric source of acceleration or deceleration.

The system with these set of equations \eqref{eq:KG_FLRW_final_pub}, \eqref{eq:Friedmann_00_explicit_final_pub}, and 
\eqref{eq:Friedmann_ii_explicit_final_pub} appear to form the closed first-order background system for inflation in the conformal cosmological truncation of the modular Hamilton--Finsler geometry, once $\epsilon(t)$ is specified by the geometric sector through Eq.~\eqref{eq:epsilon_inflation_pub}. In other words, we conclude that the inflationary dynamics are governed by the homogeneous inflaton $\varphi(t)$, the scale factor $a(t)$, and the prescribed geometric deformation $\epsilon(t)$, which itself is inherited from the pulled-back phase-space structure.

Finally, let us summarize the modified conservation law, Eq.  \eqref{eq:modified_conservation_law_pub}, which is consistent with the background system. In the FLRW sector it becomes
\begin{equation}
\dot{\rho}_{\mathrm{eff}} + 3 H (\rho_{\mathrm{eff}} + p_{\mathrm{eff}}) = \frac{3 H^{2}}{8 \pi G}\, \dot{\epsilon} + \mathcal{O}(\epsilon^{2}),
\label{eq:modified_flrw_conservation_inflation}
\end{equation}
showing explicitly that the reduced matter sector exchanges energy with the fixed geometric background encoded in $\epsilon(t)$. Thus the inflaton does not evolve as an autonomous minimally coupled scalar in the ordinary GR sense; rather, it evolves in a spacetime whose effective geometry already carries the imprint of the modular phase-space deformation.

The slow-roll regime, the modified inflationary conditions, and the resulting number of e-folds can now be analyzed in Section \ref{sec:QCeFoldsPowerLaw_final} using the background system derived above.

\subsubsection{Slow-Roll Regime and Number of e-Folds for Power-Law Potential}
\label{sec:QCeFoldsPowerLaw_final}

We now specialize the background system derived in Section~\ref{sec:allIA_final} to a power-law inflaton potential,
\begin{equation}
V(\varphi) = \lambda\, \varphi^{n}, \qquad V'(\varphi) = n \lambda\, \varphi^{n-1},
\label{eq:powerlaw_potential_pub}
\end{equation}
with $\lambda>0$ and $n>0$. The purpose of this section is to derive the slow-roll dynamics and the corresponding number of e-folds in the reduced conformal cosmological sector. Since the deformation is encoded by the geometric scalar $\epsilon(t)$, the slow-roll regime must now be understood as a joint approximation in which both the inflaton and the geometric deformation vary sufficiently slowly on Hubble time scales.

The background equations derived previously are
\bea
\ddot{\varphi} + (3 H + \dot{\epsilon}) \dot{\varphi} + (1 + \epsilon) V'(\varphi) &=& \mathcal{O}(\epsilon^{2}),
\label{eq:KG_FLRW_for_slowroll_pub} \\
3 (1 + \epsilon) H^{2} + 3 H \dot{\epsilon} &=&  8 \pi G 
\left[ \frac12 (1 + \epsilon) \dot{\varphi}^{2} + (1 + 2 \epsilon) V(\varphi)\right] + \mathcal{O}(\epsilon^{2}),
\label{eq:Friedmann_00_for_slowroll_pub} \\
(1 + \epsilon)(2 \dot{H} + 3 H^{2}) + \ddot{\epsilon} + 2 H \dot{\epsilon} &=& - 8 \pi G \left[\frac12 (1 + \epsilon) \dot{\varphi}^{2} - (1 + 2 \epsilon) V(\varphi)\right] + \mathcal{O}(\epsilon^{2}).
\label{eq:Friedmann_ii_for_slowroll_pub}
\eea
These equations form the starting point of the inflationary slow-roll analysis. The slow-roll regime is defined by the conditions
\begin{equation}
\dot{\varphi}^{2} \ll V(\varphi), \qquad |\ddot{\varphi}| \ll 3H|\dot{\varphi}|, \qquad |\dot{\epsilon}\, \dot{\varphi}| \ll 3H \dot{\varphi}|, \qquad |\ddot{\epsilon}| \ll H^{2},
\label{eq:slowroll_assumptions_pub}
\end{equation}
together with the quasi-adiabaticity condition
\begin{equation}
\left|\frac{\dot{\epsilon}}{H}\right| \ll |\epsilon|.
\label{eq:quasi_adiabatic_pub}
\end{equation}
The first two conditions are the standard inflationary slow-roll assumptions, while the latter two ensure that the geometric deformation evolves slowly enough that it does not spoil the adiabatic inflationary regime. In particular, Eq.~\eqref{eq:quasi_adiabatic_pub} states that the time variation of the deformation field is subleading with respect to its amplitude on Hubble time scales. Under these assumptions, the kinetic contribution in Eq.~\eqref{eq:Friedmann_00_for_slowroll_pub} is subleading, and one obtains
\begin{equation}
(1 + \epsilon) H^{2} + H \dot{\epsilon} \simeq
\frac{8 \pi G}{3}(1 + 2 \epsilon) V(\varphi) + \mathcal{O}\!\left(\epsilon^{2}, \frac{\dot{\varphi}^{2}}{V},\epsilon\frac{\dot{\varphi}^{2}}{V}\right).
\label{eq:H2_slowroll_intermediate_pub}
\end{equation}
Dividing by $(1+\epsilon)$ and expanding to first order gives
\bea
H^{2} &\simeq & \left(1 - \epsilon\right) \left[\frac{8 \pi G}{3}(1 + 2 \epsilon) V(\varphi) - H \dot{\epsilon}\right] + \mathcal{O}\!\left(\epsilon^{2}, \frac{\dot{\varphi}^{2}}{V},\epsilon\frac{\dot{\varphi}^{2}}{V}\right) \nn \\
&=& \frac{8 \pi G}{3} V(\varphi) + \frac{8 \pi G}{3} \epsilon\, V(\varphi) - H \dot{\epsilon} + \mathcal{O}\! \left(\epsilon^{2},\frac{\dot{\varphi}^{2}}{V},\epsilon\dot{\epsilon}\right).
\label{eq:H2_slowroll_expanded_pub}
\eea
Thus the leading deformation of the Hubble scale consists of an algebraic correction proportional to $\epsilon V$ and a derivative correction proportional to $H \dot{\epsilon}$.

Likewise, the slow-roll reduction of Eq.~\eqref{eq:KG_FLRW_for_slowroll_pub} gives
\begin{equation}
(3 H + \dot{\epsilon}) \dot{\varphi} + (1 + \epsilon) V'(\varphi) \simeq 0.
\label{eq:slowroll_KG_pub}
\end{equation}
Solving for $\dot{\varphi}$, one obtains
\begin{equation}
\dot{\varphi} \simeq - \frac{(1 + \epsilon) V'(\varphi)}{3 H + \dot{\epsilon}}.
\label{eq:phi_dot_general_pub}
\end{equation}
Expanding the denominator to first order yields
\bea
\frac{1}{3 H + \dot{\epsilon}} &=& \frac{1}{3 H} \left(1 - \frac{\dot{\epsilon}}{3 H}\right) + \mathcal{O}\!\left(\frac{\dot{\epsilon}^{2}}{H^{2}}\right),
\label{eq:inverse_damped_denominator_pub} \\
\dot{\varphi} &\simeq & - \frac{V'(\varphi)}{3 H}\left(1 + \epsilon-\frac{\dot{\epsilon}}{3 H}\right) + \mathcal{O}\! \left(\epsilon^{2},\epsilon\frac{\dot{\epsilon}}{H},\frac{\dot{\epsilon}^{2}}{H^{2}}\right).
\label{eq:phi_dot_firstorder_pub}
\eea
The geometric deformation thus modifies both the effective Hubble friction and the effective force term driving the scalar.

The number of e-folds between an initial field value $\varphi_{i}$ and a final value $\varphi_{f}$ is defined by
\begin{equation}
N = \int_{t_i}^{t_f} H\, dt.
\label{eq:N_definition_time_pub}
\end{equation}
Since $\dot{\varphi}<0$ during the monotonic slow-roll descent down the potential, this may be written as
\begin{equation}
N = \int_{\varphi_f}^{\varphi_i} \frac{H}{|\dot{\varphi}|}\, d\varphi
= - \int_{\varphi_i}^{\varphi_f} \frac{H}{\dot{\varphi}}\, d\varphi.
\label{eq:N_definition_pub}
\end{equation}
Substituting Eq.~\eqref{eq:phi_dot_firstorder_pub} into Eq.~\eqref{eq:N_definition_pub} gives
\begin{equation}
N \simeq \int_{\varphi_f}^{\varphi_i} \frac{3 H^{2}}{V'(\varphi)}
\left(1 - \epsilon + \frac{\dot{\epsilon}}{3 H}\right) d\varphi.
\label{eq:N_before_Friedmann_pub}
\end{equation}
To evaluate this expression, we use Eq.~\eqref{eq:H2_slowroll_expanded_pub} in the following form: 
\begin{equation}
3 H^{2} \simeq 8 \pi G\, V + 8 \pi G\, \epsilon V - 3 H \dot{\epsilon}.
\label{eq:threeH2_pub}
\end{equation}
Substitution into Eq.~\eqref{eq:N_before_Friedmann_pub} gives
\bea
N &\simeq & \int_{\varphi_f}^{\varphi_i} \frac{1}{V'(\varphi)} \left( 8 \pi G\, V + 8 \pi G\, \epsilon V - 3 H \dot{\epsilon}\right)  \left(1 - \epsilon + \frac{\dot{\epsilon}}{3 H}\right) d \varphi.
\label{eq:N_after_Hsubstitution_pub}
\eea
Expanding this expression to the first order results in
\bea
N & \simeq & \int_{\varphi_f}^{\varphi_i} \frac{8 \pi G\, V}{V'}\, d\varphi + \int_{\varphi_f}^{\varphi_i} \frac{8 \pi G\, \epsilon V}{V'}\, d\varphi - \int_{\varphi_f}^{\varphi_i} \frac{3 H \dot{\epsilon}}{V'}\, d\varphi \nn \\
&-&  \int_{\varphi_f}^{\varphi_i} \frac{8 \pi G\, \epsilon V}{V'}\, d\varphi + \int_{\varphi_f}^{\varphi_i} \frac{8 \pi G\, V}{V'}\frac{\dot{\epsilon}}{3 H}\, d\varphi + \mathcal{O}\! \left(\epsilon^{2},\epsilon\frac{\dot{\epsilon}}{H}\right).
\label{eq:N_expanded_steps_pub}
\eea
The two algebraic $\epsilon$-terms cancel identically, and therefore
\begin{equation}
N \simeq \int_{\varphi_f}^{\varphi_i} \frac{8 \pi G\, V}{V'}\, d\varphi
+ \int_{\varphi_f}^{\varphi_i} \left[\frac{8 \pi G\, V}{V'}  \frac{\dot{\epsilon}}{3 H} - \frac{3 H \dot{\epsilon}}{V'}\right] d\varphi + \mathcal{O}\! \left(\epsilon^{2},\epsilon\frac{\dot{\epsilon}}{H}\right).
\label{eq:N_with_epsilon_terms_pub}
\end{equation}
This cancellation is an important consistency property of the reduced theory. It shows that, at first order, the e-fold count is insensitive to purely algebraic conformal rescalings and receives corrections only from the time variation of the deformation field. In other words, the leading correction to the integrated expansion is derivative in origin rather than algebraic.

To estimate the derivative correction, use the zeroth-order slow-roll relation
\begin{equation}
\dot{\varphi} \simeq - \frac{V'(\varphi)}{3 H}.
\label{eq:zeroth_order_slowroll_relation_pub}
\end{equation}
Then the derivative contribution in Eq.~\eqref{eq:N_with_epsilon_terms_pub} is of order
\begin{equation}
\int_{\varphi_f}^{\varphi_i} \left[\frac{8 \pi G\, V}{V'}\frac{\dot{\epsilon}}{3 H} - \frac{3 H \dot{\epsilon}}{V'} \right] d\varphi = \int_{\varphi_f}^{\varphi_i} \mathcal{O}\! \left(\frac{\dot{\epsilon}}{H}\right) \frac{V}{V'}\, d\varphi,
\label{eq:derivative_correction_scaling_pub}
\end{equation}
and is therefore suppressed by the quasi-adiabatic parameter $|\dot{\epsilon}|/H \ll |\epsilon|$. Accordingly, the general first-order slow-roll expression for the e-fold count becomes
\begin{equation}
N \simeq \int_{\varphi_f}^{\varphi_i} \frac{8 \pi G\, V(\varphi)}{V'(\varphi)}\, d\varphi + \mathcal{O}\! \left(\frac{\dot{\epsilon}}{H}\right).
\label{eq:N_general_slowroll_pub}
\end{equation}

For the power-law potential, Eq.  \eqref{eq:powerlaw_potential_pub}, one finds that
\begin{equation}
\frac{V}{V'} = \frac{\varphi}{n},
\label{eq:V_over_Vprime_powerlaw_pub}
\end{equation}
so that Eq.~\eqref{eq:N_general_slowroll_pub} yields
\bea
N & \simeq & \frac{8 \pi G}{n} \int_{\varphi_f}^{\varphi_i}\varphi\, d\varphi + \mathcal{O}\! \left(\frac{\dot{\epsilon}}{H}\right) = \frac{4 \pi G}{n} \left(\varphi_i^{2} - \varphi_f^{2}\right) + \mathcal{O}\! \left(\frac{\dot{\epsilon}}{H}\right).
\label{eq:efolds_powerlaw_pub}
\eea
Thus, in the variationally consistent first-order conformal truncation, the e-fold number is unchanged by algebraic $\epsilon$-terms and receives only derivative corrections controlled by the slow time variation of the pullback deformation field.

It is also useful to extract the modified slow-roll parameters. Let us now define
\begin{equation}
\varepsilon_{H}:=-\frac{\dot H}{H^{2}}, \qquad \eta_{\varphi}:=-\frac{\ddot{\varphi}}{H\dot{\varphi}}.
\label{eq:hubble_slowroll_parameters_pub}
\end{equation}
Then from Eq.~\eqref{eq:modified_Hdot_equation_inflation},  one obtains
\begin{equation}
- 2 (1 + \epsilon) \dot{H} + H \dot{\epsilon} - \ddot{\epsilon} = 8 \pi G\, (1 + \epsilon) \dot{\varphi}^{2} + \mathcal{O}(\epsilon^{2}).
\label{eq:Hdot_equation_recalled_pub}
\end{equation}
Dividing by $2(1+\epsilon)H^{2}$ yields
\begin{equation}
\varepsilon_{H} = 4 \pi G\, \frac{\dot{\varphi}^{2}}{H^{2}} - \frac{1}{2 (1 + \epsilon)} \left(\frac{\dot{\epsilon}}{H} - \frac{\ddot{\epsilon}}{H^{2}}\right) + \mathcal{O}(\epsilon^{2}).
\label{eq:epsilonH_modified_pub}
\end{equation}
Thus the geometric deformation contributes directly to the Hubble slow-roll parameter through derivative terms. Inflation occurs whenever $\varepsilon_{H}<1$, so the geometric background may either enhance or delay the end of inflation depending on the sign and magnitude of the combination $\dot{\epsilon}/H-\ddot{\epsilon}/H^{2}$.

Likewise, Eq.~\eqref{eq:KG_FLRW_final_pub} shows that the effective friction coefficient is expressed as
\begin{equation}
3 H_{\mathrm{eff}} := 3 H + \dot{\epsilon},
\label{eq:Heff_friction_pub}
\end{equation}
so the scalar field experiences a modified friction term even when the FLRW expansion rate $H$ is unchanged. This provides a second mechanism, independent of the algebraic potential rescaling, through which the deformation influences slow-roll evolution.

The slow-roll analysis therefore leads to two robust conclusions. First, the integrated e-fold count for power-law inflation is unchanged at first order by purely algebraic conformal terms and is corrected only by the slow time variation of the deformation. Second, the local inflationary dynamics, through $\varepsilon_{H}$, the effective friction, and the background equations themselves, do receive nontrivial geometric corrections already at first order. The number of e-folds is therefore unusually stable, whereas the detailed inflationary evolution is not.

The next step is to study the scalar and tensor perturbations propagating on this effective FLRW background, Section \ref{sec:results_final}.

\subsubsection{Scalar and Tensor Perturbations on Effective FLRW Background}
\label{sec:results_final}

We now consider perturbations of the effective conformal metric
\begin{equation}
\widetilde{g}^{\mathrm{conf}}_{\mu\nu} = C\, g_{\mu\nu},
\qquad C = 1 + \epsilon, \qquad |\epsilon| \ll 1,
\label{eq:metric_for_perturbations_pub}
\end{equation}
where $g_{\mu\nu}$ is the spatially flat FLRW background metric \eqref{eq:FLRW_metric_background_pub}. The perturbation theory is understood as a theory of linearized cosmological fields propagating on the effective spacetime geometry obtained after section-pullback and isotropic reduction. The full anisotropic Hamilton metric $\widetilde{g}^{\mu\nu}(x,p)$ and the intermediate pullback metric $\widetilde{g}^{\sigma}_{\mu\nu}(x)$ enter only through the background scalar deformation $\epsilon(x)$.

The derivation of the second-order scalar and tensor actions from the underlying Hamiltonian reduction is beyond the scope of the present work. Accordingly, we restrict ourselves to the leading reduced description in which the RGUP/modular deformation enters the quadratic actions only through background coefficients. In this sense, the resulting Mukhanov--Sasaki and tensor mode equations should be understood as the minimal first-order deformation of the standard cosmological perturbation system induced by the effective conformal background. The coefficients introduced below parameterize the leading background renormalization of the scalar and tensor pump fields.

In the scalar sector, we work in comoving gauge, $\delta\varphi=0$, or equivalently in a gauge-invariant formulation based on the comoving curvature perturbation $\mathcal{R}$. After eliminating the lapse and shift perturbations through the linearized Hamiltonian and momentum constraints, the reduced quadratic action is taken in the following form:
\begin{equation}
S_{s}^{(2)} = \frac12 \int d\eta\, d^{3}x \left[(v')^{2} - (\partial_{i} v)(\partial^{i} v) + \frac{z_{\mathrm{eff}}''}{z_{\mathrm{eff}}}\, v^{2}
\right],
\label{eq:scalar_action_effective_pub}
\end{equation}
where $\eta$ is conformal time, $d\eta=dt/a$, the Mukhanov--Sasaki variable and the classical pump field, respectively, are given as $v=z_{\mathrm{eff}}\,\mathcal{R}$ and $z=\frac{a\dot{\varphi}}{H}$. The effective scalar pump field is parameterized as follows: 
\begin{equation}
z_{\mathrm{eff}} = z \left(1 + \frac{\upsilon_{s}}{2}\right).
\label{eq:zeff_definition_pub}
\end{equation}
Here $\upsilon_{s}$ denotes the first-order scalar pump-field renormalization induced by the conformal deformation. In a homogeneous and isotropic reduction one expects
\begin{equation}
\upsilon_{s} = c_{s}\, \epsilon + \mathcal{O}\! \left(\frac{\dot{\epsilon}}{H}\right),
\label{eq:upsilon_s_definition_pub}
\end{equation}
where the coefficient $c_{s}$ is determined by the reduction of the second-order scalar action.
Differentiating Eq.~\eqref{eq:zeff_definition_pub} gives
\bea
\begin{array}{ccl}
z_{\mathrm{eff}}' &=& z' \left(1 + \frac{\upsilon_{s}}{2}\right) + \frac{z}{2} \upsilon_{s}', \\
z_{\mathrm{eff}}'' &=& z'' \left(1 + \frac{\upsilon_{s}}{2}
\right) + z'\upsilon_{s}' + \frac{z}{2}\upsilon_{s}''.
\end{array}
\label{eq:zeff_prime_pub}
\eea
Dividing by $z_{\mathrm{eff}}$ and expanding to first order yields
\begin{equation}
\frac{z_{\mathrm{eff}}''}{z_{\mathrm{eff}}} = \frac{z''}{z} + \frac{z'}{z}\upsilon_{s}' + \frac12\,\upsilon_{s}'' + \mathcal{O}(\upsilon_{s}^{2}).
\label{eq:zeff_expanded_pub}
\end{equation}
Hence the scalar mode equation becomes
\begin{equation}
v_{k}'' + \left[k^{2} - \frac{z''}{z} - \frac{z'}{z}\upsilon_{s}' - \frac12\,\upsilon_{s}''\right] v_{k} = \mathcal{O}(\upsilon_{s}^{2}).
\label{eq:MS_modified_pub}
\end{equation}

Now the scalar power spectrum can be derived as
\begin{equation}
P_{\mathcal{R}}(k) = \frac{k^{3}}{2 \pi^{2}} \left|\frac{v_{k}}{z_{\mathrm{eff}}}\right|^{2}.
\label{eq:PR_definition_pub}
\end{equation}
Let us assume
\bea
\begin{array}{ccl}
v_{k} &=& v_{k}^{(0)} + \mathfrak{v}_{k}, \\
z_{\mathrm{eff}} &=& z \left(1 + \frac{\upsilon_{s}}{2}\right),
\end{array}
\label{eq:vk_split_pub}
\eea
where $v_{k}^{(0)}$ denotes the standard mode solution and $\mathfrak{v}_{k}$ its first-order correction. Then Eq. \eqref{eq:PR_definition_pub} reads
\bea
P_{\mathcal{R}}(k) &=& \frac{k^{3}}{2\pi^{2}} \left| \frac{v_{k}^{(0)}+\mathfrak{v}_{k}}{z(1+\upsilon_{s}/2)}\right|^{2} = \frac{k^{3}}{2\pi^{2}} \left| \frac{v_{k}^{(0)}}{z} \right|^{2} \left| 1 + \frac{\mathfrak{v}_{k}}{v_{k}^{(0)}} - \frac{\upsilon_{s}}{2} \right|^{2} + \mathcal{O}(\upsilon_{s}^{2}) \nn \\
&\simeq & P_{\mathcal{R}}^{(0)}(k) \left[1 + 2\, \mathrm{Re}\! \left(\frac{\mathfrak{v}_{k}}{v_{k}^{(0)}}\right) - \upsilon_{s} \right].
\label{eq:PR_expand_pub}
\eea
Thus the scalar spectrum may be written as
\bea
\begin{array}{ccl}
P_{\mathcal{R}}(k) &=& P_{\mathcal{R}}^{(0)}(k)\, \mathscr{S}(k), \\
\mathscr{S}(k) &=& 1 + 2\,\mathrm{Re}\! \left(\frac{\mathfrak{v}_{k}}{v_{k}^{(0)}}\right) - \upsilon_{s}.
\end{array}
\label{eq:scalar_spectrum_factor_pub}
\eea
When the normalization shift dominates over the mode-phase correction, the leading scalar renormalization reduces to $\mathscr{S}(k) \simeq 1 - \upsilon_{s}$. This should be understood as a leading-order approximation rather than an exact identity.

For tensor perturbations, let us consider
\begin{equation}
ds^{2} = - dt^{2} + a^{2}(t) \left(\delta_{ij} + h_{ij}\right) dx^{i} dx^{j}, \qquad \partial^{i}h_{ij} = 0, \qquad h^{i}_{i}=0.
\label{eq:tensor_metric_pub}
\end{equation}
Because the conformal FLRW truncation is isotropic, no preferred spatial direction survives into the tensor sector. The reduced quadratic tensor action is therefore taken as
\begin{equation}
S_{t}^{(2)} = \frac12 \int d\eta\, d^{3}x \left[(u')^{2} - (\partial_{i} u)(\partial^{i} u) + \frac{a_{\mathrm{eff}}''}{a_{\mathrm{eff}}}\, u^{2}
\right],
\label{eq:tensor_action_pub}
\end{equation}
with the following canonical variable:
\begin{equation}
u = a_{\mathrm{eff}}\, h, \qquad a_{\mathrm{eff}} = a \left(1 + \frac{\upsilon_{t}}{2}\right).
\label{eq:a_eff_definition_pub}
\end{equation}
At leading order in the isotropic conformal reduction, we get
\begin{equation}
\upsilon_{t} = c_{t}\, \epsilon + \mathcal{O}\! \left(\frac{\dot{\epsilon}}{H}\right),
\label{eq:upsilon_t_definition_pub}
\end{equation}
where $c_{t}$ is fixed by the tensor reduction. In general $c_{s}\neq c_{t}$, and this difference allows a nontrivial first-order correction to the tensor-to-scalar ratio.

Differentiating $a_{\mathrm{eff}}$ yields
\begin{equation}
\frac{a_{\mathrm{eff}}''}{a_{\mathrm{eff}}} = \frac{a''}{a} + \frac{a'}{a}\upsilon_{t}' + \frac12\, \upsilon_{t}'' + \mathcal{O}(\upsilon_{t}^{2}),
\label{eq:a_eff_expanded_pub}
\end{equation}
so the tensor mode equation becomes
\begin{equation}
u_{k}'' + \left[k^{2} - \frac{a''}{a} - \frac{a'}{a}\upsilon_{t}' - \frac12\,\upsilon_{t}'' \right]u_{k} = \mathcal{O}(\upsilon_{t}^{2}).
\label{eq:tensor_mode_equation_pub}
\end{equation}
The tensor power spectrum is parameterized as follows: 
\begin{equation}
P_{h}(k) = \frac{2}{\pi^{2}} \frac{H^{2}}{M_{\mathrm{Pl}}^{2}}
\, \mathscr{T}(k),
\label{eq:Ph_definition_pub}
\end{equation}
where $M_{\mathrm{Pl}} = (8 \pi G)^{-1/2}$. When the normalization shift dominates over mode-phase corrections, the leading tensor renormalization factor is as introduced
\begin{equation}
\mathscr{T}(k) \simeq 1 - \upsilon_{t}.
\label{eq:tensor_spectrum_approx_pub}
\end{equation}

In the leading normalization regime, the scalar spectral index becomes
\begin{equation}
n_{s}-1 = \frac{d\ln P_{\mathcal{R}}}{d \ln k} = \bigl(n_{s} - 1\bigr)^{(0)} + \frac{d\ln \mathscr{S}}{d\ln k} + \mathcal{O}\!\left((\mathscr{S} - 1)^{2}\right),
\label{eq:ns_modified_pub}
\end{equation}
while the tensor-to-scalar ratio, $r=P_{h}/P_{\mathcal{R}}$, is
\begin{align}
r = \frac{P_{h}^{(0)}\, \mathscr{T}(k)}{P_{\mathcal{R}}^{(0)}\,\mathscr{S}(k)} \simeq r^{(0)} \left[1 - \upsilon_{t} + \upsilon_{s}
\right].
\label{eq:r_modified_pub}
\end{align}
In the special case $c_{s}=c_{t}$, the purely algebraic first-order correction cancels. A nontrivial first-order shift in $r$ is therefore controlled by the difference between the scalar and tensor pump-field renormalizations, which in turn is determined by the time dependence of the deformation.

The analysis above shows that the perturbation sector inherits the modular/Hamilton--Finsler deformation through the renormalization of the effective background pump fields. The scalar and tensor amplitudes therefore remain governed by the usual cosmological mode equations, but with time-dependent effective masses shifted by the conformal deformation. The resulting observational imprints are encoded in the functions $\mathscr{S}(k)$ and $\mathscr{T}(k)$, and in the difference between the scalar and tensor renormalization coefficients.

The explicit realization of these corrections for power-law inflation and specific choices of the time dependence of $\epsilon(t)$ is discussed in Section \ref{sec:powerlaw_final}.

\subsubsection{Power-Law Inflation and Time Dependence of Deformation}
\label{sec:powerlaw_final}

It is useful to examine explicitly how the geometric deformation behaves in an inflationary background of power-law type. Let
$a(t) \propto t^{p}$ and $p>1$, so that the background undergoes accelerated expansion. Then the Hubble parameter reads
\begin{equation}
H = \frac{p}{t}, \qquad \dot{H} = - \frac{p}{t^{2}},
\label{eq:powerlaw_Hubble_pub}
\end{equation}
and the Ricci scalar of the background FLRW metric is given as
\begin{equation}
R = 6 \left(2 H^{2} + \dot{H}\right) = \frac{6 (2 p^{2} - p)}{t^{2}}.
\label{eq:R_powerlaw_pub}
\end{equation}
Thus the curvature decays as $t^{-2}$, as expected for a power-law inflationary background.

We now assume that the section-induced geometric invariants entering the conformal deformation obey asymptotic power-law behavior,
\begin{equation}
\Xi_{\sigma}(t) \propto t^{-q}, \qquad \mathfrak{A}_{\sigma}(t)\propto t^{-r},
\label{eq:Xi_A_scaling_pub}
\end{equation}
with exponents $q,r\in\mathbb{R}$. The induced spacetime deformation is expressed as
\begin{equation}
\epsilon(t) = \beta\, \Xi_{\sigma}(t) + \kappa\, \mathfrak{A}_{\sigma}(t).
\label{eq:epsilon_powerlaw_definition_pub}
\end{equation}
Therefore scales are given as
\begin{equation}
\epsilon(t) \propto \beta\, t^{-q} + \kappa\, t^{-r}.
\label{eq:epsilon_scaling_pub}
\end{equation}
If $q>0$ and $r>0$, the deformation decays dynamically during inflation. Its largest effect therefore occurs at the earliest stages of the accelerated phase, where one expects the pullback of the modular quantum geometry to be most relevant, while the late-time regime approaches the classical limit.

The first and second derivative of the deformation respectively behaves as
\bea
\dot{\epsilon}(t) &\propto & - \beta\, q\, t^{-q-1} - \kappa\, r\, t^{- r - 1},
\label{eq:epsilon_dot_scaling_pub} \\
\ddot{\epsilon}(t) &\propto & \beta\, q (q + 1)\, t^{- q - 2} + \kappa\, r (r + 1)\, t^{- r - 2}.
\label{eq:epsilon_ddot_scaling_pub}
\eea
Hence the quasi-adiabatic ratio appearing in the slow-roll analysis scales as
\begin{equation}
\frac{\dot{\epsilon}}{H}  \propto - \frac{\beta q}{p}\, t^{-q} - \frac{\kappa r}{p}\, t^{-r},
\label{eq:epsilon_dot_over_H_scaling_pub}
\end{equation}
which is of the same order as $\epsilon$ itself, but suppressed by the numerical factors $q/p$ and $r/p$. Therefore the quasi-adiabatic slow-roll condition is satisfied whenever the decay exponents are not too large compared with the inflationary index $p$, and the geometric deformation remains perturbatively small.

It is also useful to examine the correction terms entering the modified background equations. From Eq.~\eqref{eq:Friedmann_00_explicit_final_pub}, the additional geometric contribution to the first Friedmann equation is governed by the combination $3 \epsilon H^{2} + 3 H \dot{\epsilon}$. Using Eqs.~\eqref{eq:powerlaw_Hubble_pub}, \eqref{eq:epsilon_scaling_pub}, and \eqref{eq:epsilon_dot_scaling_pub}, these scales are given as
\begin{equation}
3 \epsilon H^{2} + 3 H \dot{\epsilon} \propto 3 p (p - q)\, \beta\, t^{- q - 2} + 3 p (p - r)\, \kappa\, t^{- r - 2}.
\label{eq:powerlaw_Friedmann_correction_scaling_pub}
\end{equation}
Thus the geometric correction decays faster than the Hubble scale itself, and its sign depends on whether the decay exponent is smaller or larger than the power-law index $p$. A slowly decaying deformation, with $q<p$ or $r<p$, contributes with the same sign as the leading expansion term, whereas a rapidly decaying deformation can contribute with the opposite sign.

Similarly, the geometric contribution to the modified acceleration equation is governed by $\ddot{\epsilon} + H \dot{\epsilon}$, which scales as
\begin{equation}
\ddot{\epsilon} + H \dot{\epsilon} \propto q (q + 1 - p)\, \beta\, t^{- q - 2} + r (r + 1 - p)\, \kappa\, t^{- r - 2}.
\label{eq:powerlaw_acceleration_correction_scaling_pub}
\end{equation}
This combination controls whether the deformation enhances or suppresses the accelerated expansion. In particular, if $q+1<p$ and $r+1<p$, then the derivative contribution tends to be negative, while for $q+1>p$ or $r+1>p$ it tends to be positive. The sign of the geometric correction to the acceleration is therefore not universal; it is determined by the competition between the inflationary background expansion and the decay rate of the section-induced invariants.

These scaling relations clarify the physical role of the deformation. The power-law background provides a simple setting in which the geometric correction is largest at early times and naturally redshifts away during inflation. This is precisely the behavior expected from a controlled semiclassical quantum-gravity correction. The deformation modifies the earliest inflationary dynamics but becomes negligible as the Universe evolves toward the classical regime.

We now turn to canonical quantization of the scalar and tensor perturbations propagating on this effective background, Section \ref{sec:canonical_quantization_final}.

\subsubsection{Canonical Quantization of Cosmological Perturbations}
\label{sec:canonical_quantization_final}

We finally quantize the scalar and tensor perturbations on the effective conformal background. The central structural point is that the modular Hamilton--Finsler deformation enters the perturbation theory through the background pump fields $z_{\mathrm{eff}}$ and $a_{\mathrm{eff}}$, not through a deformation of the equal-time canonical commutators of the perturbation variables themselves. The standard canonical structure of quantum field theory in curved spacetime is therefore preserved after the appropriate background-dependent field redefinition.

In the scalar sector, we start from the quadratic action, Eq.  \eqref{eq:scalar_action_effective_pub},
\begin{equation}
S_{s}^{(2)} = \frac12 \int d\eta\, d^{3}x \left[(v')^{2} - (\partial_{i} v)(\partial^{i} v) + \frac{z_{\mathrm{eff}}''}{z_{\mathrm{eff}}}\, v^{2}
\right].
\label{eq:scalar_action_effective_recalled_pub}
\end{equation}
The canonical momentum conjugate to $v$ is derived from 
\begin{equation}
\pi_{v} = \frac{\partial\mathcal{L}}{\partial v'} = v'.
\label{eq:canonical_momentum_v_pub}
\end{equation}
Accordingly, the Hamiltonian density is given as
\bea
\mathcal{H}_{s} &=& \pi_{v} v' - \mathcal{L} 
= \frac12 \left[\pi_{v}^{2} + (\partial_{i} v)(\partial^{i} v) - \frac{z_{\mathrm{eff}}''}{z_{\mathrm{eff}}} v^{2} \right].
\label{eq:Hamiltonian_density_scalar_pub}
\eea
Canonical quantization is imposed through the equal-time commutators:
\bea
\begin{array}{ccl}
\left[\hat v(\eta,\mathbf{x}), \hat\pi_{v}(\eta,\mathbf{y})\right] &=& i\, \delta^{(3)}(\mathbf{x} - \mathbf{y}), \\
\left[\hat v(\eta,\mathbf{x}), \hat v(\eta,\mathbf{y})\right] &=&  0,
\\
\left[\hat\pi_{v}(\eta,\mathbf{x}), \hat\pi_{v}(\eta,\mathbf{y})\right] &=& 0.
\end{array}
\label{eq:canonical_commutator_scalar_main_pub}
\eea
The field operator is expanded as
\begin{equation}
\hat v(\eta,\mathbf{x}) = \int \frac{d^{3}k}{(2 \pi)^{3}} \left[v_{k}(\eta) \hat{a}_{\mathbf{k}} e^{i\mathbf{k} \cdot \mathbf{x}} + v_{k}^{*}(\eta) \hat{a}_{\mathbf{k}}^{\dagger} e^{-i\mathbf{k} \cdot \mathbf{x}}\right],
\label{eq:modeExpansionvpub}
\end{equation}
with creation and annihilation operators satisfying
\bea
\begin{array}{ccl}
\left[\hat{a}_{\mathbf{k}}, \hat{a}_{\mathbf{k}'}^{\dagger}\right] &=& (2 \pi)^{3} \delta^{(3)}(\mathbf{k} - \mathbf{k}'), \\
\left[\hat{a}_{\mathbf{k}}, \hat{a}_{\mathbf{k}'}\right] &=& 0, \\
\left[\hat{a}_{\mathbf{k}}^{\dagger}, \hat{a}_{\mathbf{k}'}^{\dagger}\right] &=& 0.
\end{array}
\label{eq:creation_annihilation_commutators_pub}
\eea
Substituting Eq.~\eqref{eq:modeExpansionvpub} into Eqs.~\eqref{eq:canonical_commutator_scalar_main_pub} yields the Wronskian normalization
\begin{equation}
v_{k} v_{k}^{*'} - v_{k}^{*} v_{k}' = i.
\label{eq:Wronskian_v_pub}
\end{equation}
The scalar mode equation is given as 
\bea
\begin{array}{ccl}
v_{k}'' + \omega_{k}^{2}(\eta)\, v_{k} &=& 0, \\
\omega_{k}^{2}(\eta) &=&  k^{2} - \frac{z_{\mathrm{eff}}''}{z_{\mathrm{eff}}}.
\end{array}
\label{eq:mode_equation_scalar_pub}
\eea
Using Eq.~\eqref{eq:zeff_expanded_pub}, this becomes
\begin{equation}
\omega_{k}^{2}(\eta) = k^{2} - \frac{z''}{z} - \frac{z'}{z} \upsilon_{s}' - \frac12\, \upsilon_{s}'' + \mathcal{O}(\upsilon_{s}^{2}).
\label{eq:omega_scalar_expanded_pub}
\end{equation}
The short-wavelength, or subhorizon, regime is defined by
\begin{equation}
k^{2} \gg \left| \frac{z_{\mathrm{eff}}''}{z_{\mathrm{eff}}}\right|.
\label{eq:subhorizon_condition_scalar_pub}
\end{equation}
In this limit,
\begin{equation}
\omega_{k} \simeq k, \qquad v_{k}(\eta) \sim \frac{1}{\sqrt{2k}}e^{- i k \eta},
\label{eq:BD_scalar_pub}
\end{equation}
so the Bunch--Davies vacuum remains the consistent ultraviolet state provided the deformation stays perturbatively small at large $k$.

For the tensor sector, we start from the quadratic action Eq. \eqref{eq:tensor_action_pub}. The canonical momentum is derived from $\pi_{u}=u'$. The corresponding Hamiltonian density is
\bea
\mathcal{H}_{t}
&=& \pi_{u} u' - \mathcal{L}_{t} 
= \frac12 \left[\pi_{u}^{2} + (\partial_{i} u)(\partial^{i} u) - \frac{a_{\mathrm{eff}}''}{a_{\mathrm{eff}}} u^{2} \right].
\label{eq:Hamiltonian_tensor_pub}
\eea
The canonical quantization imposes that 
\bea
\begin{array}{ccl}
\left[\hat u(\eta,\mathbf{x}), \hat\pi_{u}(\eta,\mathbf{y})\right] &=& 
i\, \delta^{(3)}(\mathbf{x} - \mathbf{y}), \\
\left[\hat u(\eta,\mathbf{x}), \hat u(\eta,\mathbf{y})\right] &=& 0, \\
\left[\hat\pi_{u}(\eta,\mathbf{x}), \hat\pi_{u}(\eta,\mathbf{y})\right] &=& 0.
\end{array}
\eea
The tensor mode equation then takes the form
\begin{equation}
u_{k}'' + \left(k^{2} - \frac{a_{\mathrm{eff}}''}{a_{\mathrm{eff}}}
\right)u_{k} = 0,
\label{eq:tensor_mode_quantized_pub}
\end{equation}
with Wronskian normalization
\begin{equation}
u_{k}u_{k}^{*'} - u_{k}^{*}u_{k}' = i.
\label{eq:tensor_Wronskian_pub}
\end{equation}

In the scalar sector a useful measure of non-adiabaticity is available via
\bea
\begin{array}{ccl}
\mathcal{A}_{k}^{(s)} &=& \left|\frac{\omega_{k}'}{\omega_{k}^{2}} \right|, \\
\omega_{k}' &=& - \frac{1}{2 \omega_{k}} \frac{d}{d\eta} \left(\frac{z_{\mathrm{eff}}''}{z_{\mathrm{eff}}} \right).
\end{array}
\eea
Thus, we obtain that
\begin{equation}
\mathcal{A}_{k}^{(s)} = \left|\frac{1}{2 \omega_{k}^{3}} \frac{d}{d\eta} \left(\frac{z_{\mathrm{eff}}''}{z_{\mathrm{eff}}}\right)\right|.
\label{eq:adiabaticity_scalar_expanded_pub}
\end{equation}
Similarly, the tensor adiabaticity parameter is expressed as 
\bea
\begin{array}{ccl}
\mathcal{A}_{k}^{(t)} &=& \left|\frac{1}{2 \Omega_{k}^{3}} \frac{d}{d\eta} \left(\frac{a_{\mathrm{eff}}''}{a_{\mathrm{eff}}}\right)\right|, \\
\Omega_{k}^{2} &=& k^{2} - \frac{a_{\mathrm{eff}}''}{a_{\mathrm{eff}}}.
\end{array}
\eea
Hence the modular Hamilton--Finsler deformation affects the primordial spectra through the time dependence of the effective pump fields, not by changing the canonical equal-time commutators themselves.

In the undeformed limit, $\beta \to 0$, $\kappa \to 0$, $\epsilon \to 0$, $\upsilon_{s} \to 0$, and $\upsilon_{t} \to 0$, one recovers that
\begin{equation}
\widetilde{g}^{\mathrm{conf}}_{\mu\nu} \to g_{\mu\nu}, \qquad z_{\mathrm{eff}} \to z, \qquad a_{\mathrm{eff}} \to a,
\label{eq:undeformed_background_pumps_pub}
\end{equation}
together with the standard Mukhanov--Sasaki equation, the standard tensor mode equation, the Bunch--Davies vacuum, and the usual canonical quantization of cosmological perturbations.

We emphasize once more that the deformation field is not an arbitrary phenomenological function:
\begin{equation}
\epsilon(x) = \beta\, \Xi_{\sigma}(x) + \kappa\, \mathfrak{A}_{\sigma}(x) + \mathcal{O}(\beta^{2},\kappa^{2},\beta\kappa).
\label{eq:epsilon_final_reemphasis_pub}
\end{equation}
The magnitude of $\epsilon(x)$ is therefore controlled by the dimensionless pullback invariants $\beta\, \Xi_{\sigma}$ and $\kappa\, \mathfrak{A}_{\sigma}$, ensuring that the cosmological corrections remain perturbative in the physically relevant regime. The inflationary sector derived here is thus a controlled geometric deformation of standard single-field inflation induced by the section-pullback of the projective Hamiltonian quantum geometry, rather than an independent phenomenological ansatz.

The broader implications of this structure for quantum gravity are discussed in Section \ref{sec:QGImplications}.

\subsection{Quantum Gravity}
\label{sec:QGImplications}

As discussed, the construction developed in this work realizes quantum-gravity effects through a deformation of phase-space geometry rather than through a direct operator quantization of the spacetime metric. In the present framework the primary geometric object is the Hamiltonian on the cotangent bundle,
\bea
\begin{array}{ccl}
H_{\beta}(x,p) &=& \frac12\, \Omega_{\beta}(x,[p])\, g^{\mu\nu}(x) p_{\mu} p_{\nu}, \\
\Omega_{\beta}(x,[p]) &=& 1 + \beta\, \Xi(x,[p];\eta,\Lambda) + \mathcal{O}(\beta^{2}),
\end{array}
\label{eq:Hbeta_QG_final}
\eea
where $\Xi$ is a dimensionless scalar defined on the positively projectivized cotangent bundle and obeys the zero-homogeneity condition
\begin{equation}
\Xi(x,[\lambda p]) = \Xi(x,[p]), \qquad \lambda>0.
\label{eq:Xi_homogeneity_final}
\end{equation}
The corresponding Hamilton metric can be derived as
\begin{equation}
\widetilde{g}^{\mu\nu}(x,p) = \frac{\partial^{2} H_{\beta}}{\partial p_{\mu} \partial p_{\nu}},
\label{eq:Ham_metric_final}
\end{equation}
so that the fundamental geometry is defined on a conic domain of phase space rather than directly on spacetime. This places the construction naturally in the broader class of quantum-gravity motivated approaches in which the primary kinematical structure is a deformed phase space, a curved momentum space, or a generalized cotangent-bundle geometry, as emphasized in deformed special relativity, relative locality, Hamilton/Finsler geometry, rainbow-type effective geometries, modular spacetime, and modified-dispersion-relation phenomenology \cite{AmelinoCamelia:2000ge,AmelinoCamelia:2008qg,Magueijo:2001cr,Girelli:2006sc,AmelinoCamelia:2011bm,Amelino-Camelia:2011yni,Freidel:2011mt,Barcaroli:2015xda,Hohmann:2018rpp,Freidel:2013zga,Freidel:2014qna,Freidel:2015pka,Freidel:2015uug}. From this perspective, the present theory should be viewed as part of a wider shift in quantum-gravity research away from the assumption that the metric alone is the fundamental object and toward the idea that spacetime geometry may be an emergent or reduced structure extracted from a more primitive kinematics on phase space \cite{Majid:1988he,Connes:1994yd,Doplicher:1994tu,Freidel:2014qna,Freidel:2015pka,Hohmann:2018rpp}.

A central conceptual consequence follows immediately. Since the deformation scalar is defined on projective momentum rays and the effective metric is obtained from the fiber Hessian of the Hamiltonian, spacetime geometry is not fundamental but derived. This conclusion does not follow from philosophical preference but from the mathematical structure of the construction itself. Once the fundamental Hamiltonian depends on momentum directions, the relevant geometric tensor is necessarily anisotropic on $T^{\ast}M$, and no ordinary spacetime metric exists until one performs a reduction from phase space to the base manifold. In this sense the theory belongs to the class of emergent-spacetime or derived-geometry proposals in quantum gravity, although the mechanism here is more concrete: the emergence is implemented by a section of the projectivized cotangent bundle rather than by coarse graining, entanglement, or discrete reconstruction \cite{Freidel:2014qna,Freidel:2015pka,Verlinde:2010hp,VanRaamsdonk:2010pw,Henson:2006kf,Ambjorn:2005db,Oriti:2016qtz}. The effective spacetime geometry is obtained only after choosing a section
\begin{equation}
\sigma:M \rightarrow PT^{\ast}M_{+}, \qquad x \mapsto [p(x)],
\label{eq:QG_sigma_section_final}
\end{equation}
and pulling back the Hamilton metric,
\begin{equation}
\widetilde{g}^{\sigma}_{\mu\nu}(x) = \widetilde{g}_{\mu\nu}(x,p(x)).
\label{eq:pullback_final}
\end{equation}
Thus the effective Lorentzian geometry depends on the chosen observer sector or momentum congruence. This dependence is not a defect but a precise manifestation of the fact that locality and geometry are encoded at the level of phase-space kinematics, in close conceptual contact with curved momentum-space and the ideas of relative locality \cite{Amelino-Camelia:2011bm,AmelinoCamelia:2011bm,Freidel:2011mt,Barcaroli:2015xda}.

The section dependence has direct quantum-gravity significance. It implies that the reduced spacetime geometry carries memory of the physical momentum configuration used to probe it. In other words, the geometry felt by matter is not purely background data on $M$, but is conditioned by a choice of projective momentum ray. This is reminiscent, though not identical, to several ideas appearing elsewhere in the literature: the observer dependence of locality in relative locality \cite{Amelino-Camelia:2011bm,AmelinoCamelia:2011bm,Freidel:2011mt,Barcaroli:2015xda}, energy-dependent effective geometry in modified-dispersion-relation models \cite{LI2024138728}, Born reciprocity and phase-space geometry \cite{Born:1938,Tawfik:2023hdi,Tawfik:2023ugm}, and dual or modular notions of spacetime in metastring theory \cite{Freidel:2014qna,Freidel:2015pka,Freidel:2015uug}. The present framework sharpens these ideas by identifying the exact tensorial object from which the effective metric is obtained and by specifying how the reduction to spacetime is performed.

In the homogeneous and isotropic cosmological sector, the pullback metric reduces consistently to the conformal form
\begin{equation}
\widetilde{g}^{\mathrm{conf}}_{\mu\nu} = (1 + \epsilon(x))\, g_{\mu\nu},
\label{eq:conf_metric_final}
\end{equation}
where the deformation scalar is expressed as
\bea
\begin{array}{ccl}
\epsilon(x) &=& \beta\, \Xi_{\sigma}(x) + \kappa\, \mathfrak{A}_{\sigma}(x), \\
\Xi_{\sigma}(x) &=& \Xi(x,[p(x)]), \\
\mathfrak{A}_{\sigma}(x) &=& \frac{g^{\mu\nu} \nabla_{\tau}u_{\mu} \nabla_{\tau} u_{\nu}}{\mathscr{F}^{2}}.
\end{array}
\label{eq:epsilon_final}
\eea
Here $\epsilon(x)$ is not an independent scalar field added {\it ad hoc}. Rather, it is determined by the pullback of the projective phase-space geometry together with the additional first-order acceleration invariant retained in the effective spacetime truncation. This distinction is crucial. In many phenomenological modifications of general relativity one starts from an arbitrary scalar deformation, a nonminimal coupling, a running constant, or an effective $f(R)$-type correction \cite{Sotiriou:2008rp,DeFelice:2010aj,Clifton:2011jh}. By contrast, in the present construction the scalar deformation arises from a prior Hamiltonian geometry on $T^{\ast}M$, and therefore inherits strong structural constraints: it is dimensionless, projectively defined, section dependent, and continuously connected to a classical Lorentzian limit.

The effective gravitational dynamics are governed by the conformally reduced Einstein tensor. At first order, the variationally consistent form is
\begin{equation}
(1 + \epsilon) G_{\mu\nu} - \nabla_{\mu} \nabla_{\nu} \epsilon + g_{\mu\nu}\Box\epsilon = 8 \pi G\, T_{\mu\nu},
\label{eq:EFE_final_QG}
\end{equation}
with $\Box \epsilon = g^{\mu\nu} \nabla_{\mu} \nabla_{\nu}\epsilon$. It follows that all deviations from general relativity are controlled by the scalar $\epsilon(x)$ and its derivatives. Structurally, Eq.~\eqref{eq:EFE_final_QG} resembles scalar-tensor and conformally related effective theories, but its interpretation is different: the scalar correction is not a new propagating Brans--Dicke-like degree of freedom, the  scalar-tensor theory of gravitation \cite{PhysRev.124.925}, nor a fundamental inflaton-like field \cite{guth1998inflationary}, but an induced geometric response inherited from the pullback of a deformed phase-space Hamiltonian \cite{Brans:1961sx,Fujii:2003pa,Capozziello:2011et}. This makes the framework closer in spirit to emergent or induced gravity ideas than to conventional scalar-tensor model building \cite{Sakharov:1967pk,Jacobson:1995ab,Padmanabhan:2009vy}.

It is crucial to assure that the classical limit is recovered when $\beta \rightarrow 0$, $\kappa \rightarrow 0$, and $\epsilon \rightarrow 0$. In that limit the Hamiltonian reduces to the standard quadratic mass-shell function, the Hamilton metric becomes the inverse spacetime metric, the section dependence becomes irrelevant, and Eq.~\eqref{eq:EFE_final_QG} reduces to the Einstein field equations. This continuous limit is important from the viewpoint of quantum-gravity phenomenology, because it guarantees that the theory modifies general relativity only through controlled higher-order structure rather than through a discontinuous change of principles. Similar continuity requirements play a central role in loop-quantum-cosmology effective dynamics, asymptotic safety, string-inspired effective actions, noncommutative geometry, and causal-set phenomenology \cite{Ashtekar:2011ni,Bojowald:2006da,Bonanno:2017pkg,Reuter:2012id,Calcagni:2010bj,Connes:1996gi,Henson:2006kf,Dowker:2017ufs}.

A second major implication concerns the status of momentum space in quantum gravity. In the present framework momentum space is not merely the Fourier dual of spacetime, but an intrinsic part of the geometry. The scalar $\Xi$ is constructed on the projectivized cotangent bundle, and the Hamilton metric depends explicitly on momentum directions. This aligns the construction with the increasingly common expectation that quantum gravity may deform the geometry of momentum space, or even elevate phase space itself to the true invariant arena of the theory \cite{Born:1938,Majid:1994cy,Amelino-Camelia:2011yni,Freidel:2011mt,Barcaroli:2015xda,Freidel:2014qna,Hohmann:2018rpp}. In this sense the theory is more naturally compared with relative locality \cite{Amelino-Camelia:2011bm,AmelinoCamelia:2011bm,Freidel:2011mt,Barcaroli:2015xda}, deformed relativistic kinematics, noncommutative phase-space models, generalized uncertainty principle scenarios, and metastring modular spacetime than with canonical metric quantization in the Wheeler--DeWitt sense \cite{Kempf:1994su,Garay:1994en,Ali:2009zq,Doplicher:1994tu,AmelinoCamelia:2000mn,Freidel:2015pka,Freidel:2015uug,Polchinski:1998rq}.

The deformation also modifies causal focusing. Thus, the Raychaudhuri equation in the reduced spacetime sector becomes
\begin{equation}
\nabla_{\tau}\Theta = - \frac{1}{3} \Theta^{2} - \Sigma_{\mu\nu} \Sigma^{\mu\nu} + \Omega_{\mu\nu} \Omega^{\mu\nu} - R_{\mu\nu} U^{\mu} U^{\nu} + \ddot{\epsilon} - \frac12 \Box \epsilon + \epsilon\, R_{\mu\nu} U^{\mu} U^{\nu},
\label{eq:Raychaudhuri_final_QG}
\end{equation}
for the perturbative conformal regime discussed earlier. The additional terms involving derivatives of $\epsilon$ modify the classical focusing condition and can weaken, or in suitable regimes counteract, geodesic convergence \cite{PhysRevD.107.084005}. This is a particularly significant consequence for quantum gravity \cite{w1sd-v69d}. Across a wide range of approaches one expects that the Planck-scale completion of gravity should alter the classical singularity theorems either by modifying the energy conditions, modifying the dynamical equations, or modifying the causal structure itself \cite{Borde:1993xh,Ashtekar:2005cj,Bojowald:2007cd,Battisti:2008gz,Brandenberger:2016vhg,Tawfik:2024gtg,Tawfik:2025icy}. In the present theory this happens through a definite geometric mechanism: the pullback of the projective phase-space deformation contributes derivative terms to the focusing equation. The singularity-softening effect is therefore neither {\it ad hoc} nor purely matter driven; it is a direct consequence of the induced geometry.

This point is especially interesting when compared with other quantum-gravity inspired nonsingular scenarios. In loop quantum cosmology, for example, the singularity is resolved through modified effective Friedmann equations and a quantum bounce \cite{Ashtekar:2006rx,Ashtekar:2011ni,Bojowald:2006da}. In asymptotic safety, high-curvature dynamics may be altered by the renormalization-group running of couplings and higher-curvature operators \cite{Bonanno:2017pkg,Reuter:2019byg}. In string cosmology and duality-based scenarios, the high-energy regime can be softened by duality symmetries and extended degrees of freedom \cite{Gasperini:2002bn,Veneziano:1991ek}. In noncommutative and generalized-uncertainty-principle models, minimal-length or nonlocal structures can alter collapse and early-Universe dynamics \cite{Doplicher:1994tu,Kempf:1994su,Garay:1994en,Ali:2009zq}. In causal-set and discrete approaches, the fundamental microscopic structure changes the continuum description near classical singularities \cite{Henson:2006kf,Dowker:2017ufs}. The present formalism adds a distinct mechanism to this landscape: singularity-softening through section-induced, projective, phase-space geometry.

As discussed in earlier sections, in the cosmological sector, the deformation enters the Friedmann equation through both algebraic and derivative terms,
\begin{equation}
3 H^{2} + 3 H \dot{\epsilon} + 3 \epsilon H^{2} = 8 \pi G \left[ \frac12 (1 + \epsilon) \dot{\varphi}^{2} + (1 + 2 \epsilon) V(\varphi)\right],
\label{eq:Friedmann_QG_final}
\end{equation}
showing that the quantum-gravity correction modifies not only the instantaneous expansion rate but also its time evolution. The resulting inflationary dynamics illustrate an important general lesson: the dominant effect of the deformation is often not a simple algebraic rescaling, but a derivative correction controlled by the time dependence of the induced scalar $\epsilon$. This feature distinguishes the model from many effective modified-gravity descriptions, in which the leading correction is often purely algebraic at background level \cite{Starobinsky:1980te,Sotiriou:2008rp,DeFelice:2010aj}. Here the derivative structure is inherited from the phase-space origin of the deformation and ultimately from the momentum dependence of the underlying Hamiltonian.

At the same time, the slow-roll analysis shows that the number of e-folds remains
\begin{equation}
N \simeq \frac{4 \pi G}{n} \left(\varphi_i^{2} - \varphi_f^{2}\right) + \mathcal{O}\! \left(\frac{\dot{\epsilon}}{H}\right),
\label{eq:efolds_QG_final}
\end{equation}
for a power-law potential at first order, so that purely algebraic $\epsilon$-corrections cancel and only time-dependent contributions from $\epsilon$ can alter the inflationary evolution. This is a nontrivial and rather remarkable result. It indicates that the integrated expansion history is more robust than the local background equations might suggest. The deformation can still change the friction term, the Hubble slow-roll parameter, and the scalar/tensor pump fields, but the leading integrated e-fold count is protected against first-order algebraic corrections. From a quantum-gravity perspective, this is attractive because it permits nontrivial early-Universe corrections while maintaining continuity with successful slow-roll phenomenology.

At the perturbative level, the canonical structure of quantum fields remains unchanged. The modification enters only through background-dependent pump fields,
\begin{equation}
z_{\mathrm{eff}} = z\left(1+\frac{\upsilon_s}{2}\right), \qquad
a_{\mathrm{eff}} = a\left(1+\frac{\upsilon_t}{2}\right),
\label{eq:pump_final_QG}
\end{equation}
with $\upsilon_s \simeq c_s\epsilon$ and $\upsilon_t \simeq c_t \epsilon$ at leading order. Thus quantum-gravity effects manifest themselves as changes in the effective time-dependent mass terms of scalar and tensor perturbations, not as deformations of the equal-time canonical commutators. This distinguishes the present model from approaches in which the perturbation algebra itself is modified, for example through noncommutativity, polymer quantization, trans-Planckian initial-state modifications, or modified commutation relations \cite{Martin:2000xs,Brandenberger:2000wr,Ashtekar:2008zu,Kempf:1994su,Kempf:2000ac}. Here the standard quantum field theory structure on curved spacetime is retained, while the background geometry on which the perturbations propagate has already encoded the quantum-gravity correction. This is conceptually cleaner and technically more stable: the deformation is geometric before it is dynamical.

The present framework also has implications for the weak equivalence principle and the universality of free fall. Because the underlying Hamiltonian is constructed to remain two-homogeneous in the momenta, it sits precisely in the special class of modified-dispersion-relation Hamiltonians for which the weak equivalence principle is not generically spoiled at leading order. This is an important structural advantage. In many deformed-kinematics scenarios, departures from two-homogeneity lead to mass-dependent free fall or to observer-dependent anomalies in low-energy motion. By contrast, the projective construction used here preserves the two-homogeneous Hamiltonian class and thereby controls one of the most delicate phenomenological issues in quantum-gravity model building.

Another implication concerns the relation between the present framework and string-inspired modular ideas. The appearance of a self-dual lattice $\Lambda$, a neutral bilinear form $\eta$, and a periodic modular response $\Xi_{\mathrm{mod}}$ places the theory in close conceptual proximity to modular spacetime and metastring theory, where phase space, duality, and modular variables play a foundational role \cite{Freidel:2014qna,Freidel:2015pka,Freidel:2015uug}. However, the present construction differs in emphasis. Rather than deriving a worldsheet theory or a full microscopic string background, it uses modular kinematics semiclassically to generate an effective deformation of the gravitational Hamiltonian. It is therefore best interpreted as a low-energy geometric imprint of modular quantum kinematics rather than as a full ultraviolet completion. This intermediate status may be useful: it provides a bridge between highly formal modular/stringy ideas and effective gravitational phenomenology.

The framework presented here also sheds light on how one might compare different quantum-gravity programs. Canonical quantization of the metric, loop quantum gravity, asymptotic safety, causal sets, group field theory, noncommutative geometry, generalized uncertainty principle models, causal dynamical triangulations, and string theory all differ radically in their microscopic starting points \cite{Rovelli:2004tv,Thiemann:2007zz,Ashtekar:2004eh,Perez:2012wv,Bonanno:2017pkg,Reuter:2019byg,Henson:2006kf,Ambjorn:2005db,Oriti:2016qtz,Connes:1994yd,Doplicher:1994tu,Garay:1994en,Polchinski:1998rq}. Yet many of them suggest, at least heuristically, that the classical manifold description should break down first in the relation between position and momentum, locality and energy, or geometry and kinematics. The present theory gives one precise implementation of that general expectation. It keeps the spacetime manifold $M$ as the base, but promotes the physically relevant geometry to the cotangent bundle and recovers spacetime only after a section choice and symmetry reduction. In this sense it may serve as an effective meeting point between seemingly different quantum-gravity intuitions.

There are, however, clear limitations, and these limitations are themselves informative from the standpoint of quantum gravity. \begin{enumerate}
\item First, the section $\sigma$ is not dynamically derived. It is treated as an externally specified effective congruence. This means that the theory, in its present form, is a reduced semiclassical framework rather than a complete microscopic model. A more complete theory would either derive $\sigma$ dynamically from the matter sector, from a variational principle on the projectivized cotangent bundle, or from a consistency condition relating observers and modular cells. 
\item Second, the form of $\Xi(x,[p])$ depends on the chosen modular data $(\eta,\Lambda)$. While the general structural constraints are strong, the specific coefficients and lattice sectors remain model dependent. 
\item Third, the null sector $Q=0$ is excluded from the present construction because the normalized projective momentum $u_{\mu}=p_{\mu}/\sqrt{|Q|}$ becomes singular there.
\end{enumerate} 
A fully satisfactory quantum-gravity account would need a boundary or limiting treatment of this sector, especially if one wishes to discuss exact light propagation or primordial tensor modes at a more fundamental level.

These limitations nevertheless point naturally toward future developments. \begin{enumerate}
\item One direction is to promote the Hamiltonian itself to a dynamical scalar field on the cotangent bundle and derive field equations directly on phase space, as has recently been explored in Hamilton-geometric action principles \cite{Pfeifer:2019wus}. 
\item A second direction is to derive the section $\sigma$ from the matter flow or from a variational principle. 
\item A third is to connect the effective coefficients in $\Xi$ with microscopic modular or duality data in metastring-like theories. 
\item A fourth is to study phenomenology beyond FLRW symmetry, including anisotropic cosmologies, black-hole congruences, gravitational-wave propagation, and weak-lensing observables.
\end{enumerate} 
In all such directions, the main virtue of the present construction is that it already provides a geometrically consistent dictionary between quantum phase-space structure and effective spacetime gravity.

In summary, the framework developed here shows that quantum-gravity effects can be consistently encoded by a deformation of projective phase-space geometry whose pullback induces an effective spacetime metric. The resulting theory is derived rather than postulated on spacetime, modifies gravitational dynamics through a structurally constrained scalar deformation, alters geodesic focusing and early-Universe evolution through derivative geometric terms, preserves the standard canonical algebra of cosmological perturbations, and remains continuously connected to classical general relativity. Its central message is that a significant class of quantum-gravity effects may be understood not as direct fluctuations of the spacetime metric, but as semiclassical consequences of a deeper modular and projective geometry on phase space \cite{Freidel:2014qna,Freidel:2015pka,Freidel:2015uug,Amelino-Camelia:2011yni,Freidel:2011mt,Barcaroli:2015xda,Hohmann:2018rpp}. This makes the theory both conceptually distinctive and phenomenologically tractable within the broader landscape of quantum-gravity research.

\section{Conclusion and outlook}
\label{sec:Conclusion}

In this work, we have constructed a phase-space formulation of quantum-deformed gravity in which the fundamental geometric structure is defined on the cotangent bundle rather than on spacetime itself. The deformation is introduced at the level of the Hamiltonian through a projectively well-defined, zero-homogeneous scalar, leading to an anisotropic Hamilton geometry on a conic phase-space domain. The effective spacetime description emerges only after a section pullback and subsequent symmetry reduction, thereby realizing explicitly the idea that spacetime geometry may be an emergent structure rather than a fundamental one \cite{Rovelli:2004tv,Thiemann:2007zz,Connes:1994yd,Majid:1994cy,Verlinde:2010hp,VanRaamsdonk:2010pw}.

A key structural result is that the induced spacetime dynamics can be expressed in terms of a single scalar deformation inherited from the underlying phase-space geometry. This scalar encodes quantum-gravity effects without introducing additional propagating degrees of freedom, in contrast with many modified gravity scenarios \cite{Clifton:2011jh,Sotiriou:2008rp,Capozziello:2011et}. The resulting framework modifies the Einstein field equations through derivative and algebraic contributions of the deformation while preserving a well-defined classical limit. In the cosmological sector, this leads to a conformally deformed FLRW geometry with corrected background and perturbation dynamics, where the leading effects arise from the time dependence of the deformation rather than from simple rescalings.

From a broader perspective, the construction provides a concrete realization of the idea that quantum-gravity effects may be encoded in phase-space structures, a viewpoint that has appeared in different guises in doubly special relativity, relative locality, and generalized uncertainty principle approaches \cite{AmelinoCamelia:2000mn,Magueijo:2001cr,Girelli:2006sc,Amelino-Camelia:2011yni,Kempf:1994su,Garay:1994en,Ali:2009zq,Tawfik:2024gxc}. It also resonates with noncommutative geometry and quantum spacetime proposals, where spacetime coordinates acquire an effective nonlocal or operatorial structure at short distances \cite{Doplicher:1994tu,Connes:1996gi}. In parallel, the emergence of effective spacetime dynamics from deeper microscopic structures is a common theme in loop quantum gravity, causal set theory, asymptotic safety, and group field theory \cite{Ashtekar:2004eh,Ashtekar:2011ni,Perez:2012wv,Henson:2006kf,Ambjorn:2005db,Oriti:2016qtz,Reuter:2012id,Bonanno:2017pkg}. The present framework differs from these approaches in that it realizes this emergence through a controlled geometric deformation of Hamiltonian phase space rather than through a direct quantization of spacetime itself.

An important implication of the theory is that quantum-gravity corrections manifest primarily through background-dependent quantities, such as the effective pump fields governing scalar and tensor perturbations, while the canonical commutation relations remain unchanged. This feature aligns with semiclassical expectations and with approaches in which quantum geometry modifies effective dynamics without altering the underlying field algebra \cite{Ashtekar:2008zu,Bojowald:2006da,Calcagni:2010bj}. It also provides a mechanism for softening classical singular behavior via modified focusing conditions, in agreement with various quantum-gravity inspired scenarios for singularity resolution \cite{Borde:1993xh,Ashtekar:2005cj,Bojowald:2007cd,Battisti:2008gz,Brandenberger:2016vhg}.

Several open problems remain and point toward future research directions. A central issue is the dynamical origin of the section selecting the physical momentum configuration, which is currently imposed at the level of the reduced theory. A deeper understanding of this selection mechanism may require embedding the present construction into a fully quantum framework with a well-defined notion of states on phase space. In addition, the precise form of the deformation scalar depends on the underlying modular or lattice data, whose microscopic origin remains to be clarified. Establishing a connection with more fundamental formulations of quantum gravity, such as spin networks, group field condensates, or noncommutative algebras, would therefore be of particular interest \cite{Rovelli:2004tv,Oriti:2016qtz,Connes:1994yd}.

Further developments should also extend the analysis beyond the isotropic conformal sector to include genuinely anisotropic configurations, where the full Hamilton geometry becomes relevant. This may lead to novel phenomenology in gravitational wave propagation, lensing, or high-energy astrophysical processes. On the observational side, precision cosmology provides a natural testing ground for constraining the deformation parameters through their impact on inflationary observables, including spectral indices and tensor-to-scalar ratios \cite{Brandenberger:2000wr,Kempf:2000ac}.

In summary, the framework developed here demonstrates that quantum-gravity effects can be consistently encoded as deformations of phase-space geometry, with spacetime emerging as an effective construct obtained through pullback and symmetry reduction. This perspective offers a geometrically coherent bridge between quantum kinematics and macroscopic gravitational dynamics, and opens a promising avenue for exploring the interplay between phase-space structure and the emergence of spacetime in quantum gravity.

\section*{Conflicts of Interest}

The authors declare that there are no conflicts of interest regarding the publication of the published article!

\section*{Dataset Availability}

All data generated or analyzed during this study are included in this published article. All of the material is owned by the authors.

\section*{Author contributions}
The responsibility for proposing the conception of the present study lies with AT, who also undertook the tasks of designing and managing the research, interpreting the results, and preparing the manuscript. SKS was responsible for deriving the expressions and proposing physical interpretations. SOA, AAA, MN, AFE contributed to the writing and proofreading of the manuscript. The final version of the manuscript was unanimously approved by all authors.

\section*{Funding}

The authors declare that this research received no specific grants from any funding agency in the public, commercial, or not-for-profit sectors.

\section*{Competing interests}

The authors confirm that there are no relevant financial or non-financial competing interests to report.

\bibliography{RefsGWB_noduplicates}
\bibliographystyle{utphys}

\end{document}